\documentclass[11pt,fleqn]{book} 
\usepackage{pdfpages,amsmath}
\usepackage{natbib}

\usepackage[top=3cm,bottom=3cm,left=3cm,right=3cm,headsep=10pt,letterpaper]{geometry} 

\usepackage{graphicx} 
\graphicspath{{}} 

\usepackage{lipsum} 

\usepackage{tikz} 

\usepackage[english]{babel} 

\usepackage{enumitem} 
\setlist{nolistsep} 

\usepackage{booktabs} 

\usepackage{xcolor}
\definecolor{ocre}{RGB}{243,102,25} 

\definecolor{MidnightBlue}{RGB}{25,25,112}

\usepackage{avant} 
\usepackage{mathptmx} 

\usepackage{microtype} 
\usepackage[utf8]{inputenc} 
\usepackage[T1]{fontenc} 

\usepackage{calc} 
\usepackage{makeidx} 
\makeindex 

\usepackage{titletoc} 

\contentsmargin{0cm} 

\titlecontents{part}[0cm]
{\addvspace{20pt}\centering\large\bfseries}
{}
{}
{}

\titlecontents{chapter}[1.25cm] 
{\addvspace{12pt}\large\sffamily\bfseries} 
{\color{MidnightBlue}\contentslabel[\Large\thecontentslabel]{1.25cm}\color{MidnightBlue}} 
{\color{MidnightBlue}}  
{\color{MidnightBlue}\normalsize\;\titlerule*[.5pc]{.}\;\thecontentspage} 

\titlecontents{section}[1.25cm] 
{\addvspace{3pt}\sffamily\bfseries} 
{\contentslabel[\thecontentslabel]{1.25cm}} 
{}
{\hfill\color{black}\thecontentspage} 
[]

\titlecontents{subsection}[1.25cm] 
{\addvspace{1pt}\sffamily\small} 
{\contentslabel[\thecontentslabel]{1.25cm}} 
{}
{\ \titlerule*[.5pc]{.}\;\thecontentspage} 
[]

\titlecontents{figure}[0em]
{\addvspace{-5pt}\sffamily}
{\thecontentslabel\hspace*{1em}}
{}
{\ \titlerule*[.5pc]{.}\;\thecontentspage}
[]

\titlecontents{table}[0em]
{\addvspace{-5pt}\sffamily}
{\thecontentslabel\hspace*{1em}}
{}
{\ \titlerule*[.5pc]{.}\;\thecontentspage}
[]

\titlecontents{lchapter}[0em] 
{\addvspace{15pt}\large\sffamily\bfseries} 
{\color{MidnightBlue}\contentslabel[\Large\thecontentslabel]{1.25cm}\color{MidnightBlue}} 
{}  
{\color{MidnightBlue}\normalsize\sffamily\bfseries\;\titlerule*[.5pc]{.}\;\thecontentspage} 

\titlecontents{lsection}[0em] 
{\sffamily\small} 
{\contentslabel[\thecontentslabel]{1.25cm}} 
{}
{}

\titlecontents{lsubsection}[.5em] 
{\normalfont\footnotesize\sffamily} 
{}
{}
{}

\usepackage{fancyhdr} 

\fancypagestyle{plain}{\fancyhead{}} 

\makeatletter
\renewcommand{\cleardoublepage}{
\clearpage\ifodd\c@page\else
\hbox{}
\vspace*{\fill}
\thispagestyle{empty}
\newpage
\fi}

\usepackage{amsmath,amsfonts,amssymb,amsthm} 

\newtheoremstyle{ocrenumbox}
{0pt}
{0pt}
{\normalfont}
{}
{\small\bf\sffamily\color{MidnightBlue}}
{\;}
{0.25em}
{\small\sffamily\color{MidnightBlue}\thmname{#1}\nobreakspace\thmnumber{\@ifnotempty{#1}{}\@upn{#2}}
\thmnote{\nobreakspace\the\thm@notefont\sffamily\bfseries\color{black}---\nobreakspace#3.}} 

\newtheoremstyle{blacknumex}
{5pt}
{5pt}
{\normalfont}
{} 
{\small\bf\sffamily}
{\;}
{0.25em}
{\small\sffamily{\tiny\ensuremath{\blacksquare}}\nobreakspace\thmname{#1}\nobreakspace\thmnumber{\@ifnotempty{#1}{}\@upn{#2}}
\thmnote{\nobreakspace\the\thm@notefont\sffamily\bfseries---\nobreakspace#3.}}

\newtheoremstyle{blacknumbox} 
{0pt}
{0pt}
{\normalfont}
{}
{\small\bf\sffamily}
{\;}
{0.25em}
{\small\sffamily\thmname{#1}\nobreakspace\thmnumber{\@ifnotempty{#1}{}\@upn{#2}}
\thmnote{\nobreakspace\the\thm@notefont\sffamily\bfseries---\nobreakspace#3.}}

\newtheoremstyle{ocrenum}
{5pt}
{5pt}
{\normalfont}
{}
{\small\bf\sffamily\color{MIdnightBlue}}
{\;}
{0.25em}
{\small\sffamily\color{MidnightBlue}\thmname{#1}\nobreakspace\thmnumber{\@ifnotempty{#1}{}\@upn{#2}}
\thmnote{\nobreakspace\the\thm@notefont\sffamily\bfseries\color{black}---\nobreakspace#3.}} 
\makeatother

\newcounter{dummy} 
\numberwithin{dummy}{section}
\theoremstyle{ocrenumbox}
\newtheorem{theoremeT}[dummy]{Theorem}

\newtheorem{exerciseT}{Exercise}[chapter]
\theoremstyle{blacknumex}
\newtheorem{exampleT}{Example}[chapter]
\theoremstyle{blacknumbox}

\newtheorem{definitionT}{Definition}[section]
\newtheorem{corollaryT}[dummy]{Corollary}
\theoremstyle{ocrenum}

\RequirePackage[framemethod=default]{mdframed} 

\newmdenv[skipabove=7pt,
skipbelow=7pt,
backgroundcolor=black!5,
linecolor=MidnightBlue,
innerleftmargin=5pt,
innerrightmargin=5pt,
innertopmargin=5pt,
leftmargin=0cm,
rightmargin=0cm,
innerbottommargin=5pt]{tBox}

\newmdenv[skipabove=7pt,
skipbelow=7pt,
rightline=false,
leftline=true,
topline=false,
bottomline=false,
backgroundcolor=MidnightBlue!10,
linecolor=MidnightBlue,
innerleftmargin=5pt,
innerrightmargin=5pt,
innertopmargin=5pt,
innerbottommargin=5pt,
leftmargin=0cm,
rightmargin=0cm,
linewidth=4pt]{eBox}	

\newmdenv[skipabove=7pt,
skipbelow=7pt,
rightline=false,
leftline=true,
topline=false,
bottomline=false,
linecolor=MidnightBlue,
innerleftmargin=5pt,
innerrightmargin=5pt,
innertopmargin=0pt,
leftmargin=0cm,
rightmargin=0cm,
linewidth=4pt,
innerbottommargin=0pt]{dBox}	

\newmdenv[skipabove=7pt,
skipbelow=7pt,
rightline=false,
leftline=true,
topline=false,
bottomline=false,
linecolor=gray,
backgroundcolor=black!5,
innerleftmargin=5pt,
innerrightmargin=5pt,
innertopmargin=5pt,
leftmargin=0cm,
rightmargin=0cm,
linewidth=4pt,
innerbottommargin=5pt]{cBox}

\makeatletter
\renewcommand{\@seccntformat}[1]{\llap{\textcolor{MidnightBlue}{\csname the#1\endcsname}\hspace{1em}}}                    
\renewcommand{\section}{\@startsection{section}{1}{\z@}
{-4ex \@plus -1ex \@minus -.4ex}
{1ex \@plus.2ex }
{\normalfont\large\sffamily\bfseries}}
\renewcommand{\subsection}{\@startsection {subsection}{2}{\z@}
{-3ex \@plus -0.1ex \@minus -.4ex}
{0.5ex \@plus.2ex }
{\normalfont\sffamily\bfseries}}
\renewcommand{\subsubsection}{\@startsection {subsubsection}{3}{\z@}
{-2ex \@plus -0.1ex \@minus -.2ex}
{.2ex \@plus.2ex }
{\normalfont\small\sffamily\bfseries}}                        
\renewcommand\paragraph{\@startsection{paragraph}{4}{\z@}
{-2ex \@plus-.2ex \@minus .2ex}
{.1ex}
{\normalfont\small\sffamily\bfseries}}

\newcommand{\@mypartnumtocformat}[2]{%
\setlength\fboxsep{0pt}%
\noindent\colorbox{MidnightBlue!20}{\strut\parbox[c][.7cm]{\ecart}{\color{MidnightBlue!70}\Large\sffamily\bfseries\centering#1}}\hskip\esp\colorbox{MidnightBlu!40}{\strut\parbox[c][.7cm]{\linewidth-\ecart-\esp}{\Large\sffamily\centering#2}}}%
\newcommand{\@myparttocformat}[1]{%
\setlength\fboxsep{0pt}%
\noindent\colorbox{MidnightBlue}{\strut\parbox[c][.7cm]{\linewidth}{\Large\sffamily\centering#1}}}%
\newlength\esp
\newlength\ecart
\def\@part[#1]#2{%
\ifnum \c@secnumdepth >-2\relax%
\refstepcounter{part}%
\addcontentsline{toc}{part}{\texorpdfstring{\protect\@mypartnumtocformat{\thepart}{#1}}{\partname~\thepart\ ---\ #1}}
\else%
\addcontentsline{toc}{part}{\texorpdfstring{\protect\@myparttocformat{#1}}{#1}}%
\fi%
\startcontents%
\markboth{}{}%
{\thispagestyle{empty}%
\begin{tikzpicture}[remember picture,overlay]%
\node at (current page.north west){\begin{tikzpicture}[remember picture,overlay]%
\fill[MidnightBlue!20](0cm,0cm) rectangle (\paperwidth,-\paperheight);
\node[anchor=north] at (4cm,-3.25cm){\color{MidnightBlue!40}\fontsize{220}{100}\sffamily\bfseries\thepart};
\node[anchor=south east] at (\paperwidth-1cm,-\paperheight+1cm){\parbox[t][][t]{8.5cm}{
\printcontents{l}{0}{\setcounter{tocdepth}{1}}%
}};
\node[anchor=north east] at (\paperwidth-1.5cm,-3.25cm){\parbox[t][][t]{15cm}{\strut\raggedleft\color{white}\fontsize{30}{30}\sffamily\bfseries#2}};
\end{tikzpicture}};
\end{tikzpicture}}%
\@endpart}
\def\@spart#1{%
\startcontents%
\phantomsection
{\thispagestyle{empty}%
\begin{tikzpicture}[remember picture,overlay]%
\node at (current page.north west){\begin{tikzpicture}[remember picture,overlay]%
\fill[MidnightBlue!20](0cm,0cm) rectangle (\paperwidth,-\paperheight);
\node[anchor=north east] at (\paperwidth-1.5cm,-3.25cm){\parbox[t][][t]{15cm}{\strut\raggedleft\color{white}\fontsize{30}{30}\sffamily\bfseries#1}};
\end{tikzpicture}};
\end{tikzpicture}}
\addcontentsline{toc}{part}{\texorpdfstring{%
\setlength\fboxsep{0pt}%
\noindent\protect\colorbox{ocre!40}{\strut\protect\parbox[c][.7cm]{\linewidth}{\Large\sffamily\protect\centering #1\quad\mbox{}}}}{#1}}%
\@endpart}
\def\@endpart{\vfil\newpage
\if@twoside
\if@openright
\null
\thispagestyle{empty}%
\newpage
\fi
\fi
\if@tempswa
\twocolumn
\fi}

\newif\ifusechapterimage
 \usechapterimagetrue
 \newcommand{\thechapterimage}{}%
 \newcommand{\chapterimage}[1]{\ifusechapterimage\renewcommand{\thechapterimage}{#1}\fi}%
 \newcommand{\autodot}{.}
 \def\@makechapterhead#1{%
 {\parindent \z@ \raggedright \normalfont
 \ifnum \c@secnumdepth >\m@ne
  \if@mainmatter
   \begin{tikzpicture}[remember picture,overlay]
   \node at (current page.north west)
   {\begin{tikzpicture}[remember picture,overlay]
   \node[anchor=north west,inner sep=0pt] at (0,0) {\ifusechapterimage\includegraphics[width=\paperwidth]{\thechapterimage}\fi};
   \draw[anchor=west] (\Gm@lmargin,-9cm) node [line width=2pt,rounded corners=15pt,draw=ocre,fill=white,fill opacity=0.5,inner sep=15pt]{\strut\makebox[22cm]{}};
   \draw[anchor=west] (\Gm@lmargin+.3cm,-9cm) node {\huge\sffamily\bfseries\color{black}\thechapter\autodot~#1\strut};
   \end{tikzpicture}};
   \end{tikzpicture}
  \else
   \begin{tikzpicture}[remember picture,overlay]
   \node at (current page.north west)
   {\begin{tikzpicture}[remember picture,overlay]
   \node[anchor=north west,inner sep=0pt] at (0,0) {\ifusechapterimage\includegraphics[width=\paperwidth]{\thechapterimage}\fi};
   \draw[anchor=west] (\Gm@lmargin,-9cm) node [line width=2pt,rounded corners=15pt,draw=ocre,fill=white,fill opacity=0.5,inner sep=15pt]{\strut\makebox[22cm]{}};
   \draw[anchor=west] (\Gm@lmargin+.3cm,-9cm) node {\huge\sffamily\bfseries\color{black}#1\strut};
   \end{tikzpicture}};
   \end{tikzpicture}
  \fi
 \fi\par\vspace*{270\p@}}}

\def\@makeschapterhead#1{%
\begin{tikzpicture}[remember picture,overlay]
\node at (current page.north west)
{\begin{tikzpicture}[remember picture,overlay]
\node[anchor=north west,inner sep=0pt] at (0,0) {\ifusechapterimage\includegraphics[width=\paperwidth]{\thechapterimage}\fi};
\draw[anchor=west] (\Gm@lmargin,-9cm) node [line width=2pt,rounded corners=15pt,draw=ocre,fill=white,fill opacity=0.5,inner sep=15pt]{\strut\makebox[22cm]{}};
\draw[anchor=west] (\Gm@lmargin+.3cm,-9cm) node {\huge\sffamily\bfseries\color{black}#1\strut};
\end{tikzpicture}};
\end{tikzpicture}
\par\vspace*{270\p@}}
\makeatother

\usepackage{hyperref}
\hypersetup{hidelinks,backref=true,pagebackref=true,hyperindex=true,colorlinks=false,breaklinks=true,urlcolor= ocre,bookmarks=true,bookmarksopen=false,pdftitle={Title},pdfauthor={Author}}
\usepackage{bookmark}
\bookmarksetup{
open,
numbered,
addtohook={%
\ifnum\bookmarkget{level}=0 
\bookmarksetup{bold}%
\fi
\ifnum\bookmarkget{level}=-1 
\bookmarksetup{color=ocre,bold}%
\fi
}
}

\bibpunct[, ]{(}{)}{;}{a}{}{,}

\begin{document}

\pagenumbering{roman}

\begingroup
\thispagestyle{empty}
\begin{tikzpicture}[remember picture,overlay]
  \node[inner sep=0pt] (background) at (current page.center) {%
    \includegraphics[width=\paperwidth]{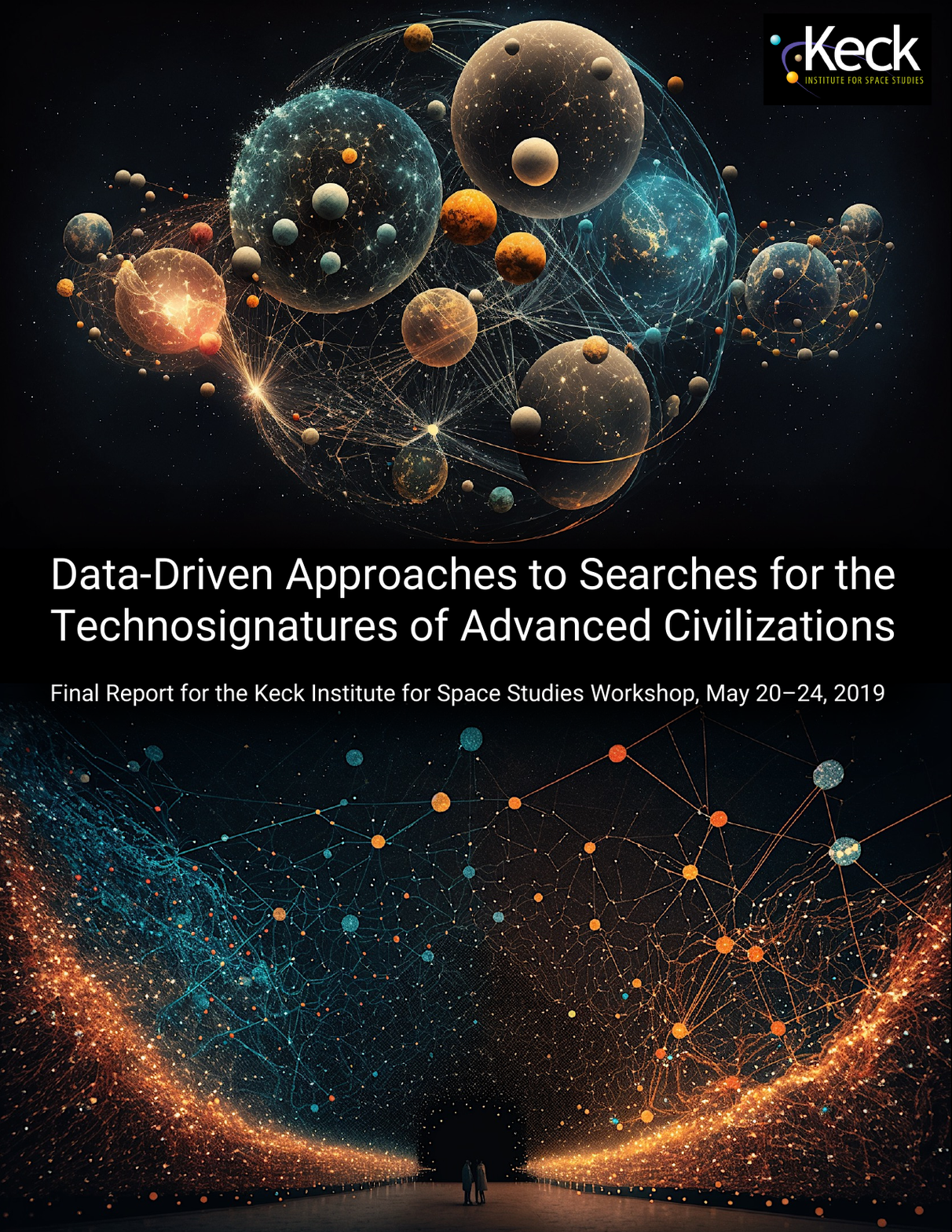}};
\end{tikzpicture}
\vfill
\endgroup

\newpage
\thispagestyle{empty}
\vspace*{0.25in}
\begin{center}
{\Large \textbf{Data-Driven Approaches to Searches for \\ the Technosignatures of Advanced Civilizations}}
\end{center}
\vfill
\noindent
Final Report prepared for the Keck Institute for Space Studies (KISS) \\
Jet Propulsion Laboratory, California Institute of Technology\\
\url{http://kiss.caltech.edu/workshops/technosignatures/technosignatures.html}\\

\noindent%
Recommended citation (long form):\\
Lazio, T.~J.~W., Djorgovski, S.~G., Howard, A., Cutler, C., Sheikh,
S.~Z., Cavuoti, S., Herzing, D., Wagstaff, K., Wright, J., Gajjar,
V.~R., Hand, K., Rebbapragada, U., Allen, B., Cartmill, E., Foster,
J., Gelino, D., Graham, M.~J., Longo, G., Mahabal, A.~A., Pachter, L.,
Ravi, V., \& Sussman, G.  2023, 
``Data-Driven Approaches to Searches for the Technosignatures of
Advanced Civilizations,'' report prepared for the W.~M.~Keck Institute
of Space Studies (KISS), California Institute of Technology, eds.\
T.~J.~W.~Lazio, S.~G.~Djorgovski, A.~Howard, \& C.~Cutler; doi:
10.26206/gvmj-sn65

\bigskip
\noindent%
Recommended citation (short form):\\
Lazio, T.~J.~W., Djorgovski, S.~G., Howard, A., \& Cutler, C.\ (eds.)
2023, 
``Data-Driven Approaches to Searches for the Technosignatures of
Advanced Civilizations,'' report prepared for the W.~M.~Keck Institute
of Space Studies (KISS), California Institute of Technology: 10.26206/gvmj-sn65

\vspace{0.1in}
\noindent
Study Workshop: 2019 May~20--24\\

\vspace{0.1in}
\noindent
\copyright\ 2023.  All rights reserved.

\newpage
\pagestyle{fancy} 
\vspace*{0.25in}
\begin{center}
{\Large \textbf{Authors and Study Participants}}
\end{center}

\vspace*{0.5in}
\begin{center}
\includegraphics[width=\textwidth]{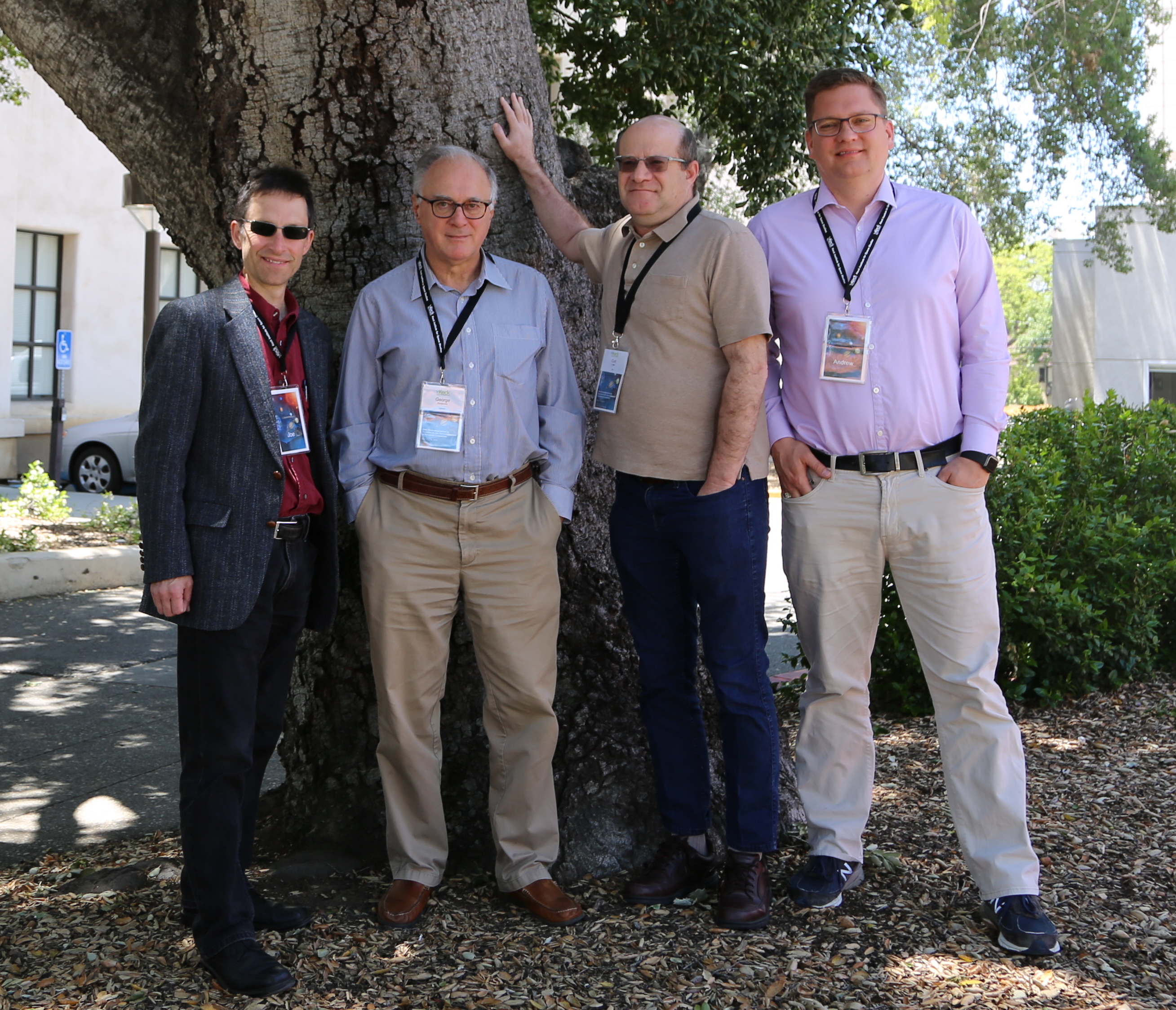}
\vspace*{0.1in}
Study leads (\textit{from left}): Joseph Lazio (JPL/Caltech), George Djorgovski (Caltech), 
Curt~Cutler~(JPL/Caltech), Andrew Howard (Caltech)
\end{center}

\newpage

\begin{center}
\includegraphics[width=\textwidth]{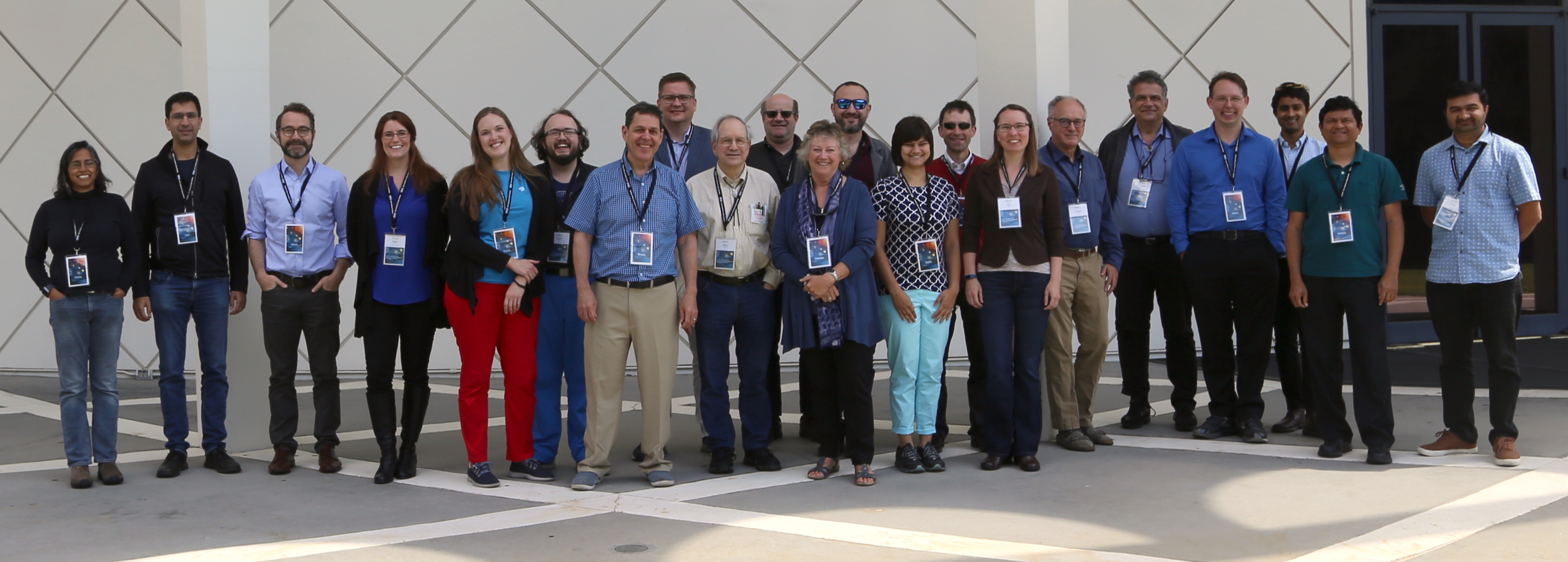}
\vspace*{0.1in}
Participants of the Data-Driven Approaches to Workshop on Searches for
the Technosignatures of Advanced Civilizations held at the Keck
Institute for Space Studies, California Institute of Technology, 2019~May~20--24\\
From Left: 
Umaa~Rebbapragada,
Lior Pachter,
Kevin~Hand,
Dawn~Gelino,
Erica~Cartmill,
Jacob~Foster,
Bruce Allen,
Andrew Howard,
Gerry Sussman,
Curt Cutler,
Denise Herzing,
Stefano Cavuoti,
Sofia~Sheikh,
Joseph~Lazio,
Kiri~Wagstaff,
George~Djorgovski,
Guiseppe~Longo,
Jason~Wright,
Vikram~Ravi,
Ashish~Mahabal,
V.~Gajjar; 
Not pictured: M.~Graham.
\end{center}

\clearpage

\begin{table}
\centering
\begin{tabular}{ll}
Curt Cutler       & Jet Propulsion Laboratory, California Institute of Technology\\
George Djorgovski & California Institute of Technology\\
Andrew Howard     & California Institute of Technology\\
Joseph Lazio      & Jet Propulsion Laboratory, California Institute of Technology\\
Sofia Z.~Sheikh   & SETI Institute \\
Stefano Cavuoti   & University of Naples Federico~II\\
Denise Herzing    & Wild Dolphin Project/Florida Atlantic University\\
Kiri Wagstaff     & Jet Propulsion Laboratory, California Institute of Technology\\
Jason Wright      & Pennsylvania State University\\
Vishal Rasiklal Gajjar & University of California, Berkeley\\
Kevin Hand        & Jet Propulsion Laboratory, California Institute of Technology\\
Umaa Rebbapragada & Jet Propulsion Laboratory, California Institute of Technology\\
Bruce Allen       & Max Planck Institut for Gravitational Physics\\
Erica Cartmill    & University of California, Los Angeles\\
Jacob Foster      & University of California, Los Angeles\\
Dawn Gelino       & NASA Exoplanet Science Institute, \\
                  & \hspace*{1em}Infrared Processing and Analysis Center, California Institute of Technology\\
Matthew Graham    & California Institute of Technology\\
Giuseppe Longo    & University of Naples Federico~II\\
Ashish Mahabal    & California Institute of Technology\\
Lior Pachter      & California Institute of Technology\\
Vikram Ravi       & California Institute of Technology\\
Gerry Sussman     & Massachusetts Institute of Technology\\
\end{tabular}
\end{table}

\newpage
\vspace*{0.25in}
\begin{center}
{\Large \textbf{Acknowledgements}}
\end{center}
\vspace{0.5in}

The study ``Data-Driven Approaches to Searches for  the Technosignatures of Advanced Civilizations'' was made possible by the W.~M.~Keck Institute for Space Studies, and by the Jet Propulsion Laboratory, California Institute of Technology, under contract with the National Aeronautics and Space Administration.

The study leads gratefully acknowledge the outstanding support of Michele Judd, Executive Director of the Keck Institute of Space Studies, as well as her dedicated staff, who made the study experience invigorating and enormously productive.  Many thanks are also due to Tom Prince and the KISS Steering Committee for seeing the potential of our study concept and selecting it.

We thank all of the workshop participants for their time, enthusiasm, and contributions to the workshop and this report.  The workshop was a memorable experience and set the stage for fruitful collaborations between people who would likely not have crossed paths were it not for the Keck Institute for Space Studies.

Cover and chapter images created by Robert Hurt, using
Midjourney \hbox{AI}.

\newpage
\chapterimage{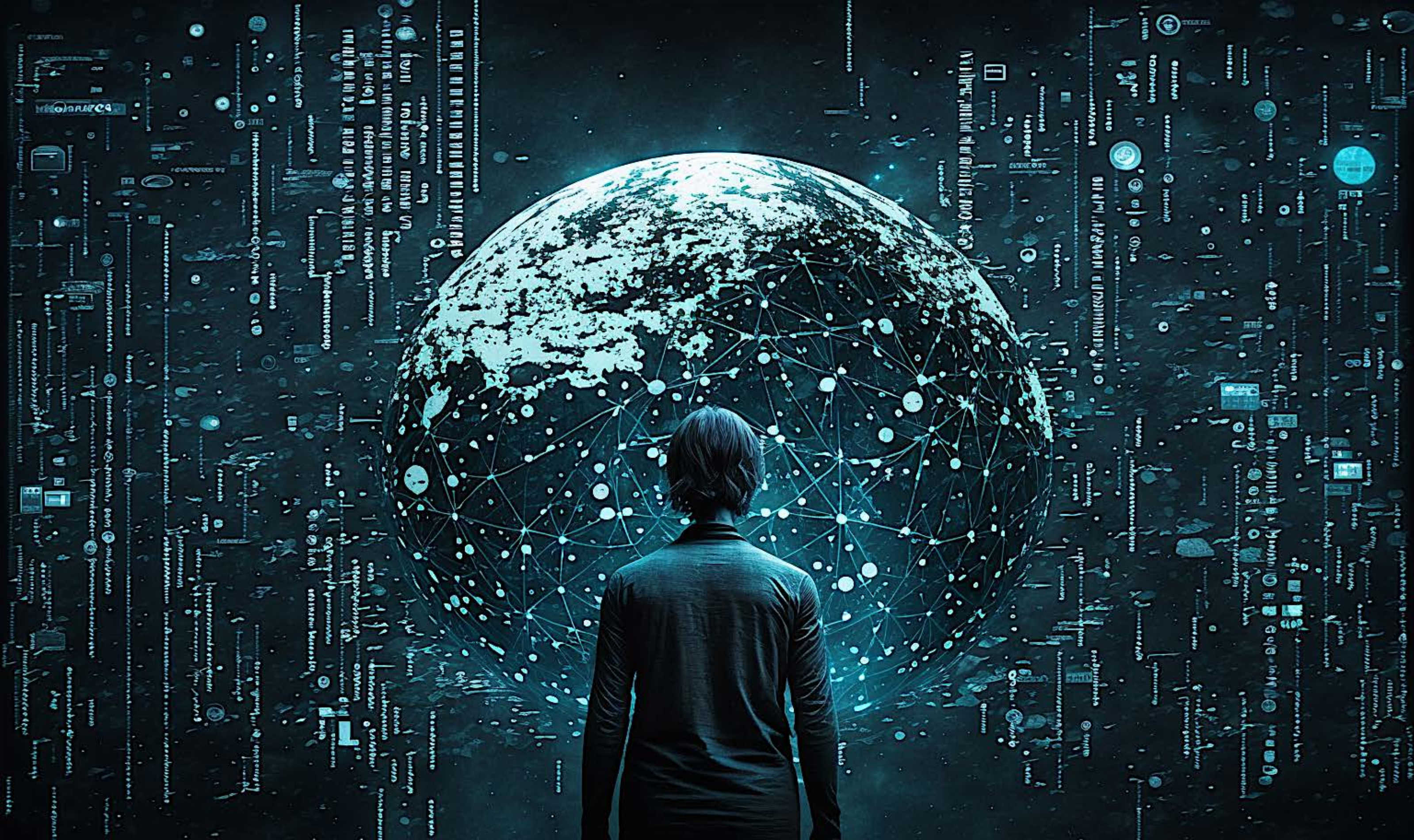}
\pagestyle{empty} 
\tableofcontents 
\cleardoublepage 
\pagestyle{fancy} 

\pagenumbering{arabic}

\chapterimage{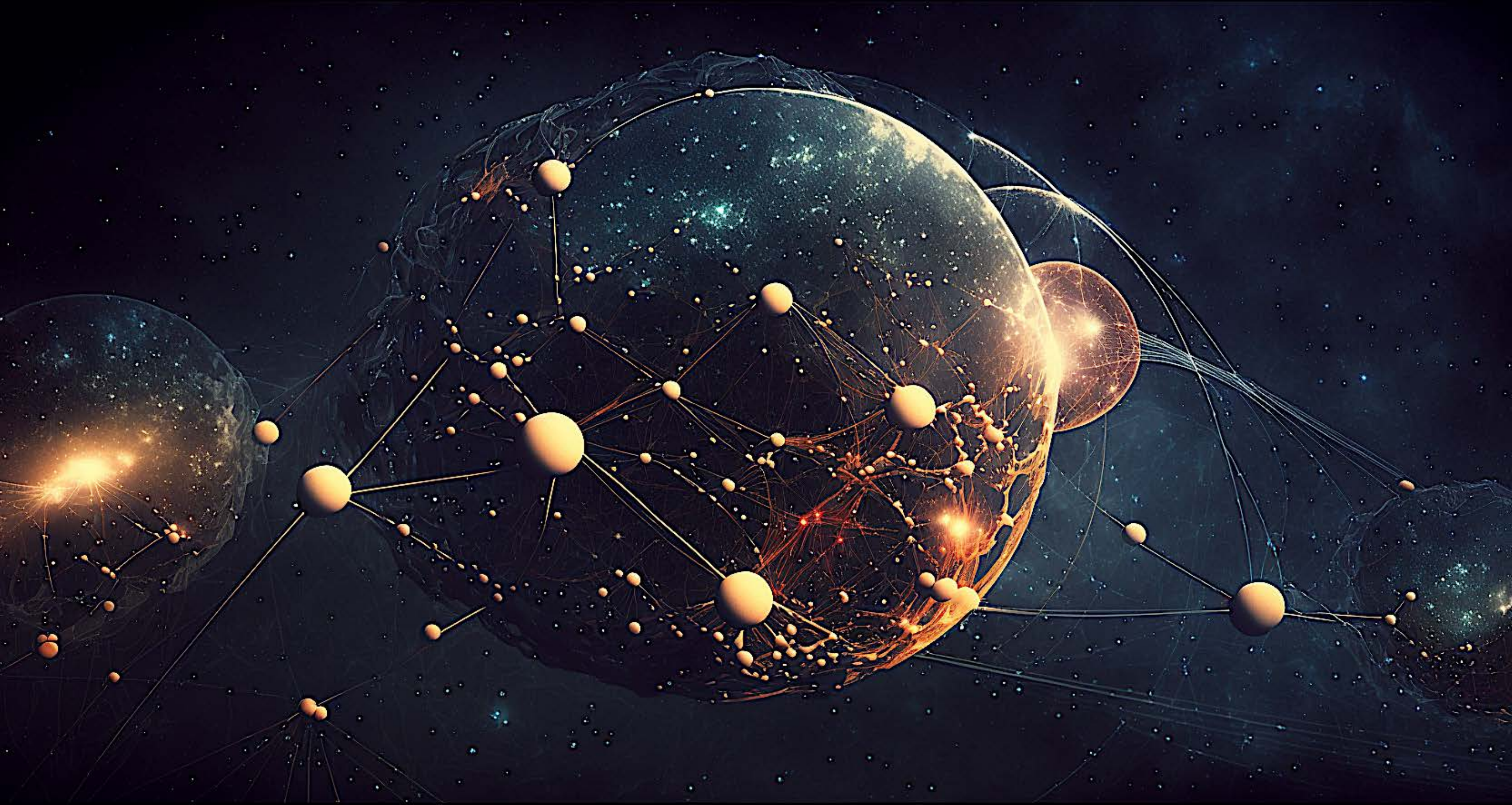}
\chapter*{Executive Summary}
\phantomsection
\addcontentsline{toc}{chapter}{\textcolor{MidnightBlue}{Executive Summary}}

\chapterimage{CoverImages/pdf/ExecutiveSummary.pdf} 

Humanity has wondered whether we are alone and about the existence of ``others'' for millennia.
The discovery of life elsewhere in the Universe, particularly
intelligent life, would have profound scientific, cultural, and
societal effects, comparable to those of recognizing that the Earth is
not the center of the Universe and that humans (\textit{Homo sapiens})
evolved from previous species.

The past two decades have witnessed rapid growths in both the fields of extrasolar planets and data-driven astronomy.
In a relatively short interval, we have seen a change from knowing of
no extrasolar planets to now knowing far more potentially habitable
extrasolar planets than there are planets in the Solar System.  In
approximately the same interval, astronomy has transitioned from a
relatively data-starved field into one in which extensive sky surveys
can generate 1~quadrillion bytes ($=10^{15}\,\mathrm{B} =
1\,\mathrm{PB}$) or more of data.

The \emph{Data-Driven Approaches to Searches for
the Technosignatures of Advanced Civilizations} study at the W.~M.~Keck
Institute for Space Studies was intended to revisit searches for
evidence of alien technologies in light of these two developments.  Experts from around the world, in a variety of disciplines, gathered
for a week to assess what new kinds of searches might be able to be
undertaken.

Of particular value for the search for technosignatures is that a
data-driven approach may be able to mitigate biases, particularly
unknown ones.  Data-driven searches, being able to process volumes of
data much greater than a human could, and in a reproducible manner,
can identify \emph{anomalies}---data that are inconsistent with a
larger sample---that could be clues to the presence of
technosignatures.

While the focus of the study was identifying technosignatures from
other civilizations, it was recognized that there are other
intelligent species on this planet, even if they do not employ
technologies capable of being detected over interstellar distances.
Learning from how various species have interacted, or coopted
interactions, may provide clues for how to search for extraterrestrial
intelligent species.  Even more tantalizing would be if universal
rules for communication among terrestrial species were to be identified.

\clearpage
A key outcome of this workshop was that technosignature searches
should be conducted in a manner consistent with Freeman Dyson’s ``First Law
of SETI Investigations,'' namely ``every search for alien civilizations should be planned to give interesting results even when no aliens are discovered.''
This approach to technosignature searches is commensurate with NASA's
approach to biosignatures in that no single observation or measurement
can be taken as providing full certainty for the detection of life.

There was broad agreement at the workshop that a variety of machine
learning techniques could be of value in searching large data
volumes.  These techniques range from extensions to the classic
matched filtering techniques to techniques in which the members of
a data set can be organized into groups based solely on the
characteristics of the individual members.  These machine learning
techniques already are being applied, with increasing success, to a
variety of problems in astronomy and other fields.  Consequently,
machine learning techniques present powerful tools for identifying
anomalies in data.

Areas of particular promise identified during the workshop were the
following:
\begin{description}
\item[Data Mining of Large Sky Surveys]%
Various large sky surveys are in the process of being conducted or
will initiate in the next decade.  Not only will these surveys be
conducted at a variety of wavelengths, many of them are introducing
a \emph{time domain} aspect, enabling rich multi-parameter searches
for anomalies to be conducted.

\item[All-Sky Survey at Far-Infrared Wavelengths]%
No technology can be perfectly efficient, because of the Second Law of
Thermodynamics.  Any technology using substantial amounts of energy
therefore will radiate some fraction of that energy as ``waste heat,''
likely to be emitted at far-IR wavelengths.  An all-sky survey at
far-IR wavelengths could be profitable for both technosignature
searches and the larger field of astronomy.

\item[Surveys with Radio Astronomical Interferometers]%
Searches at radio wavelengths have a long history in the
technosignature field.  Traditionally, these surveys have been
performed with a single large radio antenna.  Many technosignatures
have been found, but they have been \emph{interference} from
terrestrial transmitters.  The emerging suite of radio astronomical
interferometers offers new possibilities for a combination of
interference rejection and opening additional parameter space for
technosignature searches.

\item[Artifacts in the Solar System]%
Even with the number of robotic spacecraft sent throughout the Solar
System, there remains a great number of planetary bodies and vast
reaches of interplanetary space that have been surveyed poorly, if at
all.  Further exploration of the Solar System is commensurate with
larger planetary science objectives.  Moreover, in the Solar System,
there are terrestrial technosignatures on the Moon and Mars, the
product of decades of international explorations of those bodies, that
can be used as training grounds for searches for technosignatures
elsewhere in the Solar System.
\end{description}

\chapterimage{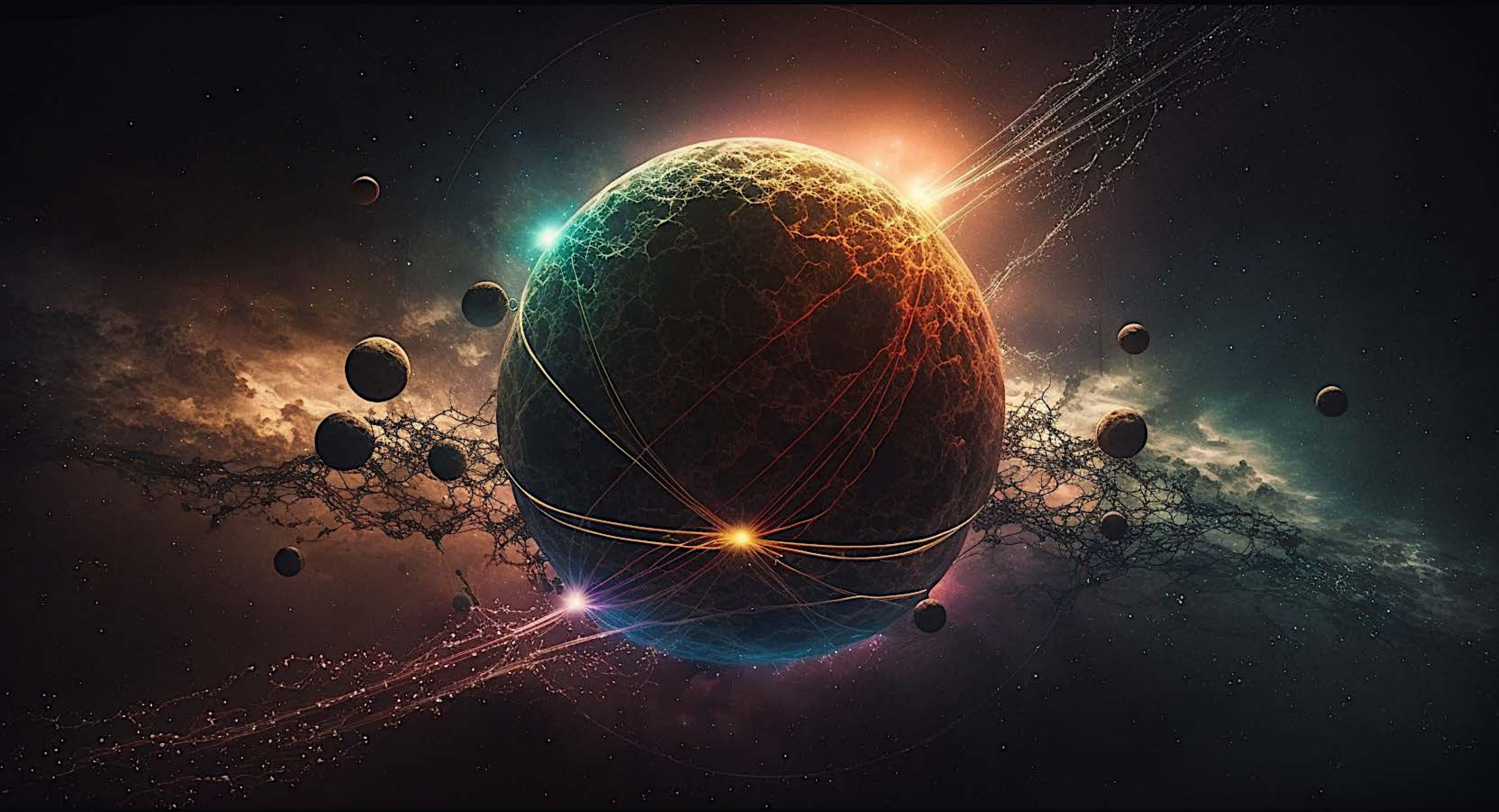} 

\chapter{Background and Motivation}\label{chap:motivate}

\section{Field Revived by Exoplanet Research and Astrobiology}\label{sec:motivate.background}

Whether other worlds exist, inhabited by other beings, is a question that has been posed for millennia, at least in Western civilization.\footnote{%
``Do there exist many worlds, or is there but a single world?  This is one of the most noble and exalted questions in the study of Nature.'' St.~Albertus Magnus, \textit{De Caelo et Mundo} ($13^{\mathrm{th}}$ Century)
}
This speculation intensified, and science fiction blossomed, about other beings on the planets in the Solar System once it became clear that they were not just lights that moved in the sky relative to the background stars but physical worlds in their own right.  In the latter half of the $20^{\mathrm{th}}$ Century, investigations by multiple spacecraft, from multiple nations, have established clearly that Mars may have had a temperate climate early in its history and prompted speculation that Venus might have as well.

In the last decade of the $20^{\mathrm{th}}$ Century, the discovery of planets around other stars changed the nature of the question entirely: There are other planets around other stars in the Milky Way Galaxy.  Moreover, the first planets to be discovered were in surprising locations---first there were planets discovered around the pulsar PSR~B1257$+$12 \citep{Wolszczan92}, then there were Jupiter-mass planets discovered orbiting extremely close to their host stars\footnote{
A discovery that resulted in M.~Mayor and D.~Queloz sharing one-half of the 2019 Nobel Prize in Physics ``for the discovery of an exoplanet orbiting a solar-type star.''}
(so-called ``hot Jupiters,'' \citealt{MayorQueloz95}).  A key element of both discoveries is that the planets were discovered serendipitously, in data being collected for other reasons.

With the data returned from the subsequent \textit{Kepler} and Transiting Exoplanet Survey Satellite (TESS) missions, the question of other beings on other worlds now has moved to a question of chemistry and biology.  The number of known exoplanets\footnote{
``Cosmic Milestone: NASA Confirms 5,000 Exoplanets''\\
\texttt{https://exoplanets.nasa.gov/news/1702/cosmic-milestone-nasa-confirms-5000-exoplanets/}
}
now exceeds 5000.  Simple extrapolation from the current data leads to the inevitable conclusion that there are ``billions and billions'' of planets in the Milky Way Galaxy.  Crucially, even in the current census, the number of potentially habitable planets, those being at distances from their host stars such that liquid water plausibly might be present on their surfaces, could be in the hundreds.

Even if the conditions are such that Earth-like life might arise on another planet, does it?  Were the conditions of the early Earth in some manner special, is life on Earth a lucky accident, or does life originate as soon as conditions allow?  Even if single-celled organisms originate easily, do complex, multi-cellular organisms evolve?  On the Earth, such complex, multi-cellular organisms have existed for only the last 10\% of its current age (last 500~million years of its 4,500~million year age).  During a small fraction of these last 500~million years, there has been a species capable of interstellar communication.  Is it likely that there are other beings on other worlds capable of interstellar communication?

As described by the \textit{Pathways to Discovery in Astronomy and Astrophysics for the
2020s} Decadal Survey report, considerable future work and investment for the next decades is ``[i]nspired by the vision of searching for signatures of life on planets outside of the solar system.''  Much of this work is focussed initially on identifying signatures of life or biological activity, so-called \emph{biosignatures}, in order to inform and motivate both current initial searches and those to be conducted with future telescopes.  In the next few decades, it is likely that we will have the answer, or at least austere constraints, on how likely it is that the chemistry on another planet will result in life originating.  Even if biosignatures are found, they may not, or likely will not address the second set of questions, whether other beings exist on those other worlds.

This report is not meant to be a comprehensive summary of the searches
for technosignatures, past, present, or future.  Rather, it covers the
subjects and ideas discussed at this workshop.  For broader surveys of
the literature on this subject, consult \textit{SETI~2020: A Roadmap for the Search for Extraterrestrial
Intelligence} \citep{2002SETI.book.....E}, 
``NASA and the Search for Technosignatures: A Report from the NASA
Technosignatures Workshop'' \citeyearpar{2018arXiv181208681N}, a
comprehensive literature list curated by \cite{2019JBIS...72..186R},
and \textit{Technosignatures for Detecting Intelligent Life in Our
Universe: A Research Companion} \citep{ResearchCompanion}.

\section{State of SETI/Technosignature Research}\label{sec:motivate.status}

The ultimate basis for the hypothesis that technically advanced
civilizations, a.{}k.{}a.\ extraterrestrial intelligences (ETIs), may
exist elsewhere in the Universe is the apparent universality of the
laws of physics and chemistry, coupled with the astrophysical
understanding of the processes of star and planet formation.  The
Earth is not unique, and the evolutionary processes that have led to
the appearance of life, followed by the intelligent life, followed by
the rise of a technological civilization, may well have operated
elsewhere.  This conclusion is reinforced by our understanding of the
ubiquity of planetary systems. Demographic studies have shown that
potentially one-third of Sun-like stars are orbited by an
approximately Earth-size planet in the Habitable
Zone \citep{2019MNRAS.487..246Z}; a small fraction of Sun-like stars
likely even have multiple planets within their Habitable Zones.
Measurements of the mass and density of some Earth-size planets have
shown that they have rocky bulk compositions.
Further, defining habitability in terms of water oceans atop rocky
surfaces is a potential Earth-centric bias, and the fraction of truly habitable planets may be higher still.

Scientific approaches to the searches for extraterrestrial
intelligence (SETI) and their technological manifestations
(\emph{technosignatures}) span a broad and diverse set of ideas,
methods, and approaches.  By far the largest effort has been in the
attempts to detect signals at radio wavelengths for which the
bandwidth-time product is of order unity \citep{Tarter01}.
There are no known astrophysical sources with this property but many
terrestrial examples; similarly, no natural radio signals are known
that contain some form of modulation.
At visible and near-infrared wavelengths, short laser pulses are
 another example of a potential signal for which no known
 astrophysical equivalents
 exist \citep{2004ApJ...613.1270H,2007AcAau..61...78H,2023arXiv230106971Z}.

Initial efforts in the mid-$20^{\mathrm{th}}$~Century focused on searches for simply modulated radio signals, mainly in the frequency range between the H\,\textsc{i} line at~1.42~GHz and the OH line at~1.67~GHz (a frequency range colloquially known as the ``water hole'').  
SETI at these frequencies both reflected the chemistry bias indicated
above and was likely influenced by the relatively rapid advances being
made in radio frequency communications at that time.  Such
centimeter-wavelength radio SETI searches have continued to the
present day, now spearheaded by the Breakthrough Listen
project
\citep{Isaacson17,2017ApJ...849..104E,2020AJ....159...86P,2020AJ....160...29S,2021AJ....162...33G}.

An alternative approach is Optical SETI (OSETI), based on searches for very short laser pulses at visible and near-infrared wavelengths.  This approach was proposed first by Charles Townes, soon after the discovery of the laser, when he recognized that laser communications might present an alternative to radio frequency communications. Pulsed optical SETI projects typically adopt a simple strategy of searching for astrophysical sources that are bright on approximately nanosecond timescales (e.g., from powerful pulsed lasers beamed toward Earth using optical telescopes as transmitters).  A simple brightness-thresholding strategy has been employed to identify candidate signals in systems built to date \citep{2004ApJ...613.1270H,2007AcAau..61...78H}. An all-sky/all-the-time optical SETI instrument in development \citep[\hbox{PANO-SETI},][]{2018SPIE10702E..5IW} opens the possibility for searching for more complicated signal types.  Searches for spectroscopic features (e.g., narrow emission lines from time-averaged laser sources) have also been attempted on thousands of stars \citep{2017AJ....153..251T}.

Motivated by the diversity of approaches and the change in landscape,
NASA held a Technosignatures Workshop, designed to review the field
and assess both the current status and how NASA might advance it (``NASA and the Search for Technosignatures: A Report from the NASA
Technosignatures Workshop'' \citeyear{2018arXiv181208681N}).

\section{New Directions, Expanding Beyond Past SETI Efforts}\label{sec:forward}

What the SETI approaches of the mid-$20^{\mathrm{th}}$ Century have in common is a set of assumptions:
\begin{itemize}
    \item A technically advanced civilization would choose to broadcast signals to other possible civilizations.
    \item It would do so using mid-$20^{\mathrm{th}}$ Century (terrestrial-analog) technology, with the same biases that we have.
    \item It would use a coding or modulation that we would be able to decipher.
\end{itemize}
Needless to say, these assumptions could reflect a broad range of cultural
and technological biases (e.g., the putative aliens would have a
similar technological level as current human civilization).
Given the diversity of human cultures, including the existence of
ancient and medieval documents that have not yet been deciphered or
translated,\footnote{
\texttt{https://en.wikipedia.org/wiki/Undeciphered\_writing\_systems}
}
there is reason to doubt the likely success of such heavily biased approaches.

Similarly, the notorious Fermi Paradox (i.e., why we haven’t seen any
evidence of ETIs, since surely they would have colonized the Galaxy by
now), which has been and is used as an argument against the existence
of ETIs, adds to the assumptions reflecting the colonial history on
our planet to the technological and cultural biases mentioned above.
It is entirely possible that we are simply not
advanced enough to understand the manifestations of a vastly more advanced technology, even if they are all around us.

The goal of this Workshop was thus to conduct a study that would be designed to minimize the cultural and anthropocentric biases that have dominated the searches for technosignatures in the past.  While eliminating such biases entirely may be very hard or even impossible, recognizing them is the first step towards their amelioration.

Specifically, this Workshop focussed on the strategies that would use
modern machine learning techniques, applied to the growing, rich data
sets from a variety of digital sky surveys, to conduct objective and
unbiased searches for sources or signals that would appear anomalous
in some well-defined way, and to investigate the possible methods of separating those of possible artificial origins from natural, albeit rare physical phenomena.

We can distinguish two cases of possible detections of the manifestations of technosignatures: intentional attempts to communicate, and byproducts of a technological activity on a large scale (e.g., waste heat, irregular eclipses, signal modulations, etc.).  The two cases may require different approaches and defining them was one of the goals of this Workshop.
Another new challenge is whether there is an objective way to indicate
the likelihood that a given detection is due to natural processes or
is artificial in origin.  This may be a challenge for the field of information theory.

\section{Opportunities from Big Data and Machine Learning}\label{sec:opportunity}

The exponential growth of data volumes and rates, as well as the growth of the data quality and information content, is transforming all sciences, including astronomy.  The data are mainly generated by the large sky surveys over the full electromagnetic (EM) spectrum, with non-EM channels (gravitational wave [GW], ultra-high energy cosmic rays [UHECRs], etc.) growing in importance. This exponential data flood has opened new scientific opportunities for the systematic studies of the Universe and its physical constituents. The sheer richness of this data (e.g., as measured in the numbers of detected sources, and the number of features measured for each one of them) enables profitable data mining, including searches for the rare or unusual objects and phenomena.  Additional leverage comes from data fusion (e.g., between different wavelengths), that may reveal new knowledge present in the data, but not recognizable in any of the data sets separately.  The initial generation of ``single pass'' panoramic sky surveys (e.g., Sloan Digital Sky Survey [SDSS], Digital Palomar Observatory Sky Survey [DPOSS], Two-Micron All-Sky Survey [2MASS], etc.) has led to the new generation of synoptic sky surveys that opened a new region of the observable parameter space, the time domain (e.g., Zwicky Transient Facility [ZTF], Catalina Real-Time Survey [CRTS], Vera C.~Rubin Legacy Survey of Space and Time [LSST], various radio astronomical facilities, etc.)

The opportunities for the systematic searches and discovery of rare or unusual types of astrophysical objects as outliers in some parameter space of measurements has been recognized as one of the key scientific drivers within the Virtual Observatory environment \citep[e.g.,][]{2001ASPC..225...52D,2001SPIE.4477...43D},
with \cite{2000ASPC..213..519D} considering the specific application to \hbox{SETI}.  Since then, the data rates and volumes increased by at least a factor of~$10^3$, with a commensurate increase in the scientific potential and technical challenges, which is why the time is ripe to open this new approach to detecting ETIs.

In general, the process involves detection of statistically significant sources or events, and their characterization in a high dimensionality data space that would include the measurements of their fluxes, spectral energy distributions, time variability, structure (if extended), etc.  The measured or derived features are typically based on the domain expertise, and these feature spaces can have hundreds of dimensions, although usually only a few tens may be independent and/or relevant for the problem at hand.

The key methodology here is unsupervised clustering, a machine
learning approach in which the data themselves can determine the numbers
of different classes of objects present, and the statistically
significant outliers in some high-dimensionality data space.  This is
the least biased approach, as compared to, say, supervised
classification, in which the number and identity of different classes of objects are pre-determined, with a suitable set of training, validation, and testing examples used to define the optimal classifications of objects.  Such objectively defined outliers in a data/parameter space (i.e., somehow unlike or anomalous as compared to the bulk of all objects) constitute candidates to be examined further.  The challenge is achieving a high completeness (not missing any interesting ones) while maintaining a low false alarm rate and doing so in a scalable manner.

\begin{figure}
\centering \includegraphics[width=0.9\textwidth]{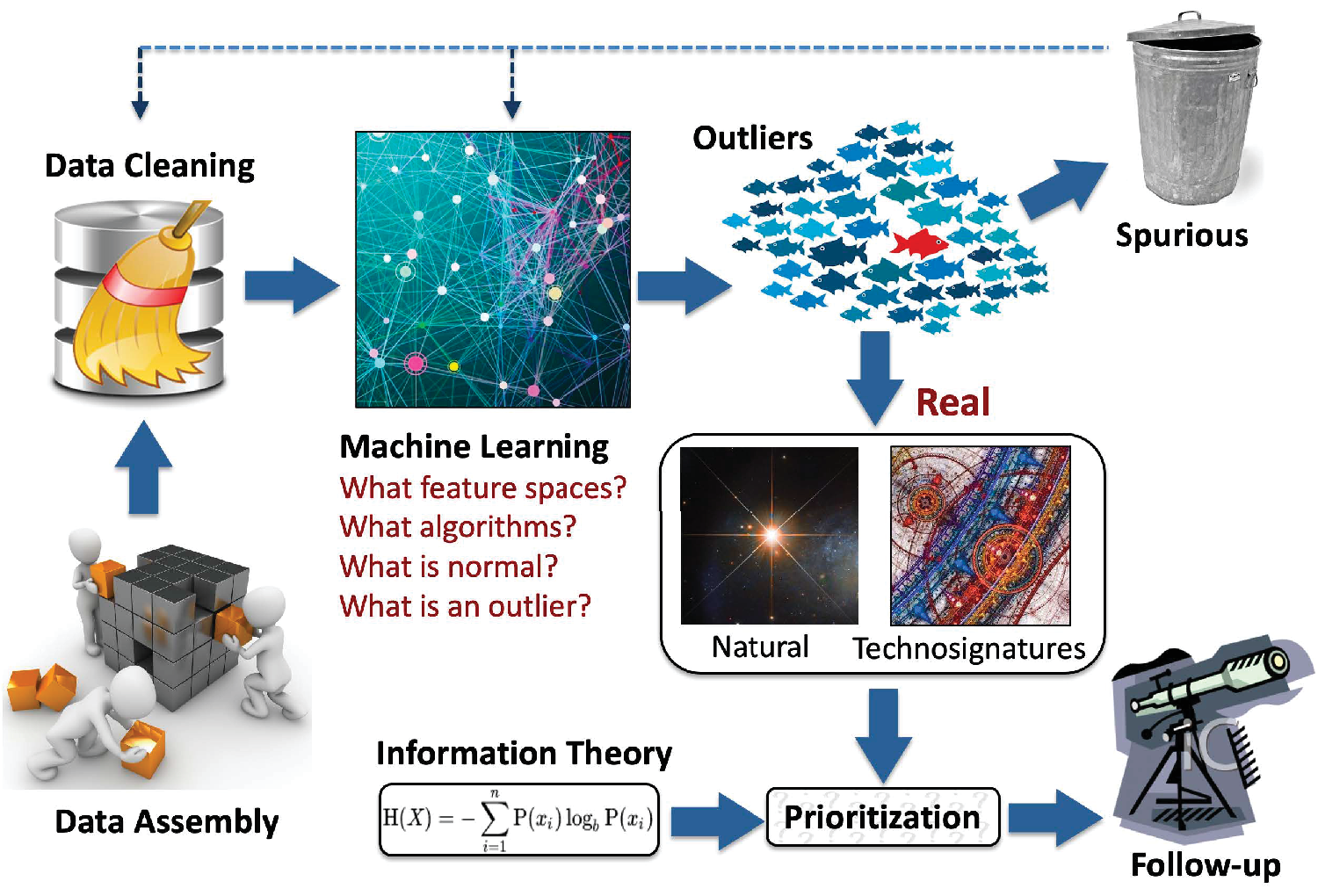}
 \caption{Schematic illustration of big data and machine learning for
 identification of technosignatures.  Data preparation and cleaning of
 the measurement glitches and artifacts can be time consuming, and
 iterated as the outlier search algorithms identify the unremoved
 ones.  The choices of the machine learning (ML) algorithms and their implementation also require an extensive experimentation and validation.  Establishing the nature of any genuine outliers in the data (defined in some objective manner as not belonging to any of the known types of objects) likely requires some follow-up observations, which may have to be prioritized on the basis of the available information.}
\label{fig:overview}
\end{figure}

Figure~\ref{fig:overview} illustrates how this task can be highly nontrivial, from the data assembly, cleaning and preparation, the choice of the right algorithms, a robust detection of outliers, many of which would be data artifacts that were not fully removed by the data cleaning process.  The surviving candidates may be natural (and possibly interesting from an astrophysical point of view), or artificial (technosignatures), and their nature would have to be clarified through the suitably prioritized follow-up observations.

These are precisely the methodologies and tools used for the astrophysical exploration of the newly opened observable parameter spaces.  In this regard, a search for technosignatures is no different than a search from new astrophysical phenomena, detectable as outliers in some observable parameter space.  Thus, a search for technosignatures becomes a collateral benefit of a systematic, astrophysically motivated exploration of large and complex data sets and data streams.

\chapterimage{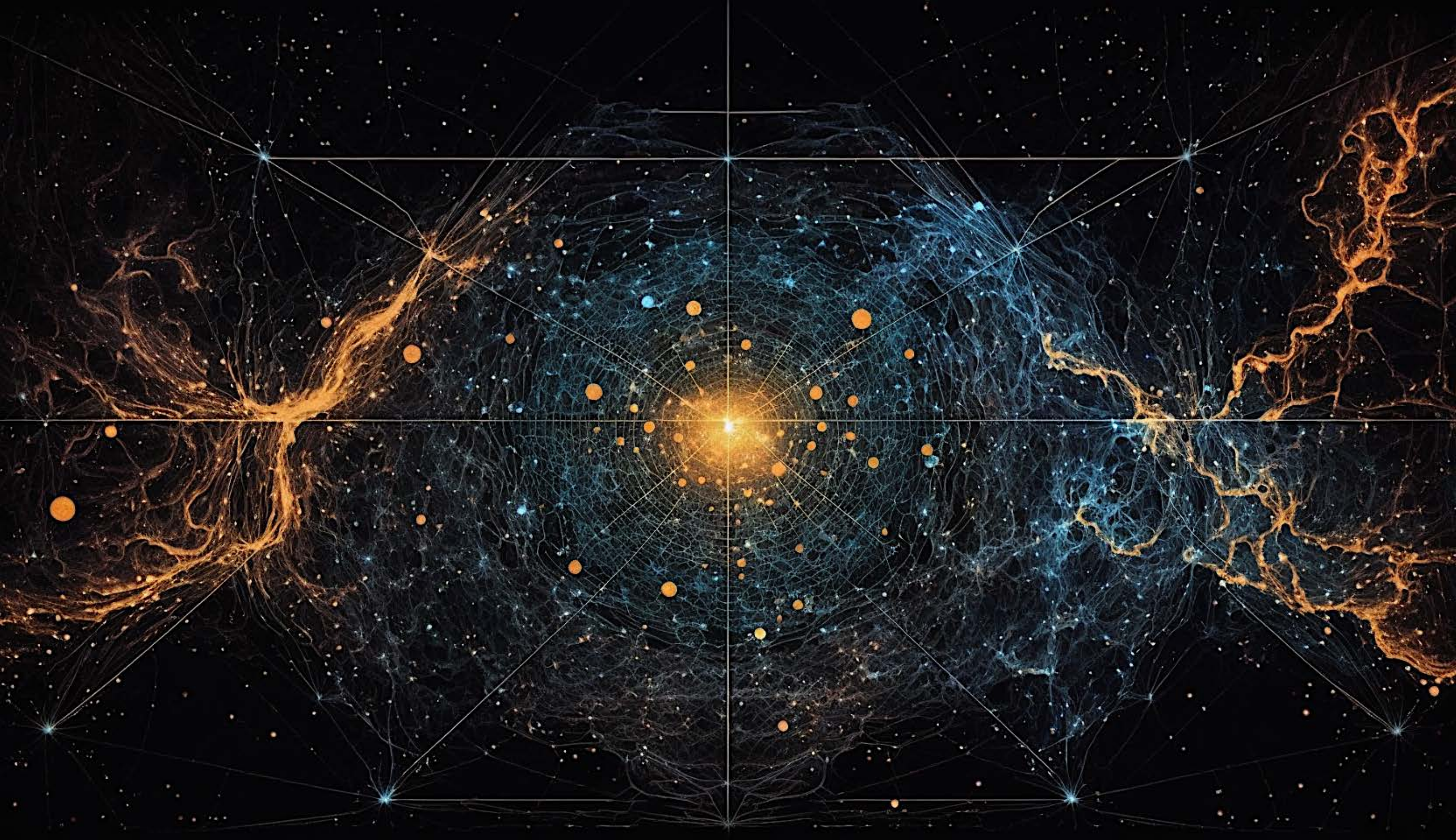}

\chapter{Methodologies}

\section{Striving for Robustness}\label{sec:method.robust}

Past proposals for technosignatures have been strongly driven by assumptions about what an ETI would do, how, and why. These assumptions, in hindsight, often reflect the imaginer's cultural and historical environment more than they reveal any fundamental truths about the behaviour of intelligent beings. For example, \cite{von1975population} muses about the ``obvious'' problems of nuclear self-destruction and overpopulation that would explain the Fermi Paradox. Recent work, such as \cite{frank2018anthropocene} instead examines climate forcings from technological intelligences; the authors conclude that the path through exo-Anthropocenes, the process of learning to live sustainably with one's planet or perishing, might represent some sort of fundamental bottleneck.

Here, we strive to consider technosignatures differently, free from as many assumptions as possible.
Strictly speaking, the only unbiased assumptions that we can make are that mathematics, physics, chemistry, and information theory are universal, and even then we recognize that there is likely more yet to be discovered in these domains---our knowledge of them is incomplete.  Anything beyond these universal assumptions may be subject to biases reflecting how life and human civilization originated and evolved, to say nothing about technology.  For example, it is in principle possible that technosignatures may involve as yet unknown physics, and thus be beyond what we can observe or interpret.

The NASA Technosignatures Report defined nine ``axes of merit'' for technosignatures searches. In that framework, we are trying to maximize merit along the axis of ``inevitable'' to ``contrived/specific'' - we want technosignatures that are inevitably produced by technology regardless of its creators or its purpose. In this report, this property is defined as robustness. Robustness relies on using fundamental laws from mathematics, physics, chemistry, astrophysics, and information theory that we can reasonably believe are universal. We would like to minimize our assumptions based on our particular biological and technological development as much possible for the design of a rigorous, scientific search.

Discussions of the motivation of an ETI affect not only theory, but also observation, creeping in to determine where and how technosignature searches have been performed in the past. This occurred partially because early searches were extremely limited in technical capability. When only small regions of parameter space can be searched at a time, a rationale about the benefits of one area over another is necessary to drive the selection of the parameters chosen for the search. However, in the era of big data and modern instrumentation, we no longer need to invoke motivation to decide where to search; we can search much larger regions at once and store all of the resulting data for later analysis. For example, much of the early radio SETI literature was devoted to the discussion of possible ``magic frequencies''---regions of frequency space that were deemed more likely to hold a signal than others, often with the additional assumption that the signal was broadcast intentionally for the purposes of communication with other civilizations \citep[e.g., the ``water hole,''][]{oliver1979rationale}.  This discussion was important at the time, because receivers only had limited bandwidths and spectral resolutions.  Now it is possible to record billions of channels simultaneously, over 6~GHz of bandwidth \citep{macmahon2018breakthrough}, so the specific choice of frequencies has become less relevant.

Application of objectively applied and statistically well-defined algorithms for the detection of outliers in large data spaces offers a path towards minimizing such human biases.
Likewise, we explore the searches for technosignatures from sources other than intentional communication. This will include searches for physical artifacts inside the solar system, remote signatures of technological activity outside the solar system, and looking for unintentional information-carrying signals.

We note that any objectively defined ``anomalies'' found in a systematic search may be due to either rare or previously unknown natural phenomena, which would be a valuable result by itself, or a manifestation of technosignatures.  The two possibilities may be indistinguishable in the primary observations; follow-up observations are likely to be necessary.

\section{Recognizing and Minimizing the Human Biases}\label{method.bias}

We have formulated the following assumptions that we believe are sufficiently robust as to minimize the possible biases:
\begin{itemize}
    \item Mathematics, physics, chemistry, and information theory are universal.  Everything else is a subject to the human biological, anthropological, cultural, and technological biases.
    \item Living organisms and civilizations consume energy and
    generate entropy within their environment, often in the form of a
    waste heat.  They modify their environments and can co-evolve with
    them. They may be sources of information, which we can detect as technosignatures.
\end{itemize}

We are also cognizant of some fundamental limitations:
\begin{itemize}
    \item We do not know whether the carbon+water chemistry is the only path to life and intelligence, although it could be.  Rapidly improving Artificial Intelligence on our planet shows than an inorganic intelligence is possible.  Science fiction provides many other potential examples.
    \item There is almost certainly physics beyond that we already know, but we cannot meaningfully speculate about the possible implications for our subject.  They may well be better mechanisms for information and matter transfer than those that we know.
\end{itemize}
However, the most important and hard-to-eliminate biases come from the human psychology, history, and culture.  The following anecdote illustrates this point:
In the early $20^{\mathrm{th}}$ Century, the three great inventors,
Nikola Tesla, Thomas Edison, and Guglielmo Marconi all thought that
they detected radio signals from Mars,\footnote{
\texttt{https://anengineersaspect.blogspot.com/2013/03/hello-earth-early-20th-century.html}}
at frequencies around 3~kHz, which we now know cannot penetrate the
 Earth’s ionosphere,\footnote{
 For a potential exception to this conclusion, see Corum \& Corum, ``Nikola Tesla and the Planetary Radio Signals,'' \texttt{https://radiojove.gsfc.nasa.gov/education/educationalcd/Books/Tesla.pdf}.}
but which they did not know at the time.  All three were firmly convinced that these were messages from the outer space, and Mars in particular.  Marconi even thought that some of these putative signals used Morse code!  These were brilliant men, and yet this was their interpretation, based on the technology they knew, unaware of the ionospheric physics and sources of radio noise, clearly influenced by then current fascination with Mars, and projecting to the putative Martians the intents that would be like their own.

\section{Defining Outliers}\label{sec:method.outliers}

An outlier is an observation that does not conform to expected behavior \citep{chandola:anom09} or observations from its resident sample or dataset \citep{goldstein2016}.

Even the use of objective tools like machine learning and statistics can be subject to biases, e.g., about the underlying probability density distributions in the data.  Such biases can lead to inappropriate choices of algorithms, and thus to misleading results.  For example, if Gaussian distributions are assumed, any apparent outliers may be simply manifestations of the ``fat tails'' of the true distributions, for example, power laws or exponentials.  Another problem is that any machine learning results are only as good as the training data sets that are used, and they can be incomplete (missing a sufficient number of training examples in some region of the feature space), unbalanced (which favors the types of phenomena that are over-represented in the training data set), or simply contaminated by bad measurements or errors that have not been caught during the initial data cleaning or preparation.  Blind use of machine learning tools can produce misleading results or miss important findings.  Multiple algorithms and data models should be used, along with very extensive testing.

Some outliers, once discovered, should be addressed or discarded.  Examples include contaminants or artifacts caused by (i)~Errors associated with telescope optics, detectors, or observing conditions, or~(ii)~Measurement or modeling errors induced by data reduction, photometric processing, or modeling software.  In some cases, one discovered, such outliers can inspire improvements to the upstream data acquisition or processing steps.

Alternately, an outlier could be a valid astronomical source or an object that presages scientific discovery, even if not a technosignature per se.  Quasars \citep{schmidt1963}, radio pulsars \citep{hewish1968}, and cosmic gamma-ray bursts \citep{klebesadel1973} were all discovered by astronomers who manually inspected aberrations in their data.

Because the former class of outliers always (dramatically) outnumbers the latter class, a well-known challenge is to distinguish between these two classes in a robust manner.  Traditionally, this discrimination is done by specifying a \emph{false alarm probability}, with intensive inspection of outliers passing such a threshold.  The key risk with such an approach is often characterized as \emph{Type~I} and \emph{Type~II} errors, in which an outlier that does not represent an actual astronomical source is identified as such (Type~I) or an outlier that is an actual astronomical source is incorrectly discarded (Type~II).  While this challenge of keeping the desired outliers and eliminating the undesired is not new, it is particularly challenging in an era of large data sets.

There are two axes of outlier categories: {\em global} versus {\em local}, and {\em individual} versus {\em clustered}.  A global outlier is an observation that is unusual with respect to its entire sample.  A local outlier is an observation that is unusual with respect to a specific subset or subspace of its sample.  Alternatively, outliers may be individual or clustered in small groups. Outlier clusters may lead to the discovery of new classes of objects. 

Unsupervised clustering is the standard methodology for outlier/anomaly detection and many algorithms exist. The choice of the optimal ones depends on the given situation, the nature of the data, and the geometry and topology of data distributions in the feature space.  One critical choice is the normalization of the input variables and the metric choice for the feature space, and a number of different possibilities exist.

\section{How Discoveries are Made}\label{sec:discovery}

\subsection{Exploration of Observable Parameter
            Spaces}\label{sec:discovery.parameter}

In general, empirical discoveries tend to be due to the technological advances, leading to measurements that were previously not possible.  A useful approach has been formulated by \cite{1975QJRAS..16..378H,2003PhT....56k..38H}, who defined observable parameter spaces with axes such as wavelength coverage; flux sensitivity; spatial, temporal, or spectroscopic resolution; etc.  As our coverage of these parameter spaces improved, new astrophysical phenomena were discovered.  In the era of big data and large digital sky surveys, this framework can be refined further \citep{2012IAUS..285..141D}.

\begin{figure}
 \centering
 \includegraphics[width=0.9\textwidth]{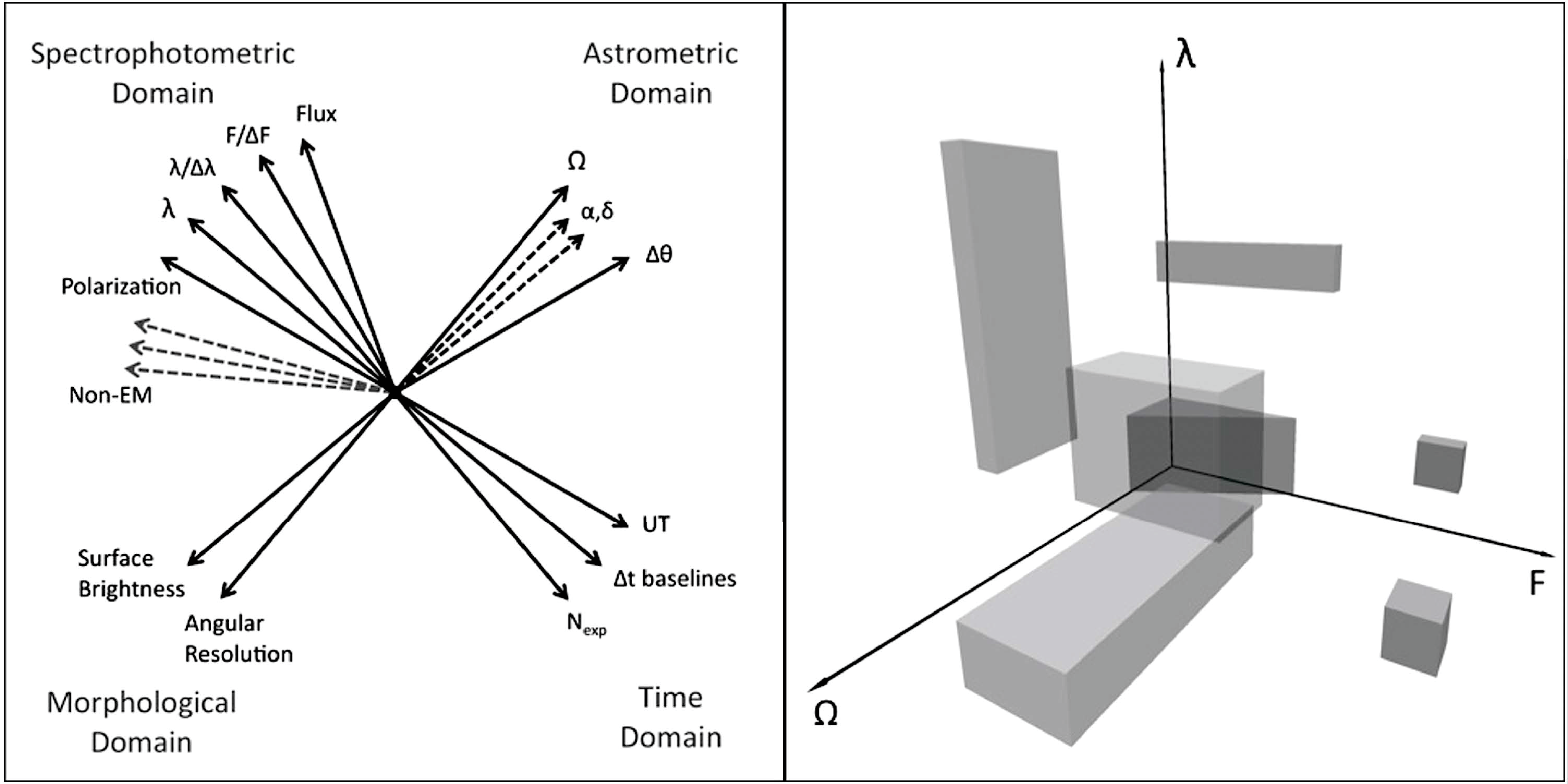}
 \vspace*{-1ex}
 \caption{Observable Parameter Space (OPS).  All axes of the OPS corresponding to independent measurements are mutually orthogonal.  Every astronomical observation, surveys included, carves out a finite hypervolume in this parameter space.
 (\textit{left})~Principal axes of the \hbox{OPS}, grouped into four domains; representing such high-dimensionality parameter spaces on a 2-D paper is difficult.
 (\textit{right})~Schematic representation of a particular 3-D representation of the \hbox{OPS}.  Each survey covers some solid angle~$\Omega$, over some wavelength range~$\lambda$, and with some dynamical range of fluxes~$F$.  Note that these regions need not have orthogonal, or even planar boundaries.  (Figure from \citealt{2012IAUS..285..141D}.)}
\label{fig:axes}
\end{figure}

Every astronomical observation, surveys included, covers some finite portion of the Observable Parameter Space (\hbox{OPS}, Figure~\ref{fig:axes}), the axes of which correspond to the observable quantities, e.g., flux wavelength and sky coverage.  Every astronomical observation or a set thereof, surveys included, subtends a multi-dimensional volume (hypervolume) in this multi-dimensional parameter space.  One example is the ``cosmic haystack'' of SETI searches, the axes of which are the area, depth, and frequency \citep[e.g.,][]{Tarter01}.  We are also starting to add the non-electromagnetic information channels: neutrinos, gravitational waves, and cosmic rays; they add their own dimensions to the general \hbox{OPS}.  The dimensionality of the OPS is given by the number of characteristics that can be defined for a given type of observations, although some of them may not be especially useful and could be ignored in a particular situation.  Some parts of the OPS may be excluded naturally, e.g., due to quantum limits, diffraction limits, opacity and turbulence of the Earth’s atmosphere or the Galactic interstellar medium (ISM) at some wavelengths, etc.  Others are simply not accessible in practice, due to limitations of the available technology, observing time, or funding.

The coverage and the dimensionality of the OPS determine what can be detected or measured; it fully describes the scope and the limitations of our \emph{observations}.  It is in principle enormous, and as our observing technologies improve, we cover an ever greater portion of it.  Selection effects due to the survey limitations are also more apparent in such a representation.   Examining the coverage of the OPS can yield insights for the optimal strategies for future explorations.

\begin{figure}[tb]
 \centering
 \includegraphics[width=0.9\textwidth]{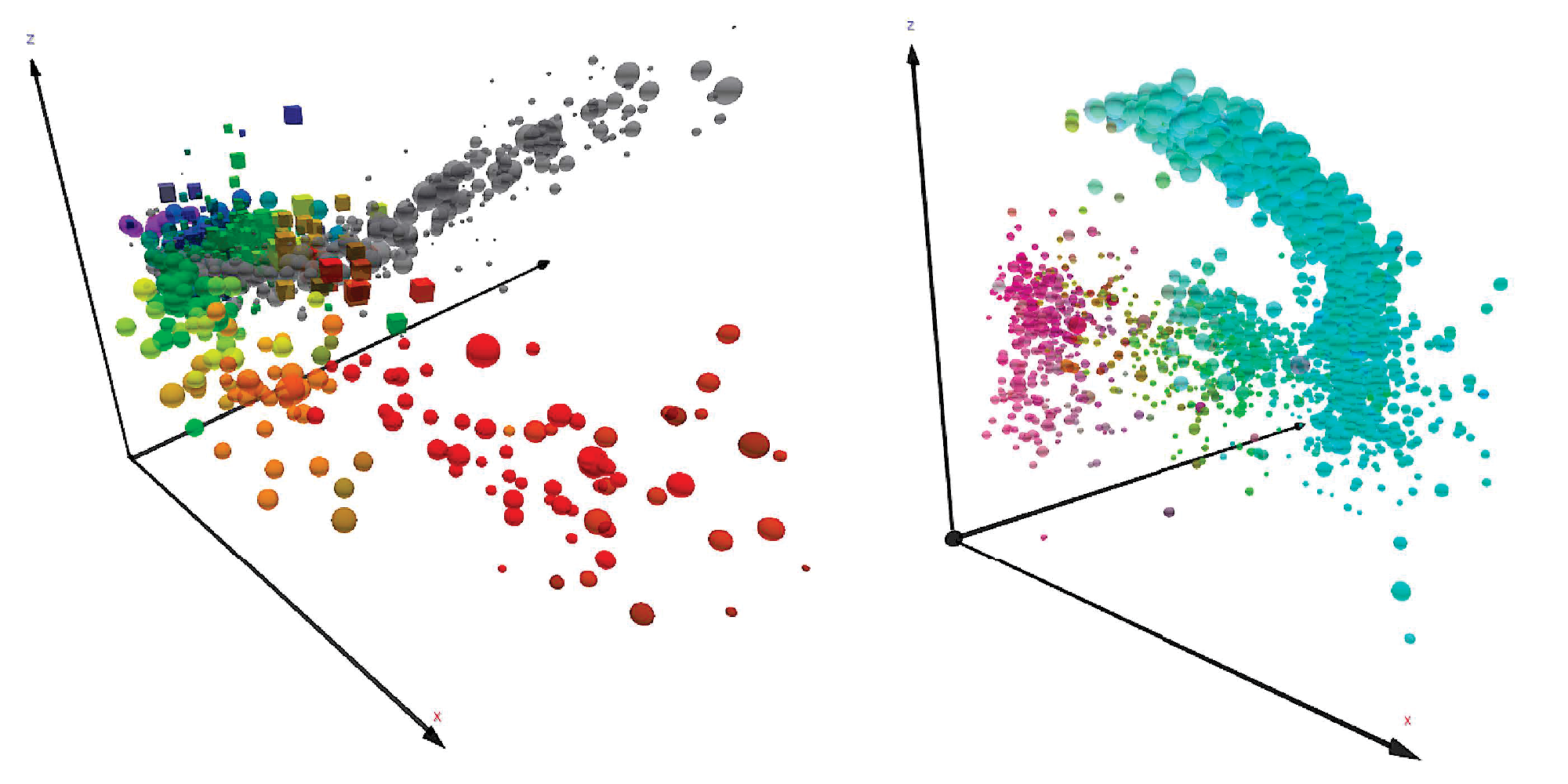}
 \caption{%
   Examples of 3D projections of the Measurement Parameter Space (MPS), specifically colors, from \protect\cite{2012IAUS..285..141D}. Only a few thousand objects are plotted, for clarity. 
   (\textit{left})~Colors of stars, quasars and galaxies in the SDSS {\it ugriz} color space. Galactic stars are represented as gray spheres, quasars as colored spheres with the color indicating the redshift, and galaxies as cubes, with the colors indicating the redshift, but on a different scale from the quasars; the sizes of the data points indicates the apparent magnitudes.
   (\textit{right})~Colors of quasars in the SDSS {\it ugr} and the \textit{Spitzer} 3.6~$\mu$m band, from Richards et al. (2009).
Objects of different types or sub-types form clusters and sequences in the MPS, that may be apparent in some of its projections, but not in others. In these examples, distributions in the space of observable parameters also carry a physical meaning, and can be used to separate the classes, or to search for rare types of objects, e.g., high-redshift quasars.
Thus, visual inspection and exploration of data represented in the MPS can lead directly to new insights and discoveries. In addition to the three spatial axes, one also could use the sizes, shapes, colors, and transparencies of the data points in order to encode additional dimensions. A more rigorous approach is to use various Machine Learning tools, such as the unsupervised clustering, classification, anomaly or outlier searches, correlation searches.
Data taken from \cite{2009AJ....137.3884R}.}
\label{fig:mps}
\end{figure}

As catalogs of sources and their measured properties are derived from imaging surveys, they can be represented as points (or vectors) in the \emph{Measurement Parameter Space} (\hbox{MPS}, Figure~\ref{fig:mps}).  Every measured quantity for the individual sources has a corresponding axis in the \hbox{MPS}.   Some axes could be derived from the primary measured quantities (e.g., fluxes), while some may be from their combinations (e.g., flux ratios).   Some parameters may not even be representable as numbers, but rather as labels; for example, morphological types of galaxies, or a star vs.\ a galaxy classification.

While the OPS represents the scope and the limitations of observations, MPS is populated by the detected sources and their measured properties.  It describes completely the content of source catalogs derived from the surveys through data reduction pipelines.  The application of  ML methods for exploration and analysis is done in the \hbox{MPS}.  Each detected source is then fully represented as a feature vector in the MPS (``features'' is a commonly used computer-science term for what are called measured parameters here).  Modern imaging surveys may measure hundreds of parameters for each object, with a corresponding dimensionality of the \hbox{MPS}; however, many axes can be highly correlated (e.g., magnitudes in a series of apertures).  It is thus a good idea to reduce the dimensionality of the MPS to a minimum set of independent axes, before proceeding to further data analysis.

Observed properties of detected sources can then be translated through the process of data reductions and analysis into their physical properties, usually requiring some additional or interpretative knowledge, e.g., distances, and/or assuming some data model.  These form the physical parameter space (PPS), where typically the scientific discoveries are made.  The MPS describes observations of individual sources found in a catalog; the PPS is populated by astronomical objects, again quantified as feature vectors.  The MPS and the PPS may partly overlap.  Some authors erroneously use the term ``phase space'' for some or all of these; a phase space is a space of possible states of a given physical system, but a supernova is not a different state of a pulsating variable star, and a pulsar is not a different physical state of a quasar, but different types of objects can and do share the same, well-defined parameter spaces.

Surfaces delimiting the hyper-volumes covered by surveys in the OPS map into the selection effects in the MPS and its projections to lower dimensionality parameter spaces.  These, in turn, map directly into the selection effects in the \hbox{PPS}.  The combination of these parameter spaces thus represents a quantitative framework for data analysis and exploration, that also defines observational biases that can affect a search for technosignatures or any other unusual types of sources.  In this framework, we search for the outliers in the \hbox{MPS}, and their interpretation (usually following some follow-up measurements) places them in the \hbox{PPS}, which is where we may distinguish technosignatures from interesting and unusual natural phenomena.

Large digital sky surveys offer both an ever increasing coverage of these parameter spaces, and also larger numbers of detected sources, which improves the chances of finding some rare and unusual types.  This, for example, is now a standard approach to discovery of quasars.

We emphasize that this approach represents a systematic exploration of
these parameter spaces.  Sometimes the term ``serendipity'' is used,
which is a misnomer: it implies stumbling upon something interesting
by a sheer dumb luck, which is very different from a systematic
exploration.  In some cases, this exploration may be guided by a model
that predicts observable properties of some type of objects, e.g.,
Dyson structures with their waste heat.  In other cases, the
exploration could be entirely unbiased by any expectations, and detect
``anomalous'' sources using some objectively defined unsupervised
clustering algorithm, reflecting the data science motto: ``let the
data tell you what is in the data.''  Doing so requires large, information-rich data sets and the correct algorithms.

\subsection{Data Fusion}\label{sec:discovery.fusion}

Another path to discovery is through data fusion.  New knowledge may be present in the individual data sets, but not recognizable as such, and revealed through data fusion.  Astronomy offers many cases, usually from combining data from different wavelength regimes.  One example is the ultraluminous starburst galaxies, originally catalogued as morphologically peculiar by H.~Arp and others.  Only with the opening of the far-infrared sky by the Infrared Astronomical Satellite (IRAS), were they identified as ultraluminous objects powered by a dust-obscured star formation, and an important link in galaxy evolution.  Quasars were first noted in the 1920s, and misclassified as variable stars (such as the BL~Lacertae, the prototypical blazar).  Only with the advent of radio astronomy in the 1960s, and the optical identification of some of the radio sources, were they revealed as a previously unknown phenomenon of nature, with many cosmological implications.

\subsection{Missed Discoveries}\label{sec:discovery.miss}

It is also of interest to understand how discoveries can be missed.  One mechanism is that the data needed to make the discovery were not available at the time, like in the examples cited above.  Another curious case is Galileo’s detection, but not discovery, of Neptune in~1612--1613; Galileo did not have any reason to think that this was not just another background star \citep{1980Natur.287..311K}.  Another example is the discovery of radioactivity: 40~years before H.~Becquerel reported it in~1896, the photographer Abel Ni{\'e}pce de~Saint-Victor made the same discovery and even reported it to the French Academy of Sciences, but it was ignored since it did not fit in the established understanding of the world.

\bigskip

Thus, in the era of big data, new discoveries can be made through a systematic exploration of information-rich data spaces, and can be missed either because they look like something already known, or because they do not fit in the established picture and expectations, and are always dependent on our expectations, experience, and the physical models of the world.

\chapterimage{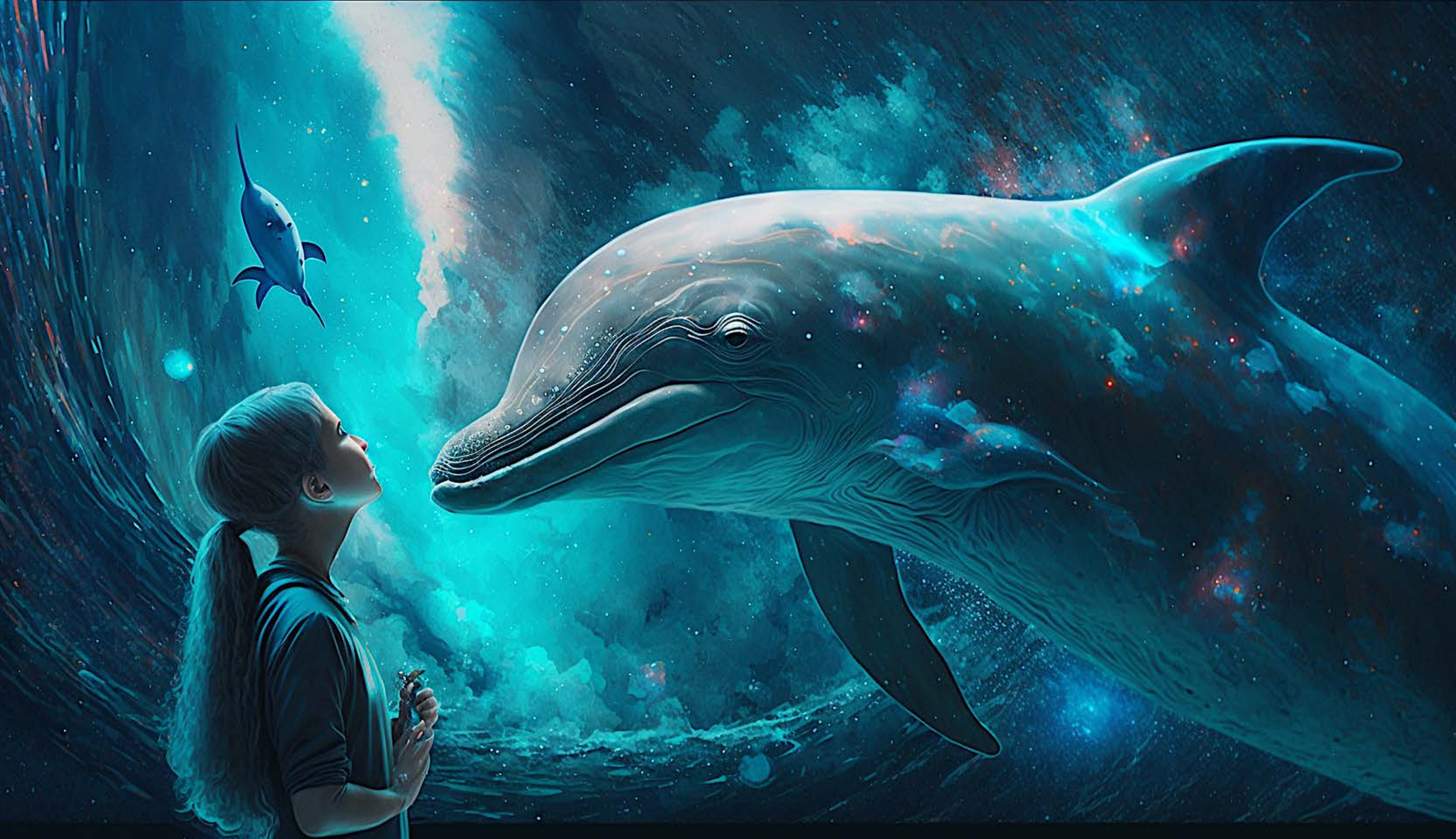}

\chapter{Insights from Other Fields}\label{chap:others}

When thinking about communicating with \hbox{ETI}, and going beyond
elementary mathematical facts, such as the existence of prime numbers
or the Pythagorean theorem, there may be insights from our own planet.
These insights includes communication between different cultures, which may be  
separated widely in time, and interpretation of their artifacts; communication with or within other biological 
species; and communication with non-biological intelligence such as
artificial intelligence (\hbox{AI}).

Indeed, it even may be that intelligence can be regarded as a planetary-scale process \citep{2022IJAsB..21...47F}.
Aspects of the different possible manifestations of intelligence are
discussed in the Santa Fe Institute's report ``Foundations of Intelligence in Natural and Artificial Systems: A Workshop Report'' \citep{2021arXiv210502198M}.

\section{Communication and Interpretation Between Different Cultures or Eras}\label{sec:others.culture}

Humans often do not communicate well even within their own species, individually or collectively.  
History abounds with examples of failed communications between independently developed cultures,
due to their different needs, beliefs, and priorities.  For the same reasons, it can be difficult
to fully understand ancient cultures and their artifacts.  Cultures evolve, are subject to many factors,
and may diverge \citep{bvdbmw21}.  Data-driven computing opens new venues to understand such
processes \citep{FOSTER2018144,doi:10.1177/1536504219883850}.

Archeologists sometimes find it difficult to decide whether a particular object, such as an unusually shaped
rock, is an artifact (a tool, an object of worship, \ldots), or a
natural object \citep[e.g.,][]{srh+97,doi:10.1126/sciadv.ade8159}.
In such cases, contextual
information is critical; for example, was it found next to ancient bones or remnants of a fire, is it
similar to other objects of an established origin (artifact typologies), does the object bear markings
that are unlikely to occur naturally, \ldots?  The use of radioisotope dating can help establish the 
temporal correlations or their absence.
Even if it is clear that we are dealing with a cultural/technological artifact, such as the Antikythera 
mechanism, the function of which is understood, we still may not be aware of the intellectual ecosystem from which it springs, or its history and provenance.
A similar situation likely will arise if we discover a potential
ETI artifact somewhere in the Solar System.

Similarly, writings on the walls, tablets, \ldots, may be hard or impossible to understand, if they 
are based on a sufficiently different model of the representation of
information.  Two classic examples\footnote{%
Other examples are discussed
at \texttt{https://en.wikipedia.org/wiki/Undeciphered\_writing\_systems}}
are
Linear~A and the hieroglyphic Minoan script from Crete, both of which
are still not understood.  

Yet, all of the above represents human cultures, including many possible anthropocentric
biases.  Should we ever encounter ETI communications or artifacts, their interpretation would 
likely be much harder and subject to the same interpretative biases.  However, these experiences
from anthropology and archeology may be a starting point.

\section{Interactions with Other (Biological) Species}\label{sec:others.animal}

In the search for \hbox{ETI}, or for intelligent species on other planets, it would be helpful to identify any
universal rules of communication in species on Earth. As early as the late 1970s, \cite{10.2307/2460385}
identified motivation rules in the acoustics of both mammals and birds. Further work with other species
has at least confirmed this finding.  If there are  other universal
patterns to the myriad of species
on this planet, and they can be decoded,
it might help develop algorithms or pattern-matching software for future encounters.
Being open to assessing other ``types'' of intelligence may be critical in this process, regardless of the
taxa type \citep{herzing_profiling_2014}.  Not only would identifing
any intentional signaling be valuable (\S\ref{sec:technosigs.kind}), but an
objective would be to determine if nonhumans had language or language-like structures in their communication,
furthering our interactions. Determining types of encoding that the animal world uses might, in the end,
help in decoding strange signals from space or those found on another planet.

The ability and ideas around communicating with nonhumans creates many challenges and equally more questions.  After almost four decades of underwater experience studying an aquatic social society, and both observing them and interacting with dolphins, the following are some of the main observations and questions directed to the search for \hbox{ETI}.  There are two main scenarios for communicating with nonhumans; real-time interactions and passive observation and data collection.  

When communicating with another species in real-time, it is essential to understand their sensory system and intraspecies communication system \citep{2010AcAau..67.1451H}.  Knowing on which modality or modalities to focus is essential---Acoustic modalities are appropriate for dolphins, birds, bats, elephants, hippopotami, prairie dogs; visual modalities are appropriate for chimpanzees, cephalopods, fireflies. Secondly, in terms of factors driving the communications, determining whether it is environmental constraints (e.g., dense vs.\ sparse forest) or the social structure (e.g., group competition, social alliances) is important.  Finally, it is important to determine how much interest and motivation another species has in communicating with humans and how another species might perceive us \citep{db92}.

Nonhuman species themselves have two strategies for interspecies communicative activity: (i)~Passive observation and coopting signals, e.g., sentinel bird alarm signals \citep{m82}; and~(ii)~Creation of mutually agreed upon signals outside of both species’ repertoires, e.g., orca pod interactions across dialects \citep{f91}.  Some species also demonstrate acute awareness of others and utilize either acoustic crypticity (orcas) to hide their signals in background noise \citep{b-lfh96} or utilize two different channels of visual signals simultaneously like cephalopods \citep{bgw12}, one for a conspecific (courtship signals) and another for a competitor (aggressive signals).

\subsection{Human and Nonhuman Communication: In the Wild and Scientific Approaches}\label{sec:others.animal.crossspecies}

Humans have a nonscientific history of interacting with other species for mutual benefit, such as deriving honey from a hive from interactions with honeyguide birds \citep{ir89} or deriving advantage for netting fish from interactions with bottlenose dolphins \citep{s-ld-jc16}, some over multiple generations of both fisherman and dolphins. 
Scientists also have a long history trying to cross the species boundary by creating mutual systems usually with referential labels like objects or names, with other primates (chimpanzees, gorillas), as well as bird and dolphin species \citep{h16}.  Most attempts have involved technology since it often requires bridging the gap across modalities (i.e., a dolphin cannot use a touch keyboard and an ape cannot make complicated vocal sounds).

\subsection{Passive Observations and Data Sets}\label{sec:others.animal.observe}

For remote collection of data and data mining for long-term
interaction, machine learning tools and other Artificial Intelligence
operations can help (\S\ref{sec:opportunity}, \S\ref{sec:ml.unsupervise}).  Developing new algorithms to data mine could be helpful in this process.  Showing proof-of-concept by testing on known data (e.g., human languages) and further tests on the myriad of nonhuman communication datasets on Earth would expand our abilities to prepare for alien communications.  Such a process might help us distinguish between attention-getting signals vs.\ referential animal signals (e.g., hand waving vs. detailed information). Like any anthropologist, the use of metadata (e.g., life history, age, relations) will be critical to interpret the meaning and functional use of signals.

\section{Communication With Non-Biological Intelligence}\label{sec:others.ai}

Intelligence need not be biological, as demonstrated by the rapidly developing fields of Artificial
Intelligence (AI).  Its substrate, architecture, and operational principles are fundamentally
different from biological intelligence, and it thinks differently.
Just like biological intelligence, it evolves, only much faster.
In \hbox{AI}, humans have created an alien intelligence on planet Earth.

Current AI systems are built by humans and trained on the data produced by humans, such as the
Web content.  The interfaces to AI are designed to be compatible with human understanding, such
as the text generated by GPT-based networks, or voice-activated assistants (Siri, Alexa, \ldots). 
Virtual human avatars are increasingly being used as interfaces to various AI-based services,
since humans like to communicate with other humans.  This is also why in the popular culture
ET aliens are usually represented in a humanoid form, which is a clear signal of anthropogenic 
biases that can affect searches for intelligence beyond Earth.  

Current AI could be regarded as symbiotic with humans, as ever more complex
tasks are outsourced to it, 
from Web searches to self-driving cars, and beyond.  AI-produced results 
that humans could not achieve on their own also are beginning to
appear.  Examples include novel strategies for the game Go by AlphaGo
\citep{10.1145/3206157.3206174} and the solution to the protein folding problem by AlphaFold \citep{DeepFold}.
The rapidly evolving AI technology offers another possibility that may
help with understanding
better the possibilities of communication with \hbox{ETI}, by training the AI systems with scientific facts
from mathematics, physics, and chemistry, rather than with human-generated content (say, the Web),
which inevitably contains deep anthropocentric biases of all kinds.

Given that AI's technological evolution, which is still controlled by humans, is already several 
orders of magnitude faster than the biological evolution that led to
primate intelligence, it is
possible that the dominant form of ETI is some form of \hbox{AI}, and that its manifestations may be
beyond the scope of human understanding.  We will refrain from any further speculations
in this regard.

\chapterimage{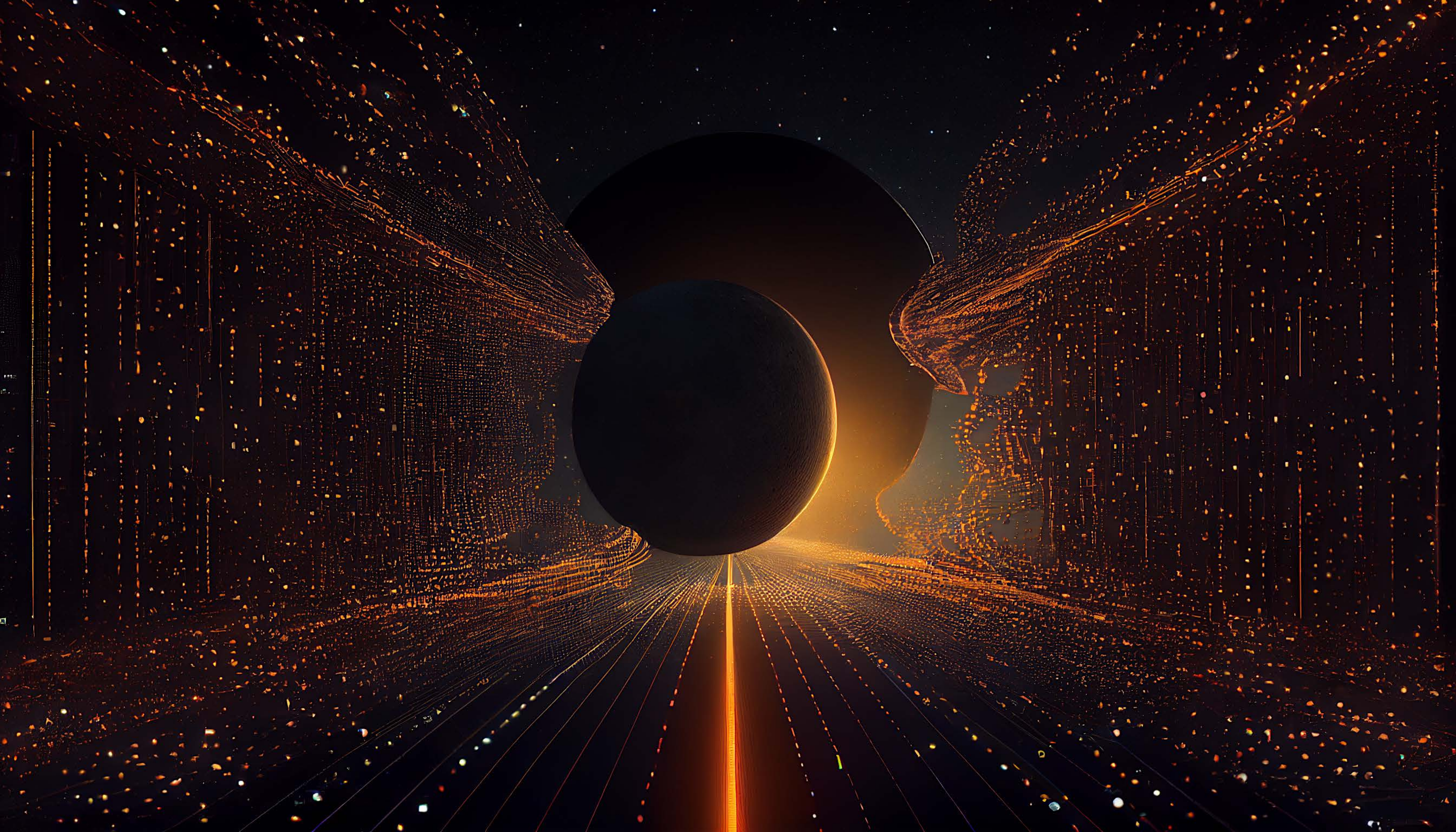}

\chapter{Types of Technosignatures}\label{chap:technosigs}

Consistent with discussions within the larger community, the KISS workshop participants focussed on two broad classes of technosignatures: those resulting from emissions produced by another civilization and those associated with structures engineered by another civilization.  The former has been the focus of much work in technosignatures or SETI in the past decades, but improvements in terrestrial technology and particularly efforts to detect extrasolar planets, and potential biosignatures from them, mean that detecting engineered structures is increasingly feasible.

\section{Intentional vs.\ Unintentional Technosignatures}\label{sec:technosigs.kind}

The search for technosignatures, particularly in the SETI context, has often focussed on \emph{intentional} technosignatures, in which a civilization undertakes an activity specifically to attract the attention of another civilization.  Often searches have been structured toward finding (electromagnetic) signals transmitted in a manner designed for reception and understanding (decoding), or even explicitly directed at Earth.  However, it has long been recognized that it might also be possible to \emph{eavesdrop} \citep[e.g.,][]{1992AcAau..26..185B}, or detect signals that a civilization might transmit for its own purpose with no expectation for reception by any other civilization.  Such transmissions are often termed ``leakage.''
Indeed, such signals might even be encoded, though with codes not intended for communication; radar signals are prime examples, and terrestrial radars for planetary science and planetary defense are used commonly as examples of signals that would be detectable over interstellar distances.
More generally, civilizations may develop technologies that are detectable over interstellar distances, even though detection by another civilization is not the intent.

As discussed earlier in this report (\S\ref{sec:forward} and Chapter~\ref{chap:others}), deciphering ETI communications that are the result of leakage or from eavesdropping is likely to be an extremely challenging activity.  Notwithstanding such challenges, determining that intelligent civilizations exist elsewhere of course would still provide ample rewards for biology and astronomy.  Less discussed, but still notable, is the possibility that an ETI might choose deliberately to encode transmissions in a manner that would make detection or decoding or both difficult.

\section{Communication}\label{sec:technosigs.comm}

The detection of technosignatures and SETI have often focussed on signals intended for communications, perhaps because there are ample examples of communications from animal life on Earth, with birds, cetaceans, and primates being species that make notable use of coded communications. 
The question of what type(s) of signals for which it makes the most sense to search, when looking for communications from ETIs, has been the subject of a good deal of study.  Indeed, the approximately 500-page book \textit{SETI~2020: A Roadmap for the Search for Extraterrestrial Intelligence}, intended to provide a roadmap for searches for the next two decades, is mostly devoted to this topic.  It summarizes the work of approximately 40 eminent scientists and engineers (from both academia and industry)  who met together for four workshops over two years.
Many of their arguments remain relevant, and here we only will summarize their findings briefly, with additional mention of a few related papers and add some very brief commentary of our own.

The following thought orients all the rest: assuming that there are some ETIs attempting to communicate with us (and presumably a host of other possibly inhabited planets), the fact that we do not already see an unmistakable beacon in the sky implies that these ETIs are operating under limiting constraints, and so communicating efficiently will be a priority.   Since we need to meet the ETIs partway, it makes sense to consider efficient communication.  A common proxy for this is the (transmitted) energy per (detected) bit of information.   From that starting point, everyone arrives at the following two conclusions.

\begin{enumerate}
    \item \textbf{It violates common sense to search for ETI signals carried by gravitational waves or neutrinos.} Both interact so weakly with normal matter that the energy cost to transmit even a few detected bits of information is \emph{many} orders of magnitude greater than for transmitting the same information via electromagnetic waves.  It is similarly impractical to send information with particles such as protons, since, e.g., the trajectories of protons with energies below about~$10^{12}$~GeV are severely bent by the Galaxy’s magnetic field, while a 1~eV photon travels straight, and with little attenuation by dust in most directions. So we should only consider EM searches.

    \item \textbf{In order to minimize the size of the background},
    and therefore maximize signal-to-noise ratio at fixed power, the
    ETI will communicate in a narrow frequency band, or in very short
    pulses, or with some combination of those two strategies.  Most
    current searches are based on just these assumptions.
\end{enumerate}
Beyond those two points, one can consider various ETI observing strategies, and then consider their ``cost''---the proxy for which is the total energy required for the search---and then selecting the waveband(s) at which to look by assuming that the ETI is attempting to minimize cost at (something like) fixed number of contacted civilizations.
At this point, observing strategies are roughly separated into two main types: all-sky and directed searches in promising-looking region (e.g., solar systems with planets in the habitable zone).
Broadly, for all-sky searches, Shannon's Theorem suggests that the optimum is at long wavelengths (e.g., radio), as the energy/photon scales as $1/\lambda$.
On the other hand, for directed searches, the fact that
shorter wavelengths can be better focused (for fixed telescope size),
by both transmitter and receiver complicates the relation between
waveband and cost, and so further considerations (e.g., the ETI
communication strategy and the cost of telescope construction, for both the ETI and us) have to be specified/guessed in optimizing the search band. 

\textit{SETI~2020} recommended feasibility studies of, or construction of test pieces for, three projects. One recommendation helped lead to construction of the Allen Telescope Array (ATA) of radio telescopes (currently containing about 10\% of number of planned number of dishes). Another suggestion was laser searches in the optical or infrared, as later demonstrated by \citet{2004ApJ...613.1270H}, and the third was the Omnidirectional Sky Survey (OSS), for which there are now examples \citep{2015PASP..127..234L,2019PASP..131g5001R,2018SPIE10702E..5IW}
though SETI has not been a primary motivation for most systems.

We shall not recapitulate here the much more complete treatments in \textit{SETI~2020} but end this section by mentioning how the assumptions of the various authors of \textit{SETI~2020} might be too restrictive.
An implicit assumption underlying much of the discussion of ``favorable bands'' is that the ETI is, in some sense, ``paying'' for all the energy it uses.
    However, one can envision the ETI making use of pre-existing
    energy sources. One example, which has been studied to some
    extent, is to use naturally occurring astronomical masers to
    amplify signals that are beamed through
    them  \citep[e.g.,][]{Cordes1993,2015NewA...34..245C}.
     Another idea, discussed
    further in Chapter~\ref{chap:search} (``Recommended Searches'') is
    that a highly advanced ETI could modulate the energy output of
    active galactic nuclei ({AGN}), e.g., by using disk instabilities
    to modulate the accretion rate. The idea is that the signal power
    comes for free; one just has to introduce some modulation that
    connotes intelligence.

\subsection{Radio Frequencies}\label{sec:technosigs.comm.radio}

The use of radio frequencies as a means of interstellar communications dates to at least the discovery of celestial radio emissions.  In its coverage of Karl Jansky's discovery of radio emissions coming from the direction of the center of the Milky Way Galaxy \citep{j33}, the New York Times noted (as part of the headline), ``No Evidence of Interstellar Signaling.''  Undoubtedly, this headline was influenced by the fact that terrestrial radio communications was still a relatively new area, and potentially by the fact that Jansky's discovery had been initiated his work in support of the Bell Labs' objective of improving long-distance communications.

The modern era of consideration for interstellar communications with radio frequencies is dated commonly to the work of \cite{1959Natur.184..844C}.  The attractiveness of radio frequencies for interstellar communications is motivated by three factors:
\begin{enumerate}
    \item Radio waves are relatively inexpensive to produce.  A single Watt of power produces of order $10^{24}$ radio photons at the frequency of the hyperfine transition of neutral hydrogen (1420~MHz), which is one of the classical frequencies for consideration of interstellar communications.  The reasoning is that any civilization capable of considering interstellar communications would be aware of this quantum mechanical effect from the most abundant element in the Universe.
    \item Radio waves can be focussed relatively effectively.  Because of the inverse square law, whether a signal is detectable over interstellar distances depends upon, in part, both the power at which it is transmitted and the distance.  A common metric is the \textit{effective isotropic radiated power} (EIRP), which can be enhanced by focussing the transmission.  With human civilization having already produced multiple radio telescopes with significant focussing capabilities (notably the now-collapsed Arecibo telescope and the Five-hundred meter Aperture Spherical Telescope [FAST]), it is plausible that other civilizations would be capable of similar constructions.
    \item Radio frequencies suffer little absorption, even on Galactic-scale distances.  For instance, radio sources in and near the center of the Milky Way have been observed since the early days of radio astronomy \citep[e.g.,][]{1960MNRAS.121..171O}, while observations of the Galactic center at visible wavelengths can suffer up to~30 magnitudes of extinction.
\end{enumerate}

\subsection{Optical Pulses}\label{sec:technosigs.comm.opticalpulse}

Only two years after \cite{SETI} suggested that extraterrestrial
civilizations could be detected by broadcasts at radio
frequencies, and only a year after the demonstration of stimulated
emission at visible wavelengths, \cite{1961Natur.190..205S} proposed
that laser transmissions would be an effective means of communicating
over interstellar distances.  

The attactiveness of lasers for interstellar communications is
motivated by the following factors:
\begin{enumerate}
\item Laser transmissions naturally have a high directivity, so that
transmissions have the potential to reach to larger distances while
remaining detectable.

\item The sky is dark on nanosecond time scales.  Over interstellar
distances, the average number of photons received from a dwarf star is
less than one per nanosecond.  By constrast, it is quite feasible to
produce nanosecond-duration pulses at visible or infrared
wavelengths.  Thus, any source of repetitive nanosecond-duration
pulses reasonably can be interpreted as a deliberate transmission.
\end{enumerate}

One potential concern with laser transmissions is that propagation
distances in the disk of the Milky Way Galaxy are not expected to be
more than a few kiloparsecs due to dust obscuration.  However, this
concern can be mitigated by considering the use of near-IR lasers,
which can propagate over Galactic-scale distances with only modest
absorption.

While the concept of searching at visible or infrared wavelengths has
had a steady presence in the
literature \citep[e.g.,][]{1993SPIE.1867...75K}, the number of
searches has been considerably smaller than those conducted at radio
frequencies.  More recently, there have been an increasing number of
efforts to search for laser pulses \citep{2004ApJ...613.1270H,2018SPIE10702E..5IW,2019AJ....158..203M}.

Finally, there also have been multiple suggestions of using lasers as
a means of propelling interstellar spacecraft, as doing so would
reduce or eliminate the need for the spacecraft to carry fuel \citep[e.g.,][]{1966Natur.211...22M,2015JBIS...68..172L}.
Such
lasers naturally would have high powers and be capable of being
detected over interstellar distances.

\section{Large-scale Manifestations of Technological Civilizations}\label{sec:technosigs.techcivs}

\subsection{Waste Heat from Dysonian Structures}\label{sec:technosigs.dyson}

At the dawn of the era of modern \hbox{SETI}, \cite{dyson60}
recognized that the exponential growth of an extraterrestrial species'
energy demand could lead it to occupy ``an artificial biosphere which
completely surrounds its parent star.''
Such a ``Dysonian structure'' could not be 100\% efficient in
capturing and using the energy emitted by the host star, and an the
observational consequence of the ``energy metabolism'' of such an
extraterrestrial technology likely would be detectable at wavelengths
near~10~$\mu$m.  
\cite{kardashev64} dubbed such a structure a ``Dyson sphere,''
although the original suggestion was quite general, and 
 \cite{dysonletters} and \cite{Dyson66} clarified subsequently that he was referring to a swarm of orbital structures, not a solid sphere or any other specific form of engineering.

The focus by \cite{dyson60} on the waste heat of technology is
probably the most ``robust'' technosignature proposed to date. It is a
completely general approach focuses on inevitable aspects of
technology (the waste heat that accompanies any energy use, as
required by the laws of thermodynamics) instead of relying on
guesswork on the motivations, purposes, or engineering methods of
extraterrestrial beings.

Searches for Dysonian strutures are thus simply searches for {\it any}
technology that commands a large energy supply.  As such, a search of
the sky at those wavelengths for point sources of emission would be
worthwhile.
\cite{2020SerAJ.200....1W} provides a more extensive discussion about Dyson structures.

\subsubsection{Previous Searches for Stellar Dysonian Structures}\label{sec:technosigs.dyson.search}

The first systematic search for Dysonian structures was
by \cite{jugaku04}, who focused on {\it IRAS} flux measurements
of~12~$\mu$m emission from fewer than 400 nearby, solar-type stars,
with upper limits reported simply as $K-[12]$ colors. The second was
by \cite{carrigan09a} who examined 11,224 point sources in {\it IRAS}
with low-resolution spectra, using those spectra to classify infrared
bright sources and reject natural confounders. He found three
interesting targets, all likely asymptotic giant branch (AGB) stars. He presented an upper limit of no Dysonian structures with more than 1~$L_{\odot}$ of waste heat within 300~pc.

\cite{Arnold05} pointed out that Dysonian structures would be easily identified in transit by observatories such as {\it Kepler}, which by design were sensitive to orbiting material of characteristic size $R_{\oplus}$ or larger. \cite{GHAT4} explored all of the anomalous signatures of such so-called ``megastructures'' (i.e., particularly large, individual components of a Dysonian structure) that would distinguish them from planets, and recommended a systematic search for them in that data.   \cite{Zackrisson2018} recommended a search strategy for Dysonian strutures composed of so many smaller components that they act as an opaque screen, using {\it Gaia} to identify apparently underluminous stars and \textit{Wide-field Infrared Survey Explorer} (\textit{WISE}) to check them for corresponding amounts of waste heat.

\subsubsection{Previous searches for Extragalactic Dysonian Structures}\label{sec:technosigs.dyson.galaxy}

There have also been a handful of searches for galaxies endemic with Dysonian strutures, on the theory that any civilization capable of building such technology could also spread throughout its galaxy. To analogize with humanity: humans learned how to spread and settle the globe before they learned to build cities, and as such there are cities across its surface. One might therefore expect that any galaxy with even a single spacefaring species would have a significant fraction of its stars surrounded by Dysonian strutures. At the very least, upper limits could be calculated on a galaxy-by-galaxy basis. 

The first search was by \cite{annis99b}, who examined the optical properties of 137 galaxies in a cluster, searching for those that appeared underluminous (from the large number of Dysonian strutures obscuring its starlight). He compared each galaxy's optical flux to the values expected from the Tully-Fisher and fundamental plane relations, finding no galaxies with $>75\%$ of their starlight obscured.  

The second was the waste heat search of \cite{GHAT3}, who examined {\it nearly all} of the $10^5$ resolved sources in the {\it WISE} all-sky survey, identifying the most extremely infrared-bright among them. This allowed them to put a rather weak upper limit of $85\%$ as the amount of starlight reprocessed as waste heat among any of them, and an upper limit of $50\%$ starlight reprocessing among all except 50 sources (all presumably starburst galaxies). They pointed out that further work to quantify the amount of dust expected from each source would allow significantly more stringent upper limits on these galaxies.

Finally, \cite{Zackrisson15} used a hybrid approach, following \cite{annis99b} in applying the Tully-Fisher relation to $\sim$1,500 disk galaxies, assessing those that appeared to be underluminous, and then diagnosing candidates with a search for infrared emission.

\subsection{Sources With Unusual Variability Patterns}\label{sec:technosigs.vary}

If a hypothetical advanced civilization is capable of building a
structure such as a Dyson
sphere, possible observable technosignatures include unusual variability patterns of their 
star, due to partial eclipses by the circumstellar structures, or other large-scale engineering 
activity.  A possible candidate for such a technosignature was
Boyajian's star \citep[KIC~8462852, TYC~3162-665-1, 2MASS~J20061546+4427248,][]{WTF}.  The star shows 
irregular partial dimmings that can be interpreted as partial eclipses by substellar objects, 
whether natural or artificial.  Multiple potential explanations of
both kinds have been proposed \citep{Wright16}.    
The cause of these events is still not fully established, but a prudent attitude may be that if a 
plausible natural explanation is available, there is no need to invoke
a  technosignature.

A more dramatic proposal is that such hypothetical stellar-scale technosignatures may be 
observable as stars that appear or disappear on decadal time
scales \citep{Villarroel16}.  These authors search for star-like sources present in the archival scans of photographic 
sky surveys (obtained from the 1950s to the 1990s) that are not present in the more modern, digital 
sky surveys such as SDSS or \hbox{PanSTARRS}.  One issue with this approach is that photographic sky 
survey plates are notorious for artifacts that are due to either contamination in the 
photographic emulsions or in the hypersensitization or development process, e.g., due to 
microdroplets of chemicals used.  Such artifacts were known as ``Kodak objects.''  During the 
processing of DPOSS (digital version of POSS-II), it quickly became clear that no detection on an 
IV-N emulsion plate (roughly the $i$~band) can be believed without a matching detection on the 
corresponding emulsions corresponding to the blue and red bandpasses.  Some of these ostensible 
transients are probably real, corresponding to known classes of
phenomena, such as supernovae in dwarf host galaxies, cataclysmic variables, flaring stars (e.g., UV~Ceti), blazars, \ldots, 
for which the low state is below the flux detection threshold in the
survey in which the star is 
ostensibly missing.  Such transients are detected in copious numbers in modern
digital sky surveys.  Indeed, at any given moment, there will be of
the order of~$10^3$ such transients at magnitudes brighter than
$18^{\mathrm{m}}$--$19^{\mathrm{m}}$ magnitude over the entire high Galactic latitude sky (and many more on the low 
Galactic latitudes), which is sufficient to explain any real
transients found by the VASCO team \citep{2020AJ....159....8V}.

Modern digital sky surveys are not immune from instrumental artifacts, due to other causes, such 
as CCD saturation residuals, head-on cosmic rays, crosstalk between different channels in the 
readout electronics, \ldots.  Survey teams develop filters that
eliminate most of such artifacts (so-called ``real-bogus'' filters),
but some can leak through.  As a rule, at least two independent detections are required in order 
to confirm transients, which may be separated by minutes to years.  They cover areas comparable 
to the old photographic sky surveys but obtain hundreds to thousand of exposures per surveyed 
field, usually deeper, and with a higher data quality and better sampling.

One of the scientific goals of the modern survey is to look for
objects with unusual variability patterns, defined broadly, but using
well-sampled light curves with at least tens of detections and
spanning time baselines from minutes to over~20~years.  The field of
astronomy is now in the regime of billions of sources with up to
trillions of measurements and growing.  To date, no objects have been
found that could not be explained by natural causes, but the
possibility remains.  Such large data sets and data streams, in combination with machine learning for the detection of outliers, 
are exactly what inspired this workshop.

\section{Physical Artifacts in the Solar System}\label{sec:technosigs.probes}

The concept that there may be physical objects in the Solar System that would constitute technosignatures has long been discussed in the literature \cite[e.g.,][]{b74,f80} and is well motivated by our own spacecraft.  Pioneer~10 and~11, Voyager~1 and~2, and New Horizons have all either left the Solar System or are on trajectories to escape the Solar System, and multiple concepts for missions designed to leave the Solar System have been studied by KISS itself, \hbox{NASA}, \hbox{ESA}, and NASA Centers, including \hbox{JPL}.\footnote{
At the time of writing of this report, a white paper describing the
Interstellar Probe
concept, \texttt{http://interstellarprobe.jhuapl.edu/}, has been
submitted for consideration as part of the Solar \& Space Physics
Decadal Survey.  The concept study was conducted by the Applied
Physics Laboratory of the Johns Hopkins University and was motivated
in part by a previous KISS study, ``Science and Enabling Technologies
to Explore the Interstellar Medium.''}
While none of these were or would be designed to be operational on the
time scales relevant for reaching a nearby star ($> 10\,000$~yr), the
existence of these spacecraft establishes an existence proof that it
is feasible to launch interstellar probes.

Further, multiple landers and rovers dot the surfaces of the Moon and
Mars,  and satellites in or near geosynchronous orbit (GEO) are
expected to remain in orbit for time scales of tens of thousands of
years or longer.  Inspired by the Pioneer plaques and the Voyager
Golden Records, one of these GEO satellites, \hbox{EchoStar~XVI},
carries a deliberate message, the Artifact \citep{wp12}.

Whether operational or not, and whether aimed intentionally at the Solar System or not, finding a spacecraft or other artifact of non-terrestrial origin in the Solar System would be an unambiguous technosignature.

Table~\ref{tab:probes} summarizes the classes of physical artifacts that might exist within the Solar System, and it is possible that a physical object could transition from one class to another (e.g., from passive to active by charging batteries or from active to passive from a component failure).
The utility of artifacts as technosignatures, be they intentional carriers of a message or merely relics of exploration, has long been recognized \citep{b60}, because of their persistence (``time capsule'' or ``message in a bottle'' nature) and time-integrated data volume.
Even for passive probes, the total amount of information stored on such objects would be limited only by the total mass and available media used to store and retrieve data. From etched drawings to gold-plated data disks, many forms of hard-coded media could be used to ensure the persistence of information over time, impervious to the deleterious effects of Galactic cosmic rays. 
Indeed, \cite{rose2004inscribed} argued that the efficiency of inscribed matter for information storage on spacecraft sent from one civilization to another made them vastly more efficient for interstellar communications than transmissions of photons.

\begin{table}
\centering
    \caption{Classes of Physical Technosignatures\label{tab:probes}}
    \begin{tabular}{l|p{0.4\textwidth}|p{0.4\textwidth}}
    \noalign{\hrule\hrule}
          & \textbf{Active} & \textbf{Passive} \\
    \noalign{\hrule}
    \textbf{Probe} & Object on orbit, either stable or hyperbolic, that uses an energy source, either internal or solar radiation, to conduct measurements or transmit signals, not necessarily aimed at Earth
          & Object on orbit, either stable or hyperbolic, that undertakes no actions \\
    \noalign{\hrule}
    \textbf{Surface Artifact} & Object on surface of planetary body that uses an energy source, internal to it, from the body, or from solar radiation, to conduct measurements or transmit signals, not necessarily aimed at Earth
          & Object on surface of planetary body that undertakes no actions \\
    \noalign{\hrule}
    \end{tabular}
\end{table}

Persistence, and thus the availability of information from a technosignature, aids in partially solving the ``problem'' of a civilization's lifetime~$L$ in the Drake equation.  Radio or optical transmissions might be fleeting signals compared to the persistence of a well-placed material object.

Below we briefly review the positive and negative attributes of each category and type of artifact.

\clearpage
\subsection{Probes}\label{sec:comm.probes}

Consideration of probes, either passive or active, is particularly timely now that multiple interstellar objects (1I/`Oumuamua and 2I/Borisov) have been identified.  It is also notable that 1I/`Oumuamua was identified initially as an object with an unexpected orbital parameter (namely an eccentricity $e > 1$).

\begin{figure}[bt]
    \centering
    \includegraphics[width=0.4\textwidth]{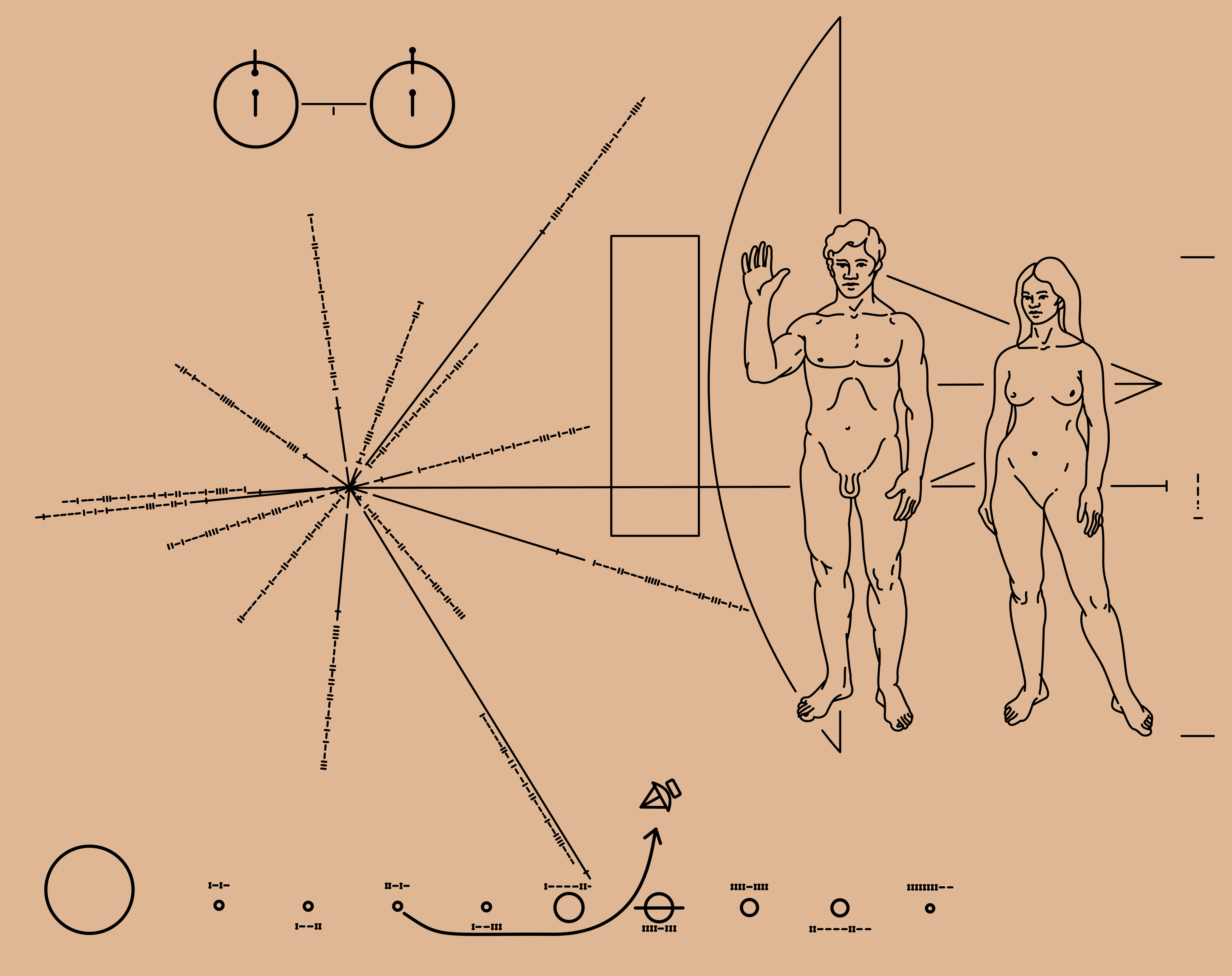}\hfil%
    \includegraphics[width=0.47\textwidth]{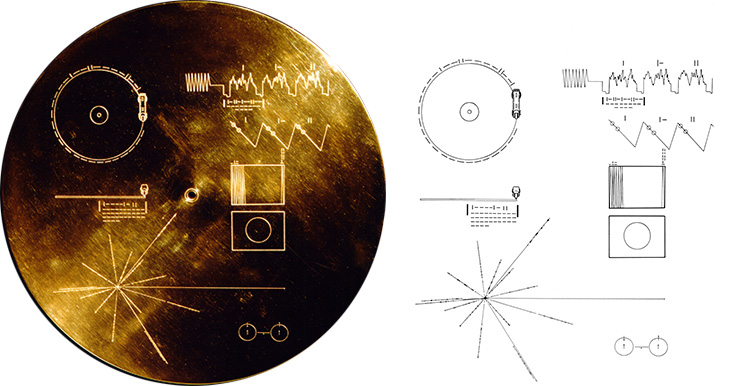}
    \vspace*{-1ex}
    \caption{Known technological artifacts in our Galaxy.
    (\textit{left})~Illustration of plaques carried by Pioneers~10 and~11.  
    (\textit{right})~Voyager Golden Record, copies of which are on Voyagers~1 and~2.
    (Credit: NASA)
    }
    \label{fig:pioneervoyager}
\end{figure}

The Pioneer~10 and~11 spacecraft each carried an identical golden plaque, on which was inscribed information about the hydrogen atom (in an effort to provide an universal context for translation of the plaque), information about the spacecraft and human beings, and lastly, a map showing the position of the Sun relative to known pulsars, and the position of the Earth within our Solar System (Figure~\ref{fig:pioneervoyager}).
The Pioneer~10 spacecraft will, in approximately two million years, approach the vicinity of Aldebaran, some 65 light-years from Earth, which may host a planet, or planets, of its own \citep{2015A&A...580A..31H}.
In approximately four million years, Pioneer~11 will pass by a star in the constellation Aquila, for which there is currently not any knowledge of candidate exoplanets.

The Voyager spacecraft, meanwhile, are still conducting their Voyager
Interstellar Missions (VIM) and sending back data about the heliopause
and edge of our Solar System.\footnote{
  They are expected to continue operations at least into the late 2020s before power levels from their radioisotope thermal generators become too low.}
Both carry ``Golden Records,'' 30~cm-diameter, gold-plated copper disks intended to describe the Earth and the life upon it, should either spacecraft be found (Figure~\ref{fig:pioneervoyager}).\footnote{\texttt{https://voyager.jpl.nasa.gov/golden-record/}}
In approximately 40,000 years, Voyager~1 will pass within~0.5~pc of a star in the constellation of Camelopardalis
  (star AC$+$79~3888). At about the same time, Voyager~2 will pass
  within~0.5~pc of 
  the star Ross~248, after which it
  will likely continue on, eventually approaching Sirius at a distance
  of~1.3~pc roughly 296,000 years from now.
Though these four missions do contain intentional artifacts for direct communication (plaques and records), the spacecraft themselves are obviously also a technosignature for any would-be civilization that these spacecraft might encounter.\footnote{%
The classic 1979 film, ``Star Trek: The Motion Picture'' captures this concept wonderfully in the denouement and revelation of \emph{V'Ger}.}

The concept of active probes in the Solar System was introduced
by \cite{b60}, if not earlier, and continues to engender
discussion \citep{2019AJ....158..150B}.  Multiple possibilities have
been discussed, both in the scientific literature and in science
fiction, for the locations of active probes in the Solar System,
ranging from co-orbiting with the Earth to Lagrange points to the Main
Belt of asteroids.  A small number of searches for such probes have
been
conducted \citep[e.g.,][]{Freitas80,1983Icar...53..453V,1985AcAau..12.1027F},
all with null results.  Additional, less obvious stable orbits in the
Solar System may be found through dynamical calculations, and, if so,
would be worth exploring observationally.

Less often discussed are the challenges of achieving a stable orbit within the Solar System.  An object entering the Solar System does so with a relative velocity of tens of kilometers per second.  For reference, the Voyager spacecraft are traveling at speeds of approximately 15~km~s${}^{-1}$, achieved in part by multiple gravity assists past giant planets.  Shedding this relative velocity violates no known laws of physics but is a substantial engineering challenge.
There are potential solutions, e.g., the spacecraft could
``hibernate'' in a manner similar to that employed by the Rosetta and
New Horizons missions during interstellar transit, but these can lead
to other engineering challenges, such as the requirements on lifetimes
of spacecraft components or the need to acquire raw materials for
fashioning replacement components.  Finally, it is notable that
``fly-by'' concepts, such as Project
Daedalus \citep{1975JBIS...28..147B} and Breakthrough
Starshot,\footnote{%
https://breakthroughinitiatives.org/initiative/3}
are envisioned as this civilization's first efforts in interstellar travel.

Whether active or passive, the search for probes is highly complementary to explorations of small bodies in the Solar System.  Characterizing the small body population, by a combination of remote sensing (radio, near-infrared, visible) and \textit{in situ} exploration, could reveal extremely anomalous objects that are determined to be probes and finding additional interstellar objects, such as 1I/`Oumuamua and 2I/Borisov, promises at least additional insights into the characters of other planetary systems.

Finally, in the spirit of this workshop, the possible conceit of assuming that an active probe would be investigating humanity is worth noting.\footnote{This topic has been explored richly in science fiction, including ``Star Trek~\hbox{IV}: The Voyage Home'' and \textit{Rendevous with Rama}.}

\subsection{Surface Artifacts}\label{sec:comm.surface}

\begin{figure}[bt]
    \centering
    \includegraphics[width=0.47\textwidth]{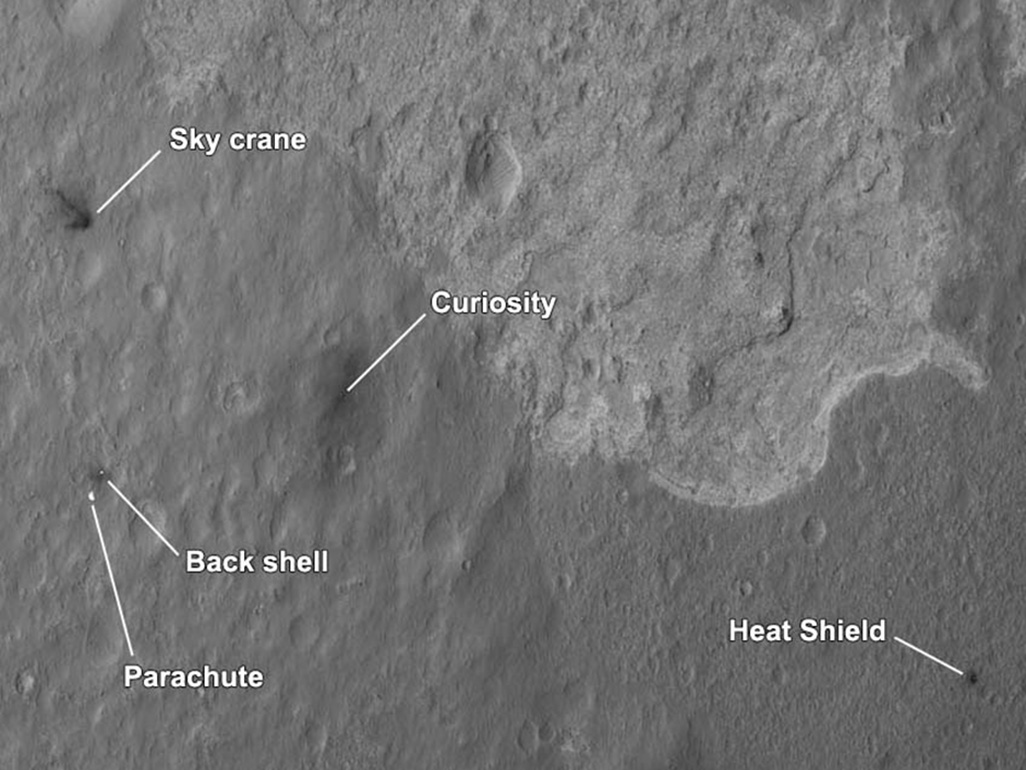}\,\hfil%
    \includegraphics[width=0.47\textwidth]{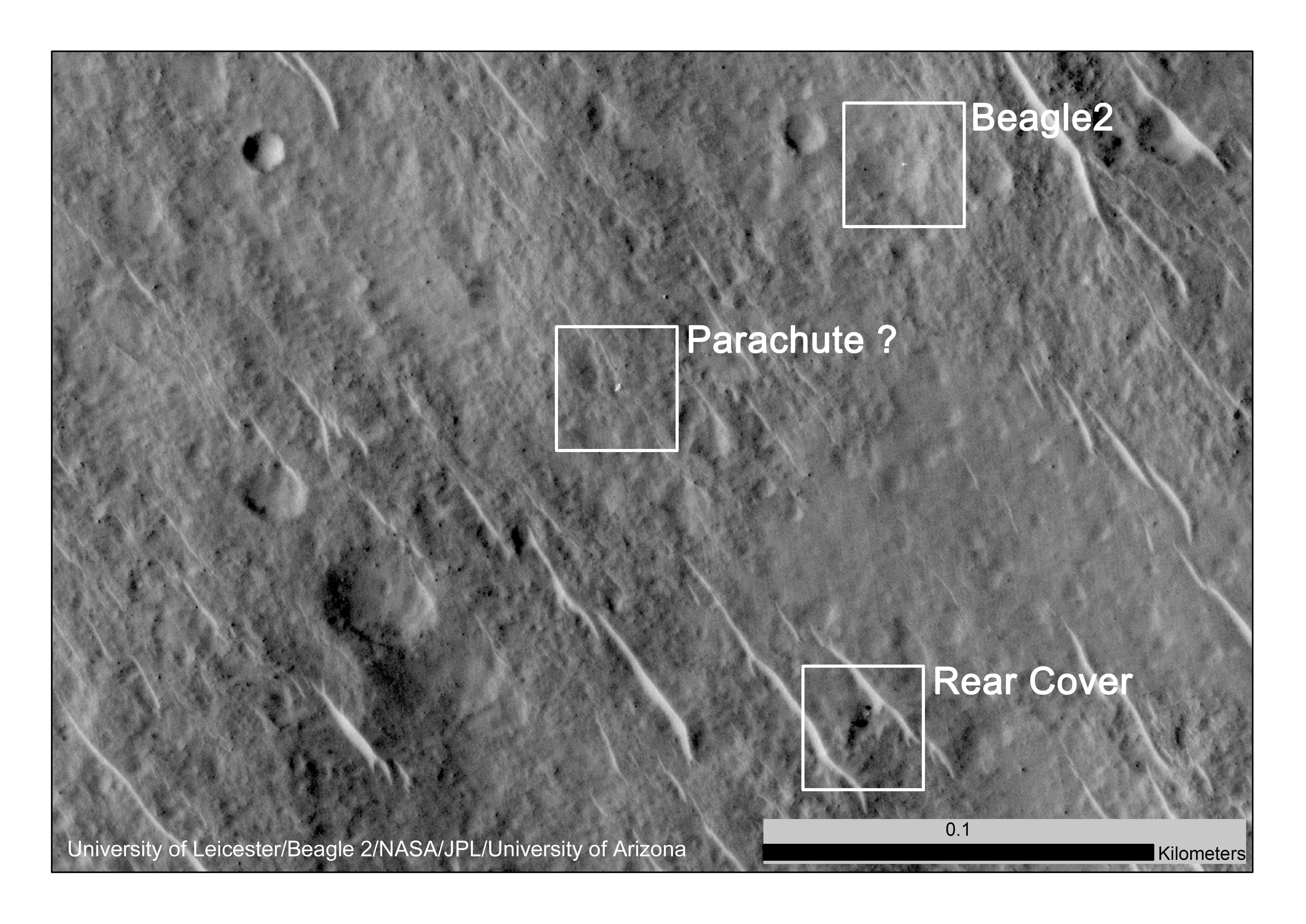}
    \vspace*{-1ex}
    \caption{Two examples of surface artifacts representing technosignatures in the Solar System, on the surface of Mars.
    (\textit{left})~Mars Science Laboratory/Curiosity rover and the components associated with its delivery to the surface of Mars, as imaged by the
    High-Resolution Imaging Science Experiment (HiRISE) camera on Mars
    Reconnaissance Orbiter (MRO).  (NASA/JPL-Caltech/Univ.\ of Arizona)
    (\textit{right})~Likely remnants of the Beagle~2 lander and the components associated with its delivery to the surface, also imaged by \hbox{MRO/HIRISE}.
    (NASA/JPL-Caltech/Univ.\ of Arizona/Univ.\ of Leicester)}
    \label{fig:surface.mars}
\end{figure}

Scattered throughout the Solar System are (terrestrial) technosignatures, with the most numerous being on the surfaces of Mars and the Moon (Figures~\ref{fig:surface.mars} and~\ref{fig:lroapollo}).  Searching for such technosignatures would be in the spirit of this KISS workshop, as only a small fraction of the surfaces of essentially all Solar System bodies have been even imaged.

Figures~\ref{fig:surface.mars} and~\ref{fig:lroapollo} illustrate some of the challenges associated with finding surface artifacts.  From a search perspective, both the spatial coverage and spatial resolution to detect any putative technosignature would be required.  For instance, the surface area of the Moon is approximately $38 \times 10^{12}$~m${}^2$; searching for technosignatures with sizes of order a few meters in size would require a search through a multi-terapixel image.  In contrast, all of the existing images of rovers or landers on the surfaces of the Moon or Mars have been facilitated by knowing approximately where to look.

The second challenge is implicit in the right panel of Figure~\ref{fig:surface.mars}.  Even with careful design, and knowledge of the characteristics of the target surface, landing on a surface can be difficult.  Moreover, if an object enters from outside of the Solar System, the relative velocity between it and a Solar System body might approach 50~km~s${}^{-1}$.  The result of an impact might be indistinguishable from that of a collision with a natural body.  Even if an extraterrestrial object were to present an initially recognizable surface feature, as a technosignature, it is less clear if it would be recognized as such over geologic time \citep{PITS}.

\chapterimage{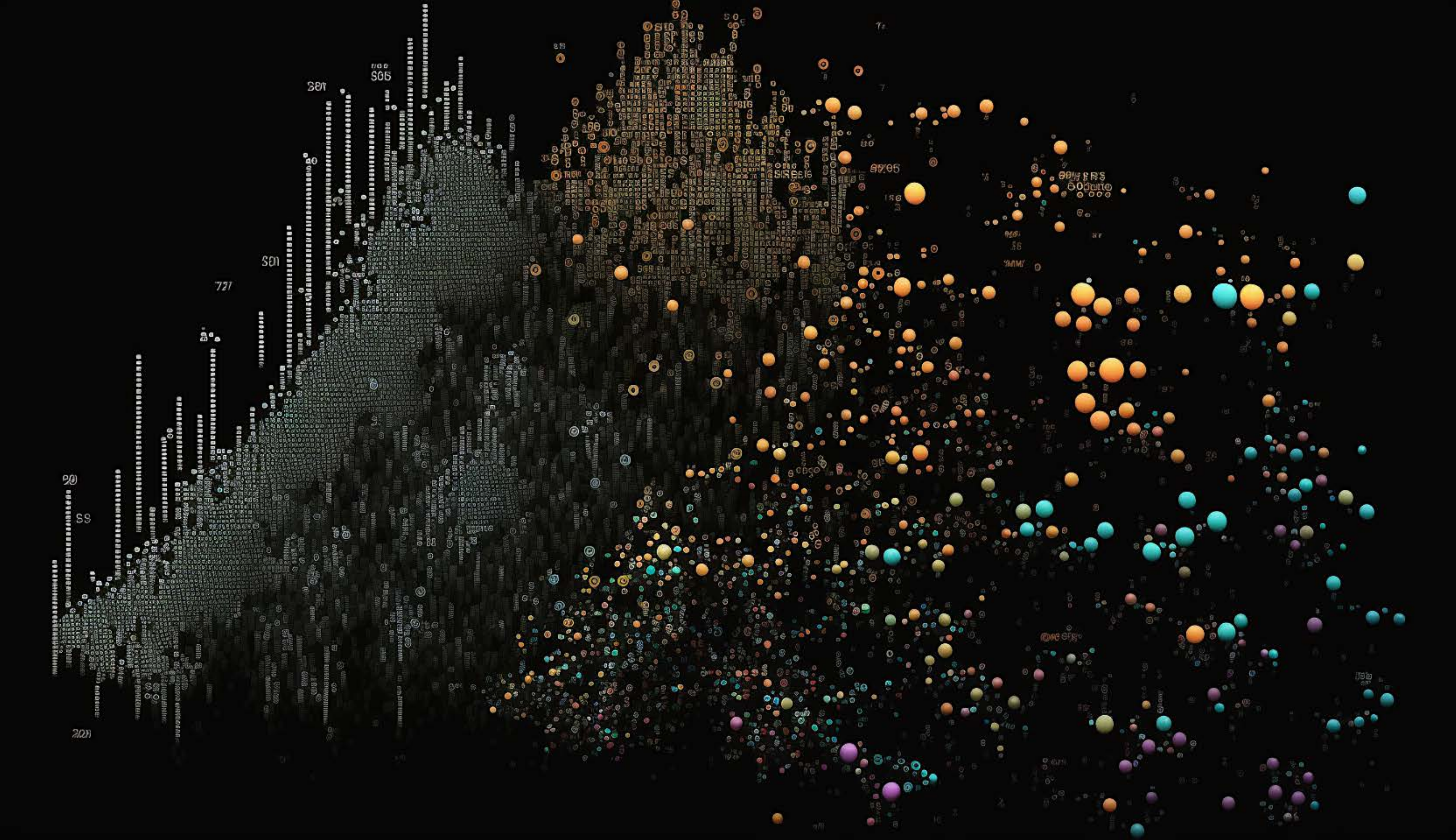}

\chapter{Applying Machine Learning}\label{chap:ml}

Machine learning methods can be used to detect and characterize signals of interest within large data sets.  They are of particular value in a ``big data'' era in which the volume of data or rate at which data appear is larger than could be analyzed by humans, and they are of value in that they can analyze data in a uniform and systematic manner without falling prey to various human frailities (e.g., fatigue, distraction).
We begin by discussing relevant methods for (i)~guided searches for
known signals of interest and (ii)~``blind'' searches for unusual or
interesting signals for which the characteristics are not known in advance.  We conclude with several caveats and limitations about the use of machine learning that are useful for any practitioner.

\section{Model-based Outlier Searches}\label{sec:ml.model}

When the signal of interest is known in advance, a guided search with a pre-specified template can highlight all matches to the signal.  This approach is equivalent to the standard template matching or matched filtering approaches in signal processing, but a variety of approaches beyond simply cross-correlating the observed data and the template for the signal of interest have been developed.

\begin{figure}[tb]
    \centering
    \includegraphics[angle=-90,width=0.95\textwidth]{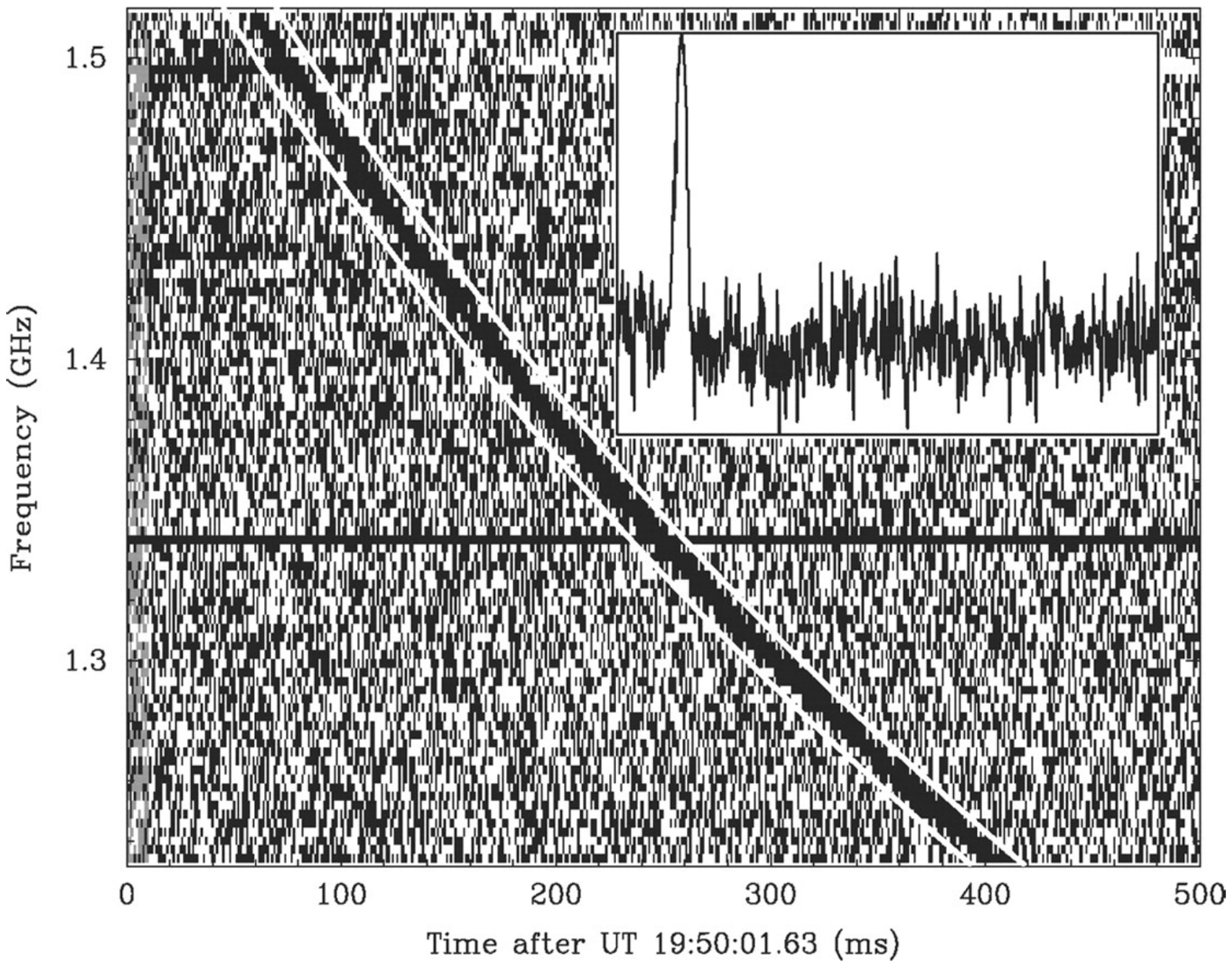}
    \vspace*{-1ex}
    \caption{Dispersed radio pulse illustrating how a model-based
    outlier search could be conducted.  The abscissa is the time of
    observation, and the ordinate is the frequency of observation.
    The dispersed radio pulse obeys the expected quadratic frequency
    sweep, indicative of propagation of a radio wave through a plasma.
    This quadratic frequency sweep is easily modeled, but this
    particular radio pulse was an outlier.  The duration (400~ms)
    required for it to sweep across the frequency band was far longer
    than expected given where the telescope was pointing.  This radio
    pulse was the first of a class of radio emitters now termed fast
    radio bursts (FRBs).  (From \citealt{2007Sci...318..777L}. Reprinted with permission from \hbox{AAAS.})}
    \label{fig:lorimer}
\end{figure}

For example, short-duration radio pulses (``impulses'') that originate from distant sources have characteristic curved shapes in time-radio frequency plots (Figure~\ref{fig:lorimer}).  The shape is marked by the pulse arriving earlier at higher radio frequencies with an increasingly longer time lag~$\Delta t$ to lower radio frequencies and having an expected radio frequency dependence $\Delta t \propto \nu^{-2}$.  This time lag or ``dispersion'' is a consequence of the plasma through which the radio pulse has propagated between the source and the Earth.  

Dispersed single pulse signals are well understood and modeled; they
are parameterized by their {\em dispersion measure}~DM or the total
electron content along the line of sight.  To search for single pulses, the radio data are de-dispersed to remove the dispersion associated with a given DM value, then summed across frequency channels to yield a one-dimensional time series, as shown in the inset in \autoref{fig:lorimer}.  When the DM value being tested matches the dispersion inherent in the signal, a strong peak in the de-dispersed time series appears.

\section{Unsupervised Learning}\label{sec:ml.unsupervise}

Large digital sky surveys, such as those mentioned in~\S\ref{sec:opportunity}, as well as any other onfoing or forthcoming surveys, ultimately produce catalogs of all detected sources, and for each of them measure a number of photometric or morphological parameters, typically tens or hundreds per source.  They form the parameter spaces, such as those described in~\S\ref{sec:discovery}, which can be explored using a variety of Machine Learning tools, and unsupervised learning and outlier searches.

If the signal of interest is not known or if even the presence of a signal is not known, an unsupervised approach can be a powerful means of identifying outliers.  This approach has a deep connection to historical approaches to developing understanding.  Scientists have long sorted objects into different classes or types, even if the underlying physical mechanism is not known.  Examples include dividing supernovae into Type~I and Type~II depending upon the absence or presence of hydrogen lines in their spectra; sorting stars into different spectral classes depending upon the particular spectral lines identified; and dividing organisms into species depending upon the physical characteristics.  As described above, the application of machine learning techniques has the potential benefit that the application of computer algorithms will not be subject to ``distractions'' to which a human might be subject.

\subsubsection{Time Series}\label{sec:ml.unsupervise.time}

The majority of astronomical time series are sparse, irregularly sampled, noisy, and with significant gaps (``gappy''). Even within a data collection, individual time series will vary in temporal baseline and number of data points. A basic approach to characterizing such data is to move to a more homogeneous representation in terms of a set of extracted features. \cite{2011ApJ...733...10R} defined what has become a standard set of statistical measures for time series including statistical moments, quantile ratios, variability measures, and period-based quantities. Libraries such as FATS \citep{2015arXiv150600010N} (and its updated version \texttt{feets} \citep{2018A&C....25..213C}) provide an easy way to calculate these. Once a uniform feature set is available, the data can be clustered and outliers identified as with catalog data (see \textit{Outlier Detection} below).

Inherent in this approach, however, are a number of unstated statistical assumptions that need to be borne in mind when interpreting any results. 
The first is that the observation errors on the measured data points are drawn from the same distribution (homoskedastic). Time series from ground-based sky surveys even using the same instrument consist of observations taken on different nights under different observing conditions, e.g., seeing, transparency, weather, etc. This means that the observing errors are drawn from different distributions (heteroskedastic) and so any feature needs to accommodate this, e.g., by using weighted quantities. Secondly, it is commonly assumed that the generating process for the time series is time invariant, i.e., stationary, so that the mean does not vary, for example. There are examples of astronomical sources, however, where this is clearly not the case: the microquasar GRS 1915+105 has at least 14 different states identified in its X-ray emission representing both deterministic and stochastic processes \citep{2017MNRAS.466.2364H}. Many features will be meaningless in such cases, for example, statistical moments. Local (weak) stationarity can be obtained by using first-order differences as the independent variable.

\paragraph{Outlier Detection}\label{par:ml.unsupervise.outlier}
Once catalog data have been pre-processed and features have been finalized, we can apply out-of-the-box ML methods, many of which are documented in multiple survey papers \citep{Hodge2004,chandola:anom09,goldstein2016}.

The most popular methods are now incorporated into major open source ML toolkits.  For example, \texttt{scikit-learn} \citep{scikit-learn} provides local outlier factor \citep[\hbox{LOF},][]{breunig00LOF}, the one-class support vector machine \citep[\hbox{SVM},][]{Scholkopf1999}, and the isolation forest \citep{liu:isoforest08}.  Many astronomy research projects now use these methods \citep{giles:dbscan-kepler19} or build custom algorithms \citep{alcock2005,Nun_2014,Nun_2016}.

A common feature of outlier detection methods is the calculation of an outlier score that is used to rank order examples by their outlierness.  However, the ML community has provided no guidance on mapping anomaly scores to human intuition until very recently.

Once an outlier detection method has been run, the next step is to inspect and review the outliers to determine which ones (if any) correspond to items of interest.  It is particularly helpful if the outlier detection method itself can provide a first step towards interpretation by indicating why a given item was flagged as an outlier.  We characterize such a method as {\em interpretable} since it provides a built-in explanation for its decisions.
Explainability for machine learning methods is an active area of research, with new methods under active development \citep[e.g.,][]{biran:expl-survey17}.

One category of interpretable outlier detection methods are those that
can highlight the features that were responsible for a given item's
identification as an outlier.  Examples include reconstruction-based
outlier detection methods such as the use of a low-dimensional data
model created via Principal Component Analysis (PCA) or Singular Value
Decomposition (SVD).  Data are projected into the space defined by the
first $k$~principal components, which discards some amount of
information, which is captured in the higher-order components.  The
data are then projected back into the original feature space and compared to the original form of the data.  The  difference between the two, which is also known as the {\em residual}, provides a per-feature explanation for why the item was selected.  The magnitude of the residual can be used as an anomaly score.  

SVD-based anomaly detection provides a ranking of all items in terms
of their global outlierness.  It is also possible to build an
incremental SVD model of the outliers themselves to aid in identifying
a {\em diverse} set of outliers rather than a globally ranked list.
The open-source Discovery via Eigenbasis Modeling of Uninteresting
Data (DEMUD) algorithm\footnote{%
\url{https://github.com/wkiri/DEMUD}}
provides this capability and generates explanations of each selected
outlier that highlights how it differs from the previously selected
outliers~\citep{wagstaff:demud13}.  DEMUD has been successfully applied
to a variety of data types, including emission
spectra~\citep{wagstaff:chemcam14} and images~\citep{lee:demud-image20}.

\subsubsection{Images}\label{sec:ml.unsupervise.image}

In principle, many of these same methods could be applied directly to the pixels of an image or images, but this very naive way is usually not feasible for two main reasons: (i)~the feature vector becames too long and therefore any outlier method based on a distance will fail and (ii)~the importance of small differences in the pixel scale becomes overloaded or exaggerated.
As such, some form of dimensionality reduction is required.

There is the possibility to deal with instruments designed to deal with images such as convolutional autoencoders \citep{autoencoders}.   Autoencoders consist of two mirrored parts, an encoder and a decoder.  The encoder is in charge of a reducing the dimensionality in a latent space while the decoder tries to reconstruct the original pattern starting from the latent space. A perfect autoencoder would result in an identity function.

Autoencoders can be used for outliers detections in two ways: (i)~We optimize the autoencoder in order to obtain the minimum reconstruction error on the whole set and then consider the patterns (images) for which such error is still big, which means that the autoencoder was not able to learn enough for this image since is somehow different from the remaining training set; or (ii)~We can use the lower dimentional space (latent space) and then apply an outlier method in this parameter space.

\chapterimage{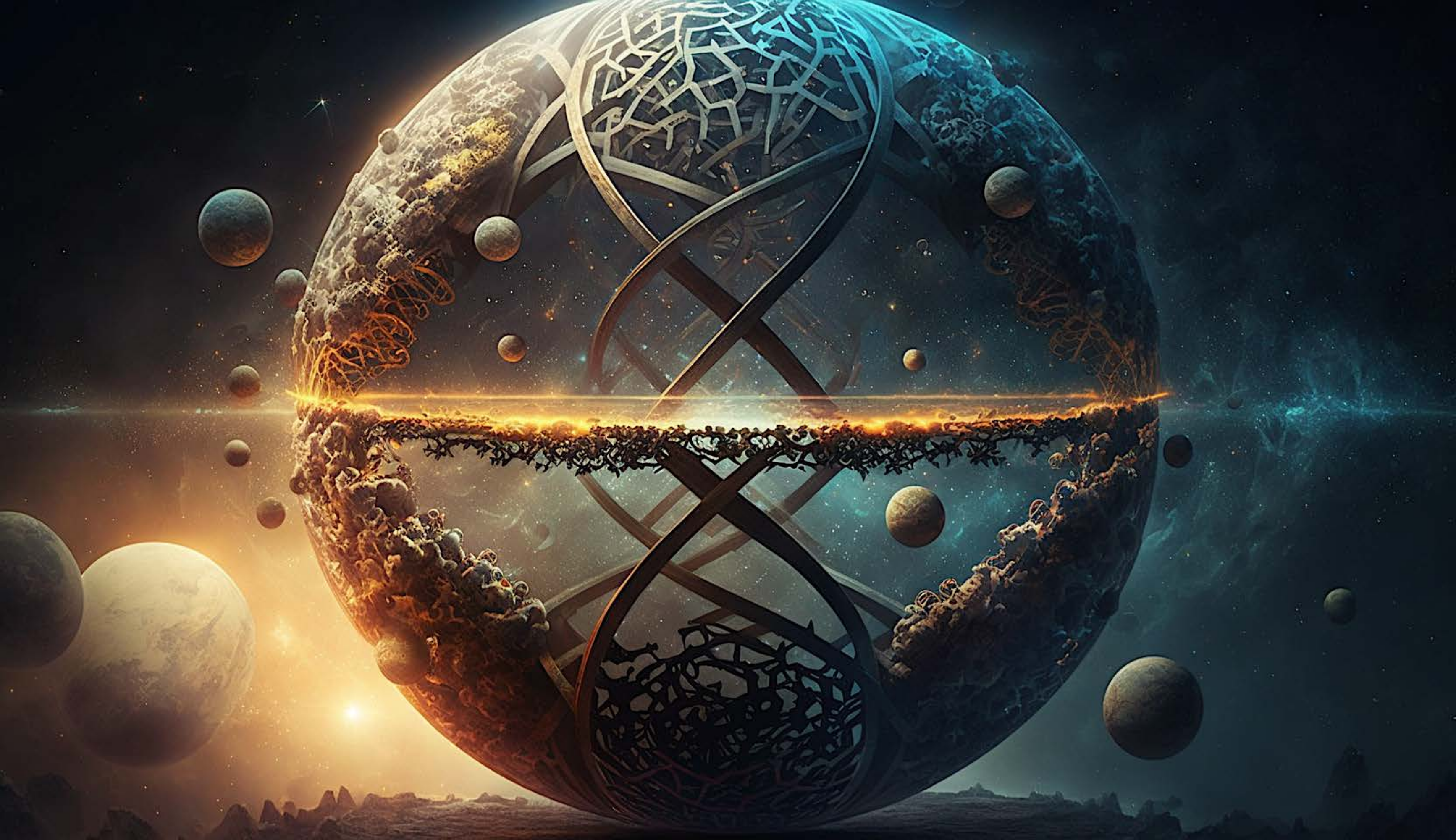}

\chapter{Recommended Searches}\label{chap:search}

During the KISS study, a number of particularly promising directions were identified by the participants, ranging from searching for objects in the Solar System to throughout the sky.  In all cases, a key element of the new search directions is that they are motivated by (much) larger data sets than were available previously and new approaches to processing the data.  However, the much larger data sets also bring an element of additional responsibility in reporting results and placing quantitative limits (or a confidence level on the detection of a technosignature!).  We summarize these directions identified at the KISS workshop, discussing them in no particular order.

\section{Mining Large Sky Surveys}\label{sec:skysurvey}

\subsection{Looking for Dysonian Structures}\label{sec:dysonsearch}

Both \cite{GHAT2} and \cite{Zackrisson2018} noted the promise that
{WISE} presented.  For Dysonian structures with effective
temperatures of $\sim$100--600~\hbox{K}, the infrared (IR) excess should be
maximized at~10~$\mu$m--30~$\mu$m, exactly wavelength range at which
the {WISE} performed an unprecedentedly sensitive all-sky survey in
its {\it W3} and {\it W4} bands. We thus now, for the first time, have
the data to conduct the search that F.~Dyson recommended in~1960.

\begin{figure}
    \centering
    \includegraphics[width=0.8\textwidth]{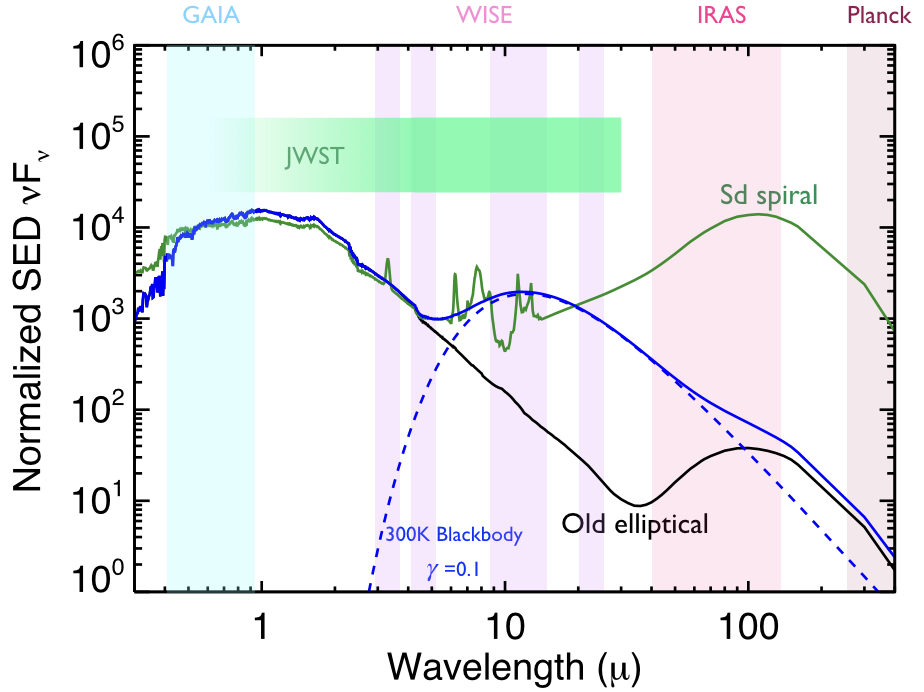}
    \vspace*{-1ex}
    \caption{Differences between the spectral energy distribution
    (SED) of an old elliptical galaxy (black, also characteristic of
    individual stars without significant circumstellar material), an
    Sd spiral galaxy (green, also characteristic of stars with
    significant circumstellar dust), and a Dyson sphere (blue, here
    parameterized as reprocessing only 10\% of the star's light
    at~300~K and transmitting the other 90\%).  Dyson spheres are
    distinguished from dust by their lack of \hbox{far-IR},
    sub-millimeter, or millimeter emission, and by their mid-IR
    spectra, which have no reason to exhibit characteristic
    polyaromatic hydrocarbon (PAH) or silicate features. Followup
    opportunities for Dyson sphere candidates include {\it JWST}
    spectra and studies with \hbox{ALMA}, many of which will be
    necessary anyway to study them as extreme and/or natural objects.
    The bandpasses of {\it JWST} and four all-sky surveys are shown
    for comparison. [From a white paper for the Astro2020 Decadal
    Survey by J.\ Wright et al.; galaxy SEDs from \cite{Silva98}]}
    \label{fig:SED}
\end{figure}

\cite{GHAT2} explored the expected spectral signatures of Dysonian structures (Figure~\ref{fig:SED}). The most obvious signature of the sphere is the emission near~10~$\mu$m for $T\approx 300$~K emission. The most obvious confounder for such emission is circumstellar or interstellar dust.

It has been argued that Dysonian structures might operate at very low
temperatures in order to optimize the Carnot efficiencies $\eta = 1 -
T_{\rm hot}/T_{\rm cold}$.  In order to improve the Carnot efficiency
for technology around a Sun-like star from~95\% to~99.5\% would mean
lowering the radiation temperature from~300~K to~30~\hbox{K}. If a
technology is optimized for efficiency, this argument goes, then it
might be expected to have a Wien peak longward of~10~$\mu$m, and be
undetectable in mid-IR surveys.

\cite{GHAT2} argued that this low temperature scenario would be unlikely.  The reason is that,
for fixed energy supplies, the relationship between the area for the
waste heat radiators and their temperature is $A \propto
T^{-4}$. Improving efficiency by~4.5\% in the example above thus
requires an increase in radiating area (and, therefore, mass) by a
factor of $10^4$. Even at the minimum surface mass density required to
avoid ejection by radiation pressure for the Sun ($\sim 0.8$
g~cm${}^{-3}$), one Jupiter mass of material would be sufficient
for ``only'' a 30~au-diameter shell with a temperature of~70~\hbox{K}.  It would thus seem to be a rather robust prediction of Dysonian structures that the temperature that optimizes the balance between efficiency and engineering difficulty is significantly hotter than 30~\hbox{K}.

The second major signature of Dysonian structures is the far-IR to mid-IR color, which should be easily distinguished as being much bluer than dust. Dust generally has a wide range of temperatures and an extremely high surface area, and so radiates significant energy from~10~$\mu$m--200~$\mu$m, while, as \cite{GHAT2} argued, Dysonian structures do not radiate there.  

Dysonian structures are thus expected to occupy a relatively empty
region of color-color space, with very red mid-IR--visible wavelength colors but neutral far-IR--mid-IR colors.  A sensitive survey of the far-IR colors of mid-IR--bright point sources and galaxies would be an outstanding Dysonian-structure search strategy.

Without a sensitive far-IR survey, the problem of natural confounders is severe. In particular, there are 100 million point sources in the {WISE} all-sky catalogs, most of them active galactic nuclei.  Fortunately, we now also have optical magnitudes and parallaxes of nearly all of the {WISE} point sources from {\it Gaia}, allowing us to identify the much smaller number of sources with {\it total} luminosities characteristic of dwarf stars.

\subsection{Signals Embedded in AGN Output}\label{sec:anomaly.agn}

If there is a Type~III civilization in the Universe, it may have the
ability to send some device to the nearest AGN and modulate its
output, in some band; this idea is an extension of ideas of modulating
the outputs of masers \citep{Cordes1993} and
pulsars \citep{2015NewA...34..245C}.  For example, it might modulate the
ultraviolet emission of the AGN by modulating the temperature of the innermost part of the accretion disk.
The idea here is that the AGN already has the luminosity to be seen
throughout the Universe; all the ETI has to do modulate that signal,
analagous to how a transistor modulate the current from some power
source.  In this concept, in addition to the intrinsic AGN signal,
there could be some simple modulation pattern, e.g., luminosity
fluctuations with beats (with rests in between) grouped as consecutive prime numbers up to some value, and then repeated, over and over.  This would be something like lighthouse beacon, announcing ``look here.''

The search idea broadly, is to use the Virtual Observatory (or similar) to look for repeating patterns in AGN output, or place an upper limit on such embedded signals.  
If such a beacon were ever found, one would know where on the sky to look for other, weaker signals that are much weaker, but have a much higher bit rate, to convey more information than just, ``We're here.''

\section{Future Directions for Searches at Radio Wavelengths}\label{sec:futureradio}

As highlighted in \S\ref{sec:technosigs.comm.radio}, radio frequencies provides one of the best possible ways to communicate across large interstellar distances. An advanced ETI might intentionally choose to transmit a powerful beacon or may ``leak'' radio emission as a byproduct of their own activities.  Such radio signals are plausibly detectable across large distances and hence should be considered for all future technosignature surveys. Although substantial efforts have been invested in the field of radio technosignature surveys over the past 50~years, it has only just scratched the surface by covering a very limited multidimensional parameter space \citep{Haystack}. For example, almost all of the earlier radio SETI surveys have focused on searching for narrowband signals (i.e., bandwidths $\le 1$~Hz spectral occupancy) drifting due to relative motion between the transmitter and receiver. Here, we highlight some of the novel signal types that could open new parameter space for future radio technosignature surveys. 

\subsection{Non-Physical Signals}\label{sec:negativedm}

\begin{figure}[tbh]
 \centering
 \includegraphics[width=0.97\textwidth]{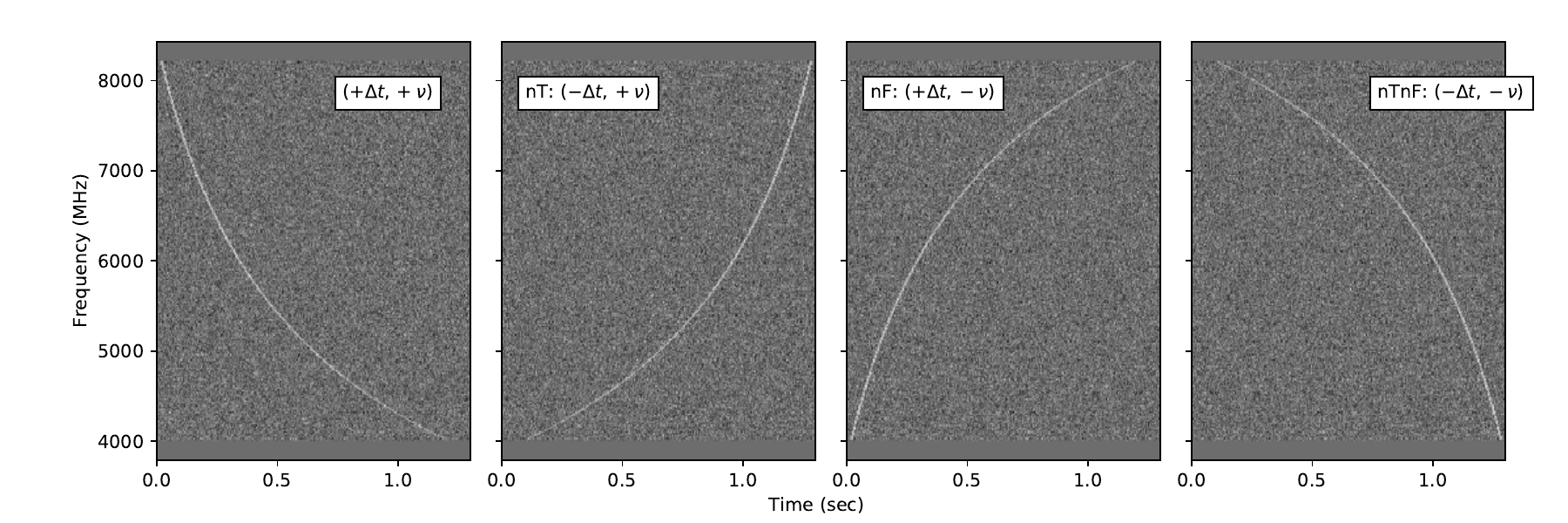}
 \vspace*{-1ex}
  \caption{Four different types of transient signals, three of which represent potential technosignatures.
  All panels show dynamic spectra of dispersed pulses as function of time (abscissa) and frequency (ordinate).
  The left panel shows a naturally dispersed pulse, occurring due to its propagation through the ionized interstellar medium.
  The other three panels show pulses that have dispersions in time and frequency that are not expected naturally and that may represent examples of signals that would be evidence for a technosignature.  
  The Breakthrough Listen SPANDAK pipeline searches for all of these signals.
   \citep[Adapted from][]{2021AJ....162...33G}.}
   \label{fig:negativedm}
\end{figure}

A radio pulse traversing the interstellar medium undergoes a frequency-dependent dispersion, with higher frequencies arriving at the observer earlier than lower frequencies.  This effect was recognized at the time that pulsars were discovered and formed part of the evidence that led \cite{Hewish68} to conclude that pulsars are sources outside of the Solar System; compensating for this dispersive effect is a standard process in searching for radio pulsars.  A signal with ``negative dispersion,'' i.e., the lower frequencies of a pulse arrive before the higher frequencies \citep{2010AcAau..67.1342S}, would challenge standard models for interstellar propagation at radio wavelengths and could constitute powerful evidence for a technosignature.
Figure~\ref{fig:negativedm} illustrates how a ``negative dispersion'' or other non-physical signals at radio wavelengths could appear in technosignature searches.

Further expanding the types of signals that are searched to include more such non-physical signals would open up a new uncharted parameter space. 

\subsection{Deep Neural Networks and Modulation Classifications}\label{sec:dnn}

The biggest challenge in radio SETI is to discriminate between truly ETI signals from signals of anthropogenic origins. One of the methods adopted in order to discriminate between ETI and likely anthropogenic signals is to conduct on- and off-source observations towards targets of interest.  The motivation is that ETI signals would appear only while observing the target of interest (``on''), while anthropogenic signals could occur in many directions (``off'').  Implicit in this method is that the ETI signal is ``on'' over a duration that is longer than on-off switching time \citep{1991ApJ...376..123C,1997ApJ...487..782C}, or the on-off switching time must be designed so that this assumption is not violated.

However, due to the growing complexity of terrestrial interference and their sporadicity, such techniques may not always provide a clean method to scrutinize large numbers of false positives.  Machine Learning (ML) has opened new opportunities to identify truly rare signals from the plethora of interference seen during typical observations. For example,
\cite{8646437,2018ApJ...866..149Z} described and applied different machine learning approaches toward the identification of technosignature candidates.  

\begin{figure}[tbh]
    \centering
    \includegraphics[width=0.97\textwidth]{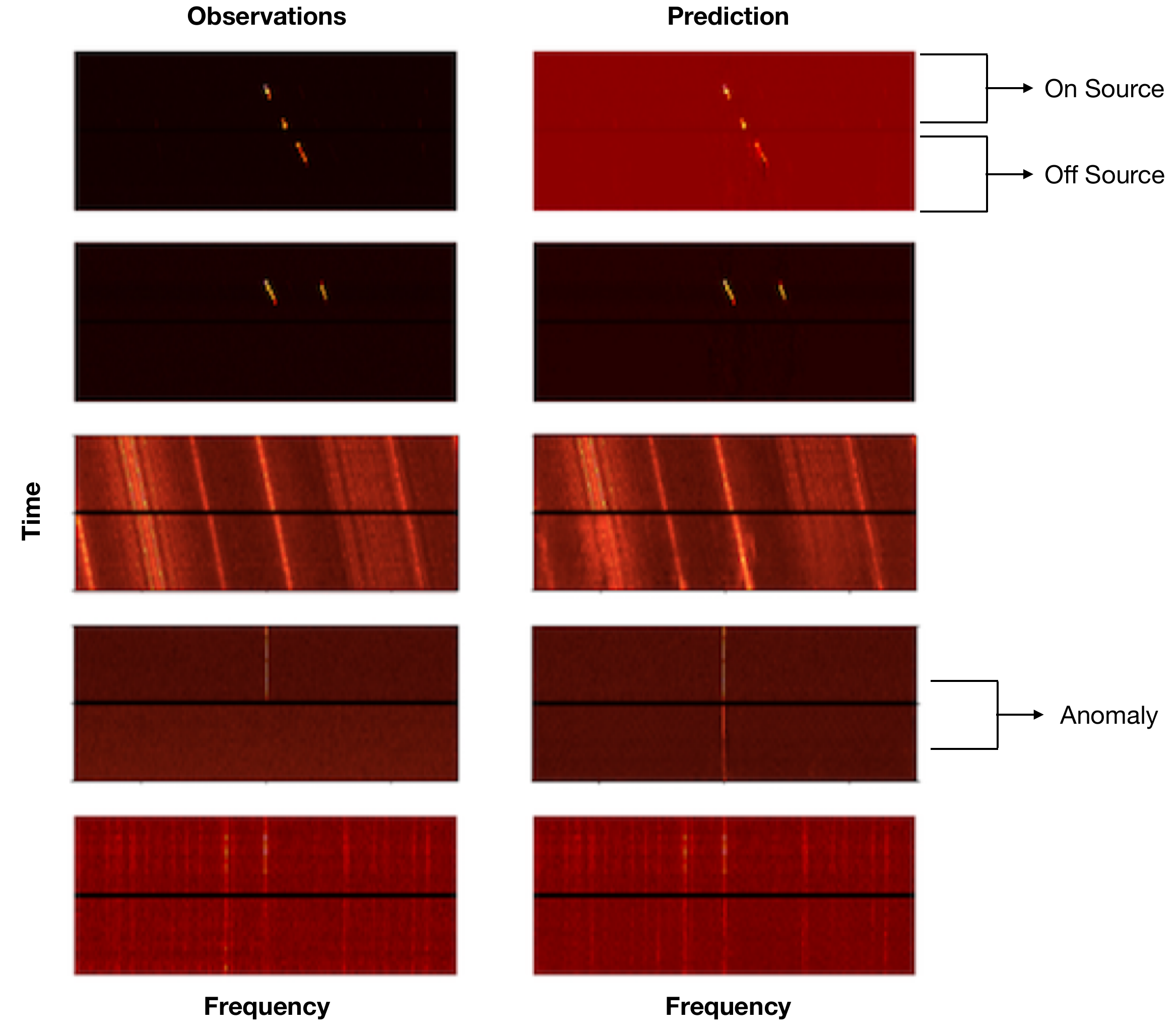}
    \vspace*{-1ex}
    \caption{Waterfall plots from five sets of on-off source pair observations with the Green Bank Telescope (GBT) that were used in a machine-learning approach toward technosignature identification.
    The left column shows the GBT observations, and the right column shows the prediction from a convolutional neural network (CNN).  Each panel shows the on-source observation or prediction stacked on top of the off-source observation or prediction, as indicated in the top row.
    \citep[Adapted from][]{2018ApJ...866..149Z}
    \label{fig:zhang_waterfall}}
\end{figure}

\cite{8646437} described an adversarial convolutional long short-term memory (ConvLSTM) network that is capable of predicting sets of on-off source pair observations. 
\cite{2018ApJ...866..149Z}
trained a convolutional neural network (CNN) on a large quantity of observations conducted with the Breakthrough Listen program from the Green Bank Telescope (GBT); Figure~\ref{fig:zhang_waterfall} shows a sets of on-off source pairs from those observations.  While no technosignatures were identified, 72 new FRBs were found, consistent with the workshop's philosophy that technosignature observations should be structured to provide astrophysical results even if no technosignatures are found.

Although \cite{2018ApJ...866..149Z} discussed only narrowband signals, they also indicated that a similar technique can be used for other types of signals. One such example of signals of interest are signals with inherent modulations.
Modulation schemes are methods of encoding information onto high-frequency carrier waves, making the transmission of that information more efficient.  Most of these methods modulate the amplitude, frequency, and/or phase of the carrier wave.  Such modulated signals are used routinely for terrestrial purposes, including for commanding interplanetary spacecraft (and the Voyager Interstellar Mission).
Such signals are of interest for detecting leakage radiation, but they are unlikely to be detected using standard filter-based approaches.

There are on-going efforts to classify various modulation types using CNNs and deep neural networks
\citep[][]{2018ISTSP..12..168O,2019Senso..19.4042Z}.
We emphasize that we are not assuming that ETI will transmit signals with any of humanity's current modulation schemes, only that they represent an energy-efficient approach to communicating information via radio signals and tools exist to try to detect such modulated signals in radio telescope observations \citep[e.g.,][]{2020RAA....20...78L}.

\subsection{Radio Technosignature Surveys with Interferometers}\label{setiinterferometers}

In the preceding decades, radio SETI has been dominated by observations
with single radio telescopes, or pairs of single radio telescopes for
the purposes of interference rejection \citep{1997abos.conf..633T}.
The Allen Telescope
Array \citep[\hbox{ATA},][]{2004ExA....17...19D} was an effort to
design an interferometric radio array explicitly for the purposes of
technosignature searching and characterization.  Interferometric
arrays offer multiple approaches toward interference rejection, and a
renewed ATA combined with commensal programs on large arrays such as
the Karoo Array Telescope (MeerKAT) and the Very Large Array (VLA)
offer the potential for much more sensitive observations and new
parameter space \citep{2021PASP..133f4502C,2021aoo..confE..13T,2021aoo..confE..14N}.

\section{Searching for Artifacts in the Solar System}\label{sec:artifacts}

The workshop participants identified that searching for artifacts within the Solar System would be a likely profitable approach.  Artifacts within the Solar System are a plausible extension of current (human) technology, and they could result either from intentional or unintentional actions of an extraterrestrial civilization or civilizations.  This section highlights two areas of considerable interest among study participants, but it is far from a complete list.  Other topics discussed included searching for objects with anomalous colors or albedoes.

\subsection{Searches for Artificial Structures or Objects on Planetary Surfaces}\label{sec:surfacesearch}

While \cite{davies2013searching} describe leveraging lunar satellite imagery as a technique to perform a technosignature search on the surface of the Moon, no algorithmic anomaly search has ever been executed on such data. Deep learning algorithms have been applied to this data in support of crater identification \cite[e.g.,][]{silburt2019lunar} but not (yet) for anomaly detection. 

There are multiple motivations for conducting a search for artificial
objects on or just underneath the surface of the Moon. A potential
limitation, however, is that it can be hard to evaluate the
completeness of an unsupervised anomaly detection algorithm.  In this
specific case, however, there is a set of almost forty
``ground truth''  artifacts, artificial (human-made) technology on the
surface of the Moon: the six \textit{Apollo} landing sites, the four
intentional impact sites of the Saturn~V launch vehicles from
the \textit{Apollo} missions, five Surveyor sites, four Ranger sites,
six landing sites from the Luna missions, and two Lunokhod rover
sites.\footnote{
\url{http://www.lroc.asu.edu/featured\_sites}
}
The results of an anomaly detection algorithm can be compared to what is known about these sites and images to assess its performance.
 
\begin{figure}
\centering
\includegraphics[width=0.47\textwidth]{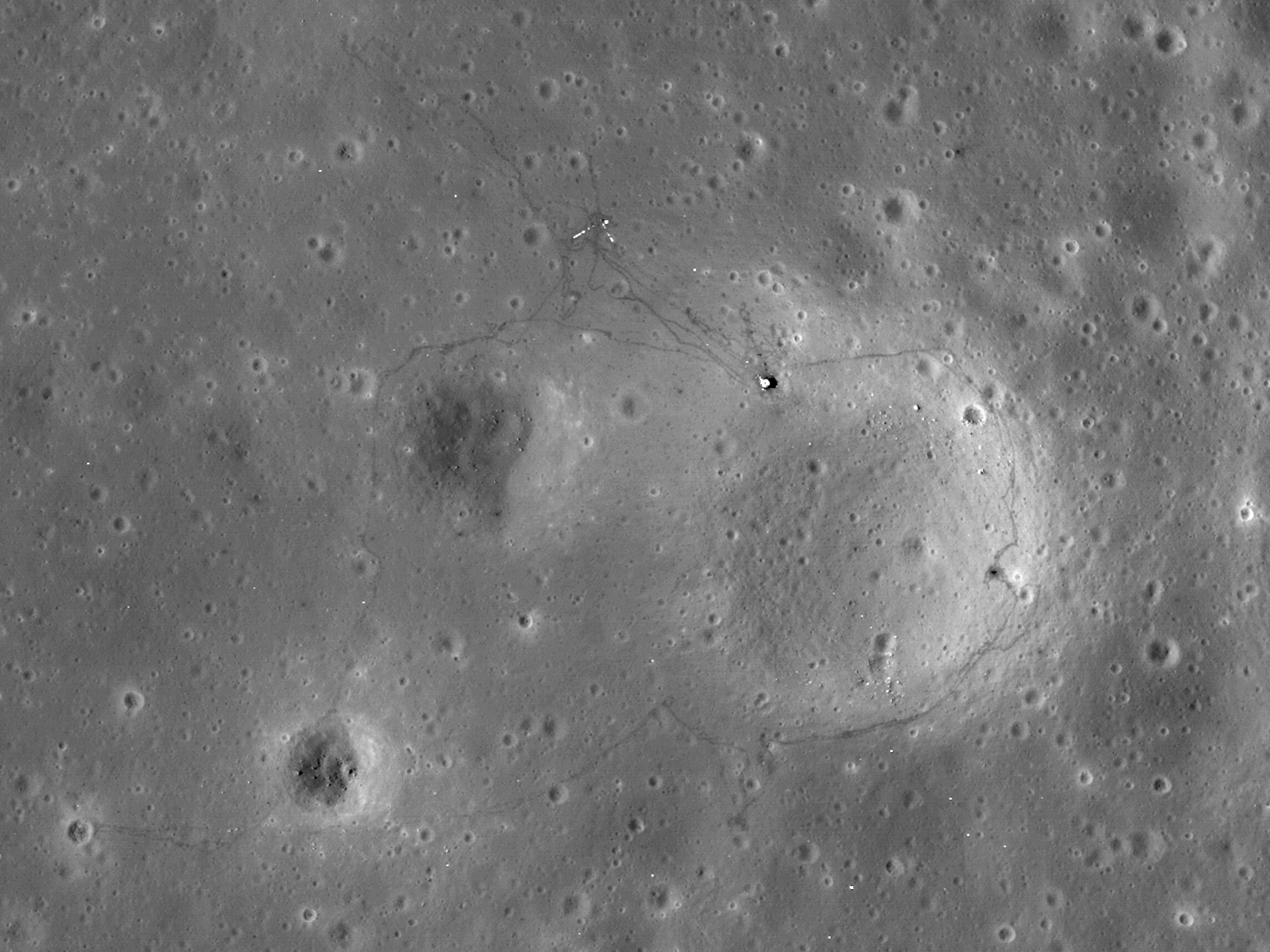}\hfil%
\includegraphics[width=0.47\textwidth]{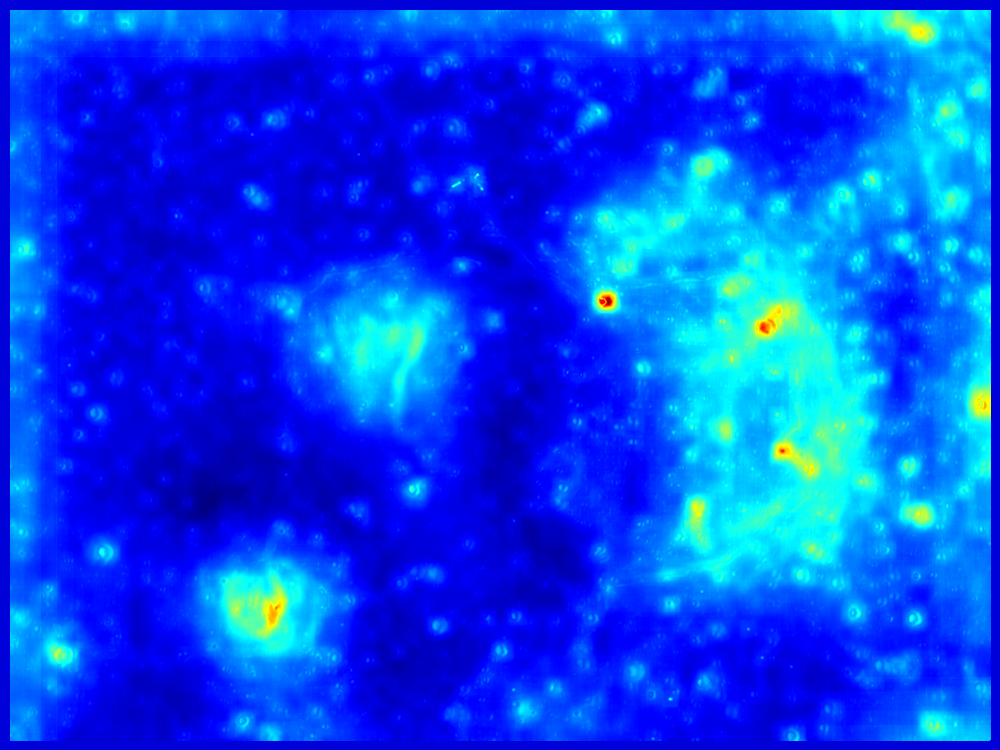}
\caption{%
(\textit{left}) LRO Narrow Angle Camera (NAC) image of
the \textit{Apollo~12} landing site.  Clearly visible are the rover
tracks and the astronaut foot paths.  The Intrepid Lunar Exploration
Module is just above the center of the image, and the \textit{Apollo}
Lunar Surface Experiment Package (ALSEP) is the \texttt{L}-shaped
object above and to the left of the Intrepid.  (NASA Goddard/Arizona State University)
(\textit{right})~Salience map covering the \textit{Apollo~12} landing site.  Red pixels have a relatively high salience, or should be ones that stand out to the human eye, while blue pixels have a relatively low salience.  The Intrepid lander is among the pixels having a high salience.  The ALSEP appears as a set of moderate salience (cyan colored) pixels.}
\label{fig:lroapollo}
\end{figure}

There is a substantial database of lunar images, provided by the Wide Angle Camera (WAC) and two Narrow Angle Cameras (NAC) on-board the Lunar Reconnaissance Orbiter (LRO) \citep{chin2007lunar}.  The former provides a spatial resolution on the Moon of~100~m per pixel (at visible wavelengths) and the two NACs provide spatial resolution approaching 0.5~m per pixel.  Figure~\ref{fig:lroapollo} illustrates the ``ground truth'' provided by the \hbox{LRO/NAC}. The datasets provided by the NACs include not only static images of the surface, but also time series of these images (often taken at different angles and illuminations) and stereoscopic topography of the lunar surface derived from correlating the images from the two NACs \citep{robinson2010lunar}. All three products could be potential inputs for anomaly detection and change detection algorithms such as those described below.

During the workshop, in order to illustrate the potential of machine learning techniques for searching for technosignatures on planetary surfaces, two approaches were discussed.  These approaches and algorithms are not intended to be exhaustive but to stimulate further work.

In order to conduct an assumption-free search for artificial technology, one approach is to apply the Isolation Forest \citep{liu:isoforest08}, a popular anomaly detection method from \texttt{scikit-learn} \citep{scikit-learn}, an open-source machine learning toolkit.  The Isolation Forest has the advantage of being computationally inexpensive and capable of finding global outliers and local outliers in arbitrary subspaces of the data.  The proposed 
experiment would be to extract features from postage stamps that are tiled across the image.  These features would characterize shape, texture, and colors within the stamp, from which an unsupervised search using the entire data set to identify candidate anomalies could be performed.  Those anomalies would be organized by similarity and compared to the set of artificial structures known to exist in these images in order to determine a missed detection rate. Analysis of candidate outliers that are certain to be non-artificial features of the Moon would determine the false positive rate.

Candidate outliers would be identified by one of two methods:\\
\begin{tabular}{p{0.45\textwidth}cp{0.45\textwidth}}
 \begin{itemize}
    \item Cut images into small tiles;
    \item Find images that are most discrepant; 
    \item Investigate;
 \end{itemize}
& or &
 \begin{itemize}
    \item Find anomalous pixels;
    \item Cluster; 
    \item Investigate.
 \end{itemize}
\end{tabular}

A second approach would be to identify image patches that are anomalous with respect to their spatial context.  In this approach, a sliding window is used to compute the {\em visual salience} at each pixel, which quantifies the degree to which a given region ``stands out'' to the human eye~\citep{wagstaff:landmarks12}.  Figure~\ref{fig:lroapollo} shows the salience map obtained for the \textit{Apollo~12} landing site.  The top four objects detected with this method include the Intrepid lander, a crater with heavy astronaut tracks (lower right), and two other craters; The ALSEP installation is also highlighted, albeit at a lower salience.  This test illustrates the potential of unsupervised searches to identify a small fraction of the lunar (or other planetary) surface for further investigation.

\subsection{Outliers in the Orbital Parameter Spaces}\label{sec:orbital}

Active probes may be able to be identified or distinguished from asteroids by anomalous or unexpected changes in their orbital parameters.
The simplest model is that the orbit of an asteroid is affected only by the gravitational force of the Sun and gravitational perturbations from planets and other asteroids.
Even this simple model, however, is known to be inaccurate in at least some cases, due to \emph{non-gravitational forces}.  
Two examples of non-gravitational forces would be collisions and the Yarkovsky effect, the latter being (subtle) changes in an object's orbit due to its anisotropic thermal radiation \citep[e.g.,][]{2020AJ....159...92G}.
Detection of orbital changes resulting from either of these non-gravitational effects can reveal insight into asteroid populations or compositions.

Extreme changes in the orbital parameters of an object, however, might
not be readily explained by either of these effects, motivating
consideration of the object being an active probe.  A much-discussed
case at the workshop was that of an active probe conducting a
propulsion maneuver.\footnote{
There is one instance in which the Wind spacecraft was classified
mistakenly as a near-Earth asteroid.  During a Goldstone Solar System
Radar (GSSR) observation, it performed a trajectory course maneuver,
producing a change in its radar return and orbit properties that were
consistent only with it having an active propulsion system.}
As the census of asteroids continues to increase, the opportunity to
identify such extreme orbital changes may also increase, particularly
if the growth in the asteroid census is accompanied by sufficient
orbital parameter determinations.

However, a confounding factor is the extent to which orbital parameters are determined well enough to permit the conclusion of an anomalous change.  In the worst case, asteroids can be ``lost,'' that is, initial observations are not sufficient to constrain the orbit much into the future and subsequent observations fail to recover (redetect) the asteroid simply because of the large uncertainties in the orbital parameters.  In more moderate cases, distinguishing between an anomalous change in orbital parameters and large uncertainties in the original orbital solutions may be challenging. 

A related possibility is whether there are objects where they are not
supposed to be.  For instance, there are areas of planetary orbital
resonances in the Solar System, due to mutual gravitational
attractions.  Asteroids are not expected in these areas as they should
be chaotically perturbed and ejected from those orbits on time scales
much shorter than the age of the Solar System; a notable example of
such locations are the Earth's Trojan locations, the Sun-Earth~L4
and~L5 points \citep{2023arXiv230211086Y}.  
Objects in those orbits, especially if they have other anomalous characteristics, could be worthy of further study.

Thus, in the spirit of the workshop, measuring and maintaining high-quality orbital parameter determinations for the census of asteroids would enable at least astrophysically-interesting detections of collisions or other non-gravitational forces, if not the identification of active probes.

\subsection{A Corner Reflector Floating in the Solar
              System}\label{sec:anomaly.reflector}

A corner reflector (sometimes called a retroreflector or corner cube)
reflects radiation directly back in the direction from which it
originated. This handy property has many applications, e.g., lifeboats
often carry them, and U.{}S.\ astronauts placed corner reflectors on
the Moon to permit precise laser ranging to the Moon, which is an
ongoing activity \citep{2019JGeod..93.2195M}.

An advanced civilization could have placed one or more corner
 reflectors in orbit in the Solar System with the idea that, at some
 time in the future, a civilization capable of searching for it might
 arise in the Solar System.  For an ETI wishing to leave such a
 ``calling card,'' an orbiting corner reflector would present several
 attractive features:~(i) It is relatively wavelength-agnostic, as
 it works for any wevelength much smaller than the reflector but not
 so short as to photoionize the atoms in the reflector, and therefore
 the ETI would not have had to guess the wavelengths at which a future
 civilization might transmit; (ii)~The corner cube ``plays back'' any
 message that was sent, which would be unmistakable evidence for the
 former presence of an \hbox{ETI}; (iii)~It requires no power; and
 (iv)~For a reflector in orbit, the ETI would not have had to worry
 about the reflector getting covered with dust, such as would happen
 if it were placed on a surface, though it may still be susceptible to
 dust impacts, as the \textit{JWST}\footnote{
 \texttt{https://blogs.nasa.gov/webb/2022/11/15/nasa-webb-micrometeoroid-mitigation-update/}}
 has been.
 
In order to search for a corner reflector (or perhaps a group of
reflectors oriented in different directions), it would be necessary to
scan the sky at some wavelength, with some clearly identifiable
signal, and search for the same signal reflected back, though at much
weaker amplitude. One possiblity is an all-sky search, but a first
strategy would be to search in some obvious places, such as in a
slow orbit around some moon, or a highly elliptical orbit the  major
axis of which points in some preferred direction, such as towards $\alpha$~Centauri.  Searching a few such ``preferred locations'' might provide a good proof of principle for some wider search.

Of course, a corner reflector by itself gives no information except that 
some ETI exists (or existed), visited the Solar System at some time in
the past, and left behind this calling card, but, if an 
ETI had gone to the trouble of leaving a corner reflector, it also
might have attached some object or device that transmits some
decipherable message, which an astronaut or robotic spacecraft could bring back to Earth.

\section{Reporting of Results and Follow-up Studies}\label{sec:onreporting}

\cite{Bradbury11}, \cite{GHAT1}, and \cite{Teodorani14} echoed the
perspective of \cite{dysonletters} that searches for Dysonian
structures could not, by themselves, discover conclusive proof of
extraterrestrial technology but that any anomalies discovered would be
good targets for more dispositive forms of SETI (such as searches for
radio or laser communication), allowing those searches to be more
focused.  \cite{Bradbury11} generalized the approach of \cite{dyson60} to include searches for any kind of extraterrestrial technology {\it other than} deliberate communicative transmissions, in what they call ``Dysonian SETI'' \cite[and others call ``artifact SETI,'' see also][]{WrightTaxonomy,WrightAdHoc}. Similarly, general anomaly hunts will be expected to turn up a huge menagerie of extreme but natural objects, among which some fraction might be indicative of technosignatures.

This approach is consistent with NASA's strategic plan for the search
for life, which states: ``There is no single measurement or experiment
that will definitively reveal the presence of extant or past life on a
body in our solar system or a planet around another star.'' Rather,
many measurements are required, with each increasingly restricting the
set of possibilities and resulting in a ``Ladder of Life Detection.''
In a similar manner, one can envision a ``Ladder of Technology
Detection,'' in which results increasingly constrain alternate
possibilities.

\begin{figure}[tbh]
    \centering
    \includegraphics[width=\textwidth]{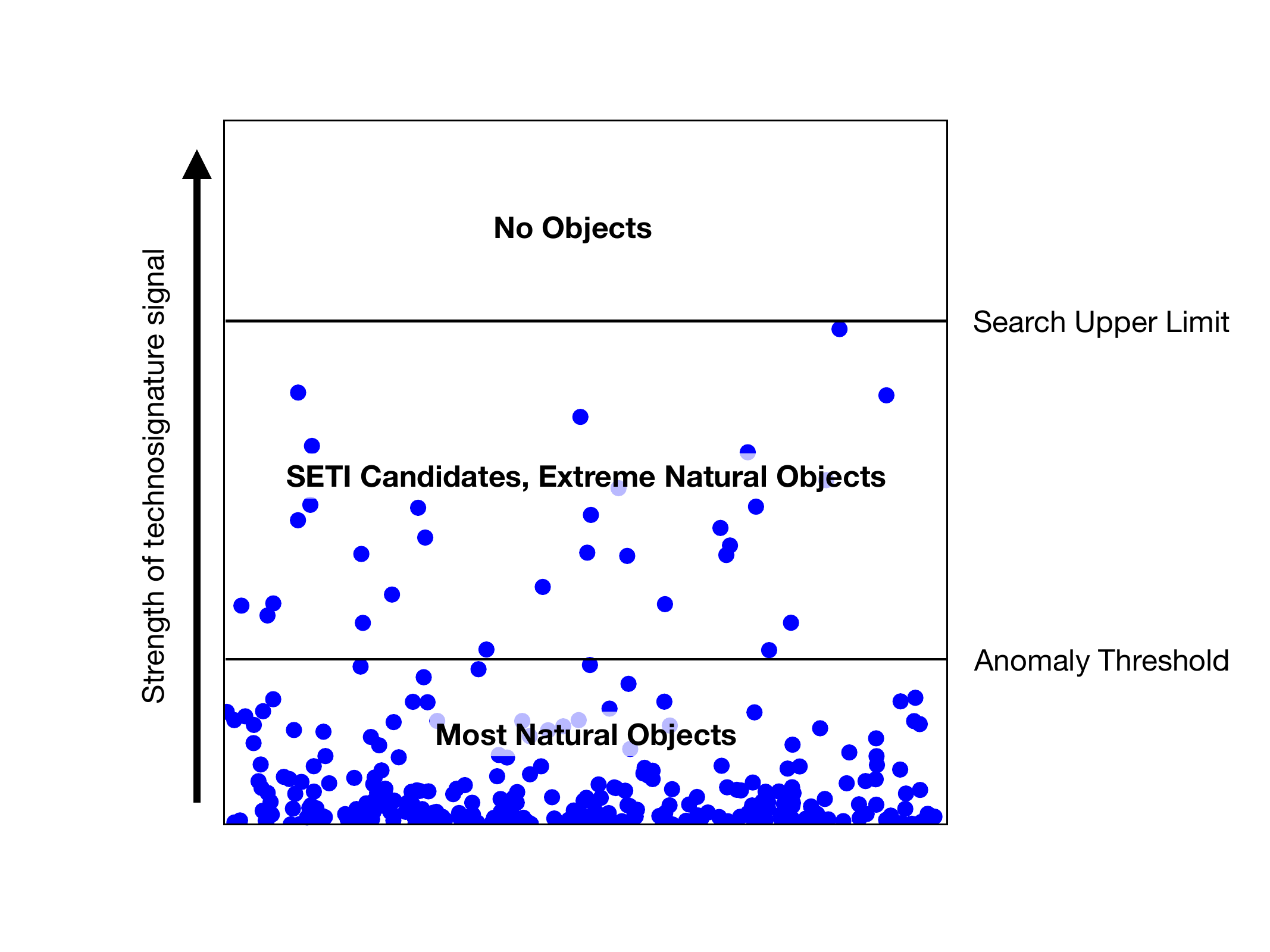}
    \vspace*{-11ex}
    \caption{We parameterize the strength of a technosignature by its observational consequence: for instance, we would use infrared excess for waste heat in a supervised search for that particular technosignature, or a metric of deviance in an unsupervised search. Because most technosignatures are not dispositive (i.e., they are consistent with a natural and instrumental confounders, such as dust and data artifacts), we must proceed by setting upper limits and identifying candidates for further study. If we rank all sources by that parameter, the most extreme source sets the upper limit in our search space: no sources have any technosignatures stronger that limit. We can then set our anomaly threshold to a value that produces a reasonable number of candidates. These are by definition the most extreme natural objects, and so are inherently worthy of study. They are also the strongest SETI candidates, and can enrich the target lists of dispositive forms of \hbox{SETI}, such as radio or laser searches.  As an object-by-object analysis of these sources establishes the nature of each one (i.e., determines that they are entirely natural or data artifacts) the technosignature upper limit moves down. In this way, SETI searches can be made robust and quantitative, and result in real progress while producing a simultaneous stream of ancillary science. \cite{Haystack} expand upon approaches to setting upper limits in high-dimensional SETI search spaces. See also \cite{2022ApJ...927L..30W}.
    }
    \label{fig:anomalies}
\end{figure}

Such an approach is consistent with Freeman Dyson’s ``First Law of SETI
Investigations'': ``every search for alien civilizations should be
planned to give interesting results even when no aliens are
discovered'' (``NASA and the Search for Technosignatures: A Report from the NASA
Technosignatures Workshop,'' \citeyear{2018arXiv181208681N}, p.~3).  That is, the
objects of a search should be phenomena that are inconsistent with
known natural sources. If the threshold for detection is set
sufficiently restrictively, only a small number of ``interesting''
candidates will result.  These  candidates necessarily will be either alien technology or the most extreme or anomalous natural sources in the sky, worthy of study in their own right. Figure~\ref{fig:anomalies} illustrates this search philosophy.

A particular challenge of unsupervised search techniques will be that there may be no parameter directly connected to technosignatures in which to rank objects, and so no way to express the result in terms of upper limits. In these cases, a search should be conducted with a specific goal in mind of identifying new classes of rare or unique astrophysical phenomena, and of specific follow up strategies that could be employed to elucidate their nature. Such studies are not explicitly searches for technosignatures, but might nonetheless uncover anomalies that prove, someday, to be technosignatures.

\section{Astrophysical Benefits}\label{sec:benefits}

The types of searches we have described, namely the use of objective data analytics tools to sift through vast quantities of data and look for ``anomalous'' signals, are precisely the type of studies done in the analysis of modern sky surveys, in searches for rare or new types of natural phenomena.  In that sense, searches for technosignatures can be seen as a collateral benefit of the systematic exploration of observable parameter spaces for astrophysical purposes.

It is also possible that new natural phenomena that were missed by the previous analyses may be found by the studies specifically designed to look for technosignatures.
An example of unexpected astrophysical discovery in the course of SETI is the \^G technosignature survey \citep{GHAT3}, which identified several classes of anomalous objects, including an unexpected nebula associated with a Be star, a previously uncataloged cluster of IR sources with no optical counterparts, and five apparently ``passive'' galaxies with unusually high UV emission. In next phase of the \^G search, \cite{GHAT4} identified the anomalous object now known as Boyajian's Star \citep{WTF,Wright16,WrightReassessment}. 

Previous, more dramatic examples of anomalous objects considered in a SETI context include the identification of radio-loud galaxies as potential hosts to communicative civilizations \citep{kardashev64} and the consideration of pulsars as planet-bound radio beacons \citep{Hewish68}.

\chapterimage{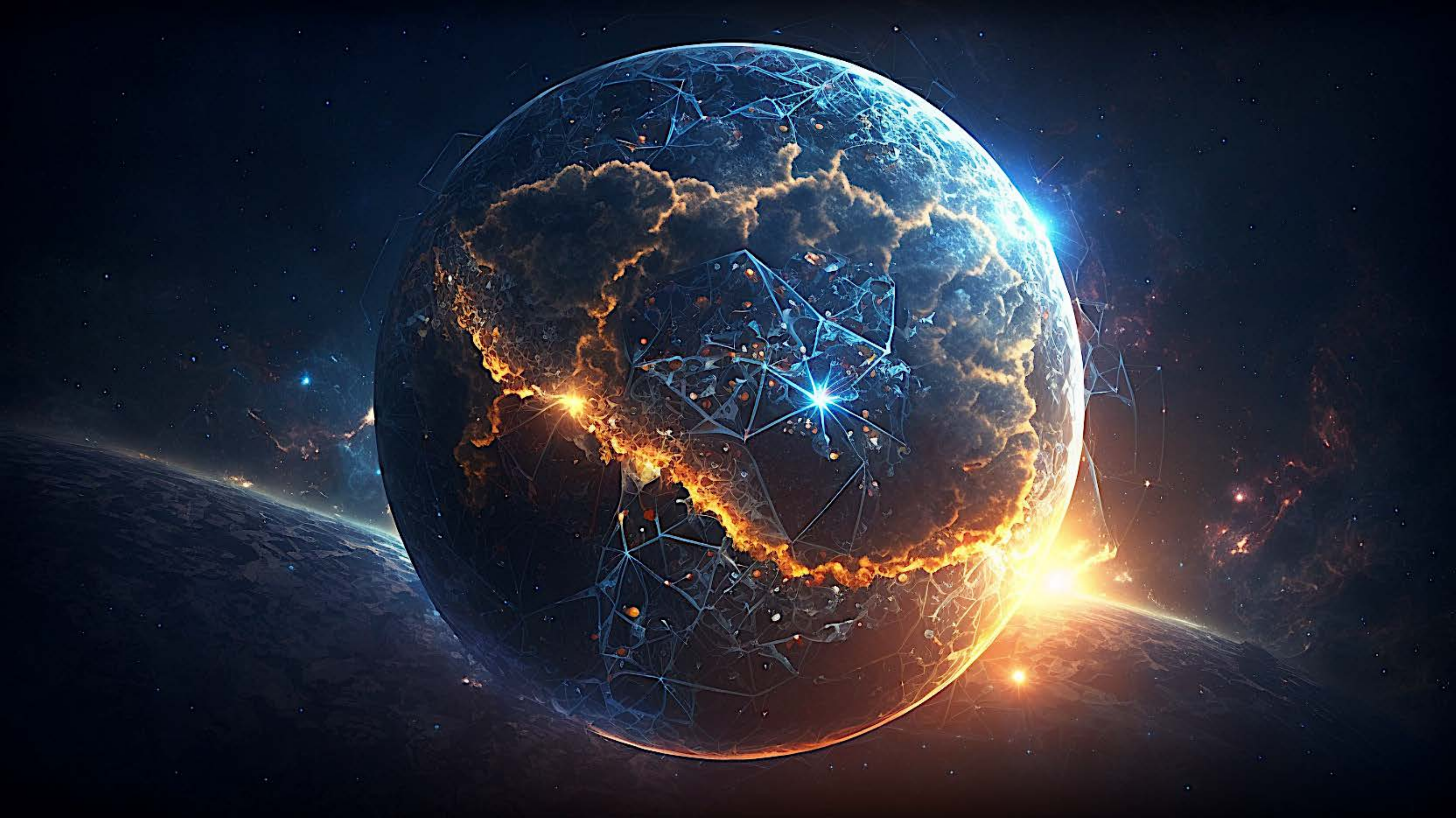}

\chapter{Recommendations and Future Opportunities}\label{chap:future}

\section{Community Building and Outreach}\label{sec:outreach}

Every scientific endeavor needs a certain amount of support, be it public, cultural, financial, or otherwise.  Astronomy has, in general, received much public support, and the exoplanet subfield has ignited the curiosity and imagination of the public like almost no other field has.  As a result of finding thousands of exoplanets and developing the technology to begin to characterize a subset of planets that have been found, astrobiology and the search for biosignatures, has gained popularity.  As noted earlier in this report, technosignatures is the next step in the search to answer the question, ``Are We Alone?''

At the moment, SETI and the search for technosignatures is met with mixed positive and negative reactions.  Many past papers and claims were erroneous, and have led to others making outrageous claims that were not scientifically verified. The same situation occurred when the search for exoplanets began.  It took many verified detections to mainstream the field of exoplanets.  It is important to have financial support (thus public support) to be able to search for technosignatures in earnest.  Therefore, what do we need to do to mainstream the search for technosignatures?

Right now, due to lack of steady funding, most technosignature work is commensal work with other science projects.  An ultimate goal for this type of technosignature search is to get every astronomer to routinely consider the possibility of technosignatures in any anomalies they may find in their data.  If we can get to this point, then we would have significant coverage of the sky in both spatial and wavelength domains.  This would be very powerful!

To mainstream the search for technosignatures, we need to take different approaches for different audiences.  In all cases, a public relations professional could help to clarify and socialize, both in the US and abroad, a brand for the entire field, rather than leaving it to individual institutions.  Breakthrough Listen already does this, so perhaps the community could build off of that.  It would help to educate all audiences on the changes that the field is making as it moves forward. For example, something like: ``SETI 2.0: From radio signatures to technosignatures through data science and AI'' may help people realize that we are using hard science and data to do more than look for little green men.

There are three main types of audiences for this field: public, scientists, and funding agencies.  In all cases, we need to clearly explain our results or non-results in a quantitative way.  This may help parts of the press to not ``cry wolf'' at every presentation of anomalies in the scientific literature, and help battle a perception among professional astronomers that SETI is simply an exercise in finding the ``aliens of the gaps.'' 

Below, we outline some suggestions on ways to mainstream the technosignature field for these three types of audiences.
\begin{description}
        \item[The public]\phantom{line}\\
        \begin{itemize}
        \item Modify the NASA Technosignatures Workshop Report into a popular form that can be referenced and understood by the curious public.
        \item Have SETI practitioners write articles about the
        philosophies of SETI for the popular technical press, such as
        Scientific American, Tom's Hardware, \hbox{WIRED}, The
        Atlantic, or Sky \& Telescope.
        \item Engage the technically-minded public beyond astronomy
        with articles and talks emphasizing, for instance, the
        implications of SETI success for humanity; connections with
        interspecies communication; explorations and imaginings of
        human futures; or the variety of intelligence and tool use among terrestrial species past and present.
        \item Engage or continue to engage the public with
        SETI-related citizen science projects.  Obvious existing
        projects with SETI-relevance include Disk Detective (as
        ``Dyson-sphere detective''); Galaxy Zoo (identifying
        anomalies); SETI@Home (already well established); or Planet Four (investigating human-identified surface anomalies).
\end{itemize}

\item[Scientific Peers]%
The perception of the field in the eyes of astronomers generally is crucial for several reasons.  Resources in astronomy are generally allocated via peer review, so our peers must have at least a rudimentary understanding of the field, its assumptions, past results, and potential future directions in order to fairly judge our proposals. Also, junior researchers may be reluctant to participate in a field that has poor standing among potential future employers.

A perhaps cynical approach to mainstreaming the field is to focus on
two qualities of research that tend to motivate broader scientific
research: resources and results.  The former means establishing
funding opportunities for researchers working in technosignature
search. A good start in this direction is the recent inclusion of
technosignatures as an allowed topic for proposals to the Exoplanets
Research Program (XRP), Exobiology, and
Astrophysics Data Analysis (ADAP) elements under NASA's
omnibus Research Opportunities in Space and Earth Sciences (ROSES) program.  The latter can be accomplished via successful research into anomaly detection that is driven by technosignature searches but results in new classes of astrophysical objects that attract widespread attention.

One difficulty is that the existing SETI literature is extremely heterogeneous in quality, the field having attracted the attention of many diletantes and hobbyists with little training in the underlying physics and other underlying disciplines.  An increase in quality in papers in the field, published in mainstream peer-reviewed journals, will raise the stature and profile of the field across astronomy. It may be worthwhile to compile papers into a special issue to attract attention.

The writing of good papers requires a good background in the field. It would be useful to have a review article of the field that defines the foundational documents and papers of the discipline, describes the underlying philosophies of technosignature search, and provides newcomers to the field with a guide to the extant literature. This should be part of a broader effort to create a standard curriculum and lexicon in the field that students can learn from as they enter the field. One such effort is that of the graduate course in SETI at Penn State University.

Astronomers in the field or entering the field should also make an effort to train young students, postdocs, and junior faculty in its method to ensure that it grows and has a bright future.  

Another way to involve astronomers is to invite them to lend their expertise to the problem of technosignature search at workshops such as this KISS workshop. By spending days learning the basics of the field and applying their specialization to the problem, they will appreciate its underlying rigor and see how they can contribute in the future.

Finally, the field needs to be a regular feature of the astronomical and astrobiological landscape, from funding proposals to telescope proposals to conferences to review articles to colloquia series.  Researchers should strive to present their research at IAU, AAS, and AbSciCon meetings; to propose for funding from NASA and the NSF; and to give talks internally and externally on the subject.

\item[Funding Agencies]%
The obvious way to increase funding of the field is to flood proposal calls with solid science proposals that prominently include technosignature searches.
In general, however, funding agencies respond to community perception of the merit of a scientific direction. This can be influenced via:
\begin{itemize}
        \item More papers in respected journals, especially astronomy journals.  
        \item A higher flux of proposals to telescope, fellowship, and funding calls. Ideally, there would  be enough that technosignature searches would get their own panel, virtually guaranteeing funding.
        \item Support from Decadal surveys
        \item Established, respected astronomers doing good science in the field, giving talks, writing papers, and communicating with the public.
\end{itemize}
\end{description}

\noindent%
Changing how people think about SETI and the search for technosignatures will take time, but we need to start now.

\section{Possible Future Space Missions and Ground-based Facilities}\label{sec:futurefacilities}

\subsubsection{A Modern Far-Infrared All Sky Survey}\label{sec:firtelescope}

IRAS and {\it AKARI} both performed far-infrared surveys of the entire sky down to a point-source sensitivity of approximately 1~Jy. This sensitivity was sufficient to provide \cite{carrigan09a} with a weak upper limit on nearly-complete Dyson spheres within 300~pc. The sensitivity of these missions was largely set by their small telescope diameters ($\sim$0.6 m).

A successor to these missions using modern infrared detector technology and with a larger aperture would allow for good discrimination between dust and Dysonian structures. Because FIR observations of faint point sources are generally background limited by the infrared cirrus, point source sensitivity scales as $D^2$. A 3~m-class telescope would thus have approximately 25 times the sensitivity of {\it AKARI} and {IRAS}, in addition to significantly better confusion limits in the Galactic plane and other crowded fields. 

The natural astrophysics case for such a mission is already quite
strong, and would be bolstered by the addition of a strong upper limit
on a robust technosignature to its science case. Indeed, the
3.5~m \textit{Herschel} telescope, although primarily a pointed
mission, performed a survey of~660~deg${}^2$ of the Galactic caps
(less than 2\% of the sky) achieving point source sensitivities better
than 10~mJy \citep{Smith17}. This data set is an excellent precursor
and proof-of-concept for an all-sky successor to \textit{Herschel}.

The recommendation from the \textit{Pathways to
Discovery} Decadal Survey report for a far-IR mission to be one of the
candidtes for a new Probe-class mission line, and NASA's subsequent
actions to execute that recommendation, mean that new limits or
detections may occur in the next decade or so.  At the time of
writing, whether a far-IR Probe mission concept will be selected as
the first Probe-class mission, and, if so, the characteristics of that
mission are not known.  Nonetheless, this recommendation from
the \textit{Pathways to Discovery} Decadal Survey report may produce
new opportunities.

\subsubsection{Planetary Science Missions}\label{sec:planetaryscience}

Humanity has been exploring our solar system with robotic spacecraft since the Mariner~2 flyby of Venus in~1962, and, though incredible discoveries and advances have been made, much remains to be explored. Here we illustrate the potential for leveraging past, present, and future solar system missions in an effort to reveal any possible signs of technosignatures within our own planetary neighborhood.

First, it is important to consider the possibility that the signal or
 message has already been collected and resides in the vast wealth of
 data returned from missions that have explored our solar system. The
 Planetary Data System (PDS)\footnote{
 \url{https://pds.nasa.gov/}
 }
 ontains all of the data, from all of the instruments, from all NASA Planetary Science missions. The PDS includes specific archival nodes dedicated to atmospheres, geosciences, cartography and imaging sciences, planetary plasma interactions, ring-moon systems, and small bodies. To the best of our knowledge, no comprehensive archival search has been conducted in an effort to reveal possible technosignatures.

Imagery sufficient to reveal a large ($> 10$~km) structure exists for parts of the Mercury, Venus, Moon, Mars, asteroids, Jupiter, Saturn, Neptune, Uranus, and Pluto systems. For small spacecraft artifacts ($< 100$~m) existing archives would likely only be useful for Moon and Mars. 

The Mars and Moon are both  well-imaged and mapped, at  near-global
resolutions of better than 100~m per pixel, with some regions imaged
at better than 1~m per pixel. While nothing has been legitimately identified as a possible technosignature, an automated search, implemented with machine-learning, is certainly warranted.   (Section~\ref{sec:surfacesearch} illustrates an example using Lunar Reconnaissance Orbiter imagery).

Radio transmissions from active probes could, potentially, have been recorded during the receptions and transmission of data to and from spacecraft and the Deep Space Network. Echos, such as those postulated by \cite{b60}, are an interesting possibility. Indeed, a search for anomalous echoes from our transmissions to operating spacecraft might be warranted. Frequencies of communication include S band (2 to 4 GHz), X band (8 to 12 GHz) and Ka band (27 to 40 GHz). The earliest and most widely used frequency is S band; X and Ka bands were implemented later, but have higher data rates. The PDS, however, does not store original, raw, transmissions, and thus may not serve as a useful resource for such signals. Raw data for specific instruments, such as magnetometers, radar, and atmospheric conductivity sensors (e.g., the Atmospheric Structure Instrument on Cassini Huygens), does exist and could be an interesting resource in the search for anomalous signals. Radar has the added benefit of potentially yielding strong signals from artifacts containing passive systems, such as corner cube reflectors.

Beyond the data archive, many existing and future missions offer opportunities for serendipitous science related to the search for technosignatures.

At present, NASA is operating two rovers on Mars (\textit{Curiosity}
and \textit{Perseverance}), one of which is in the process of caching
samples for potential return to Earth. The international community is also pursuing a strong program of Mars and Moon exploration: ESA and the Chinese National Space Administration (CNSA) will both launch rovers towards Mars during the same launch window cited above. The CNSA has established a successful lunar program that includes the Chang'e~3 and~4 missions, which landed rovers on the Moon in~2013 and~2019, respectively.

These missions serve an interesting ``dual use'' for
technosignatures:~(i) they enable direct observation of the surface
and could reveal meter- to sub-meter-scale surface artifacts that are not visible in remote imagery, and~(ii) as shown in~\S\ref{sec:artifacts}, these surface vehicles provide a proof-of-concept for search strategies.

Beyond the Moon and Mars, NASA has selected Dragonfly, a rotorcraft vehicle for
Titan exploration, as part of the New Frontiers Program. This mission
will provide an incredible direct view of Titan's surface, which is
otherwise concealed by its thick (1.5~bar) and hazy atmosphere. As a
site for surface technosignatures, Titan is perhaps interesting in
that its thick atmosphere and low gravity (1.352~m~s$^{-2}$) make for
a relatively benign entry, descent, and landing phase; parachutes and
a soft landing are possible. The ISP and $\Delta v$ challenges discussed in~\S\ref{sec:comm.probes} are reduced. Furthermore, Titan orbits a gas giant (Saturn) which may assist in capturing an incoming probe. Finally, depending on the timing and trajectory, Titan could also be the first such world encountered by a probe entering our Solar System, perhaps making it an attractive site for landing.

In the next few decades, landed spacecraft also hopefully will reach
the surfaces of Europa, Venus, Enceladus, Ceres, Triton, and perhaps
several others. Unfortunately, part of what makes these worlds
scientifically interesting is also a limiting for for
technosignatures: all of these worlds, as well as Titan, have surfaces
that are geologically young (tens to hundreds of millions of years in surface age) and thus would not likely be stable environments in which to park a long-lived, persistent probe that does not have some means for repositioning itself over the eons.

Considering orbiters and flyby missions that could contribute to our knowledge of potential technosignatures in our backyard, numerous missions offer interesting opportunities. 

The Juno mission is currently in orbit around Jupiter and though it
has completed its prime mission, this solar-powered spacecraft could
remain active in the jovian system for many years to come. Juno is
designed to study the interior structure of Jupiter and as part of its
payload it has a relatively simple camera, UV and infrared cameras,
and a microwave radiometer, all of which could help identify anomalous
artifacts. Of particular interest might be the Juno Microwave
Radiometer (JMW), which permits active sounding of the jovian
atmosphere at six frequencies: 21.9~GHz, 10.0~GHz, 5.2~GHz, and~2.6~GHz (all with 12$^{\circ}$) beamwidths) and 1.25 and 0.6 GHz (with 20$^{\circ}$ beamwidths). Such an instrument could potentially find Bracewell ``echoes'' from probes in the jovian system.

In the coming decade, at least two more spacecraft will arrive at Jupiter, the ESA Jupiter Icy Moons Explorer (JUICE) and the NASA Europa Clipper. These two spacecraft are focused on Ganymede and Europa, respectively, but will provide data on many of the large moons, and possibly some of the smaller moons. Serendipitous technosignature science utilizing the imagery, spectroscopy, and radar from these missions will be possible as all data will be made available rapidly through the \hbox{PDS}.

Lastly, Discovery-class Lucy mission has launched and, over the course
of its 12~year mission, it will study six Trojan asteroids. Trojans
asteroids lead and trailing Jupiter by~60$^{\circ}$.  In the near
future, the Discovery-class Psyche mission will launch for its target
of the unusually metal-rich asteroid 16~Psyche. 
All of the objects being studied by these two missions are interesting in the context of Solar System formation and planet formation. The Trojan asteroids in particular may be some of the oldest objects in our Solar System. Both the asteroid belt and the two regions of Trojan asteroids could be useful ``parking orbits'' for probes that have long ago entered our solar system. These missions will provide a unique glimpse with the possibility of serendipitous technosignature science.

\subsubsection{Deep Space Network Exploration of the Earth Transit Zone}\label{par:etz}

\begin{figure}[tb]
  \centering
  \makebox[0.66\textwidth][c]{\includegraphics[width=0.66\textwidth]{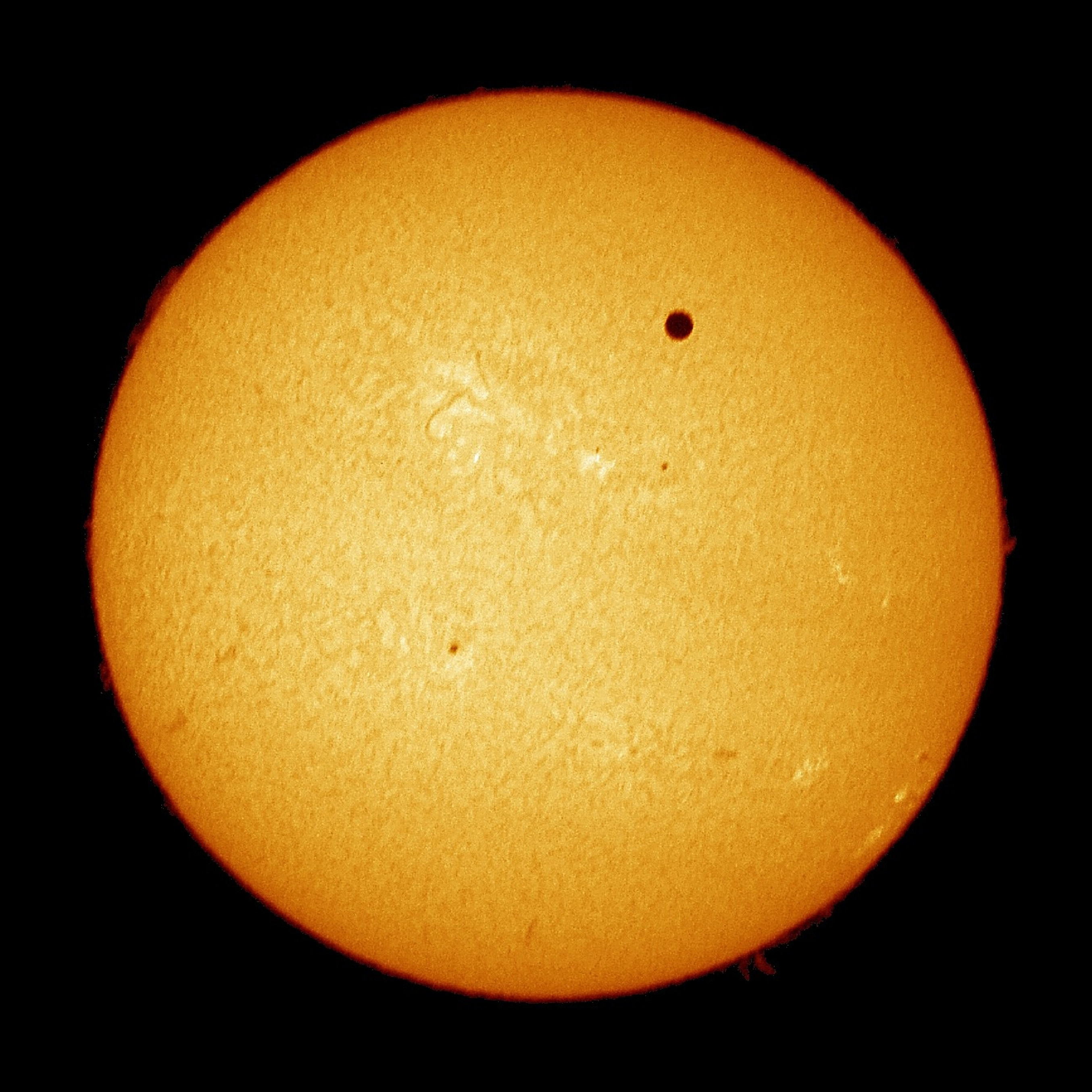}}\hfil%
  \parbox[b]{0.32\textwidth}{%
    \caption{The transit of Venus, as seen from Earth.  An extraterrestrial civilization located in the ecliptic plane should already know about the Solar System by observing Solar System planetary transits.  (Credit: ESA)}
    \label{fig:venus}
  \vspace*{1ex}
  }
\end{figure}

Any civilization located close to the ecliptic plane with the capability to conduct observations similar to those conducted by the Kepler mission should know about the existence of the solar system, due to being able to observe transits of one or more of the solar system planets (Figure~\ref{fig:venus}).  This region, termed the Earth’s transit zone (ETZ) by \cite{hp16}, is a few degrees wide and is predicted to be a prime region in which to expect that ET transmitters directed at Earth might be found---a civilization located in the ETZ that has decided to transmit or make a dedicated effort to contact Earth already knows of our existence.

\begin{figure}[tb]
\centering
\includegraphics[width=0.97\textwidth]{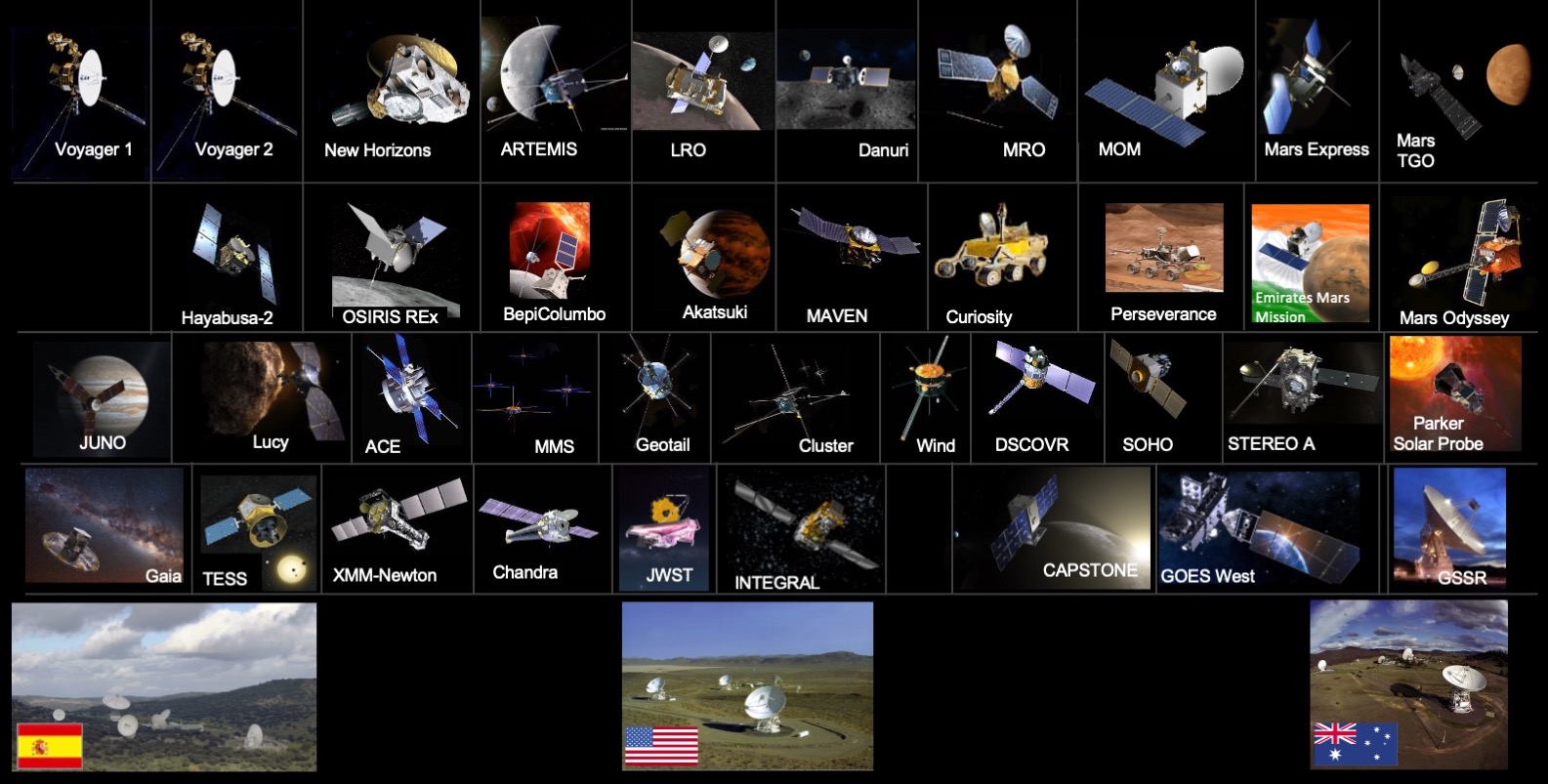}
\caption{Deep space missions enabled by
telemetry, tracking, and commanding from the Deep Space Network (DSN).
Many of these missions are in or near the ecliptic plane.  The typical
spacecraft telemetry uses only a fraction of the available bandwidth,
and searches for directed technosignatures within the Earth Transit
Zone could be conducted in a commensal manner by scanning the spectrum
not used for the spacecraft telemetry for signals.  Not only is this
approach efficient, as only a modest augmentation of the DSN would be
required, it would enable searches at frequencies that have not
commonly been searched in the past for technosignatures.  (Image
credits: \hbox{NASA}, \hbox{ESA}, \hbox{JAXA}, \hbox{ISRO}, \hbox{KARI}, \hbox{UAESA}) }
\label{fig:dsn}
\end{figure}

One potential approach to searching for directed technosignatures would be to monitor the \hbox{ETZ}.
The Deep Space Network (DSN) is the spacecraft tracking and communication infrastructure for NASA's deep space missions (Figure~\ref{fig:dsn}).  It consists of three sites, approximately equally separated in (terrestrial) longitude, with multiple radio antennas at each site.  At each site is a 70~m diameter antenna, with a surface suitable for observations as high as 25~GHz, and three to four 34~m antennas, with surfaces suitable for observations as high as 40~GHz.  (In actuality, approximately 1/3 of the missions for which the DSN provides telemetry, tracking, and command [TT\&C] are from other space agencies, such as the European Space Agency.)

Most of the time on the DSN antennas is spent either transmitting or receiving signals to deep space spacecraft, the vast majority of which are located in the ecliptic.  The standard telecommunications band has been 8.2~GHz to~8.6~GHz (X~band), but the band used for telecommunications with the \textit{James Webb Space Telescope} will be 25.5~GHz to~27~GHz (K~band) and there is an effort to move deep space spacecraft telecommunications to the band of 31.8~GHz to~32.3~GHz (Ka~band).  In all of these bands, the spacecraft downlink occupies only a small fraction ($\sim 10$\%) of the total received bandwidth.

Thus, there is an opportunity to conduct commensal searches for directed technosignatures.  A new backend for the DSN could scan the portion of the radio spectrum not being used for the spacecraft telemetry, searching for signals.  This approach is efficient, in that the DSN antennas are in some sense monitoring the sky already and potential data are not being processed, and it would involve conducting a search for technosignatures at radio frequencies that have not been commonly searched previously.

\subsubsection{Next-Generation Interplanetary Radar}\label{sec:radar}

Interplanetary radar studies have been conducted since the dawn of the Space Age, most notably determining the range to Venus \citep{mg61,1962AJ.....67..191M}, from which the astronomical unit could be determined to a precision that allowed interplanetary navigation.  Since that time, planetary radars have been used to study all of the terrestrial planets, many of the major moons, and hundreds of small bodies within the solar system.  Most of the current use of the world's planetary radars is for observations of near-Earth asteroids, both for scientific purposes and for planetary defense (most specifically, to obtain their orbits).

Beyond the fact that determining the orbits of potentially hazardous asteroids may affect our civilization's lifetime~$L$, one of the factors of the classic Drake equation, planetary radars could also be used to characterize potential interstellar probes.  As a simple example, objects with anomalously high radar cross sections might warrant further characterization to assess whether they have a substantial metallic composition, which might be an indication of a constructed object rather than a typical near-Earth asteroid.

While the current planetary radar infrastructure could serve this
purpose, a dedicated planetary radar facility would allow more
frequent studies of near-Earth asteroids and more opportunity for
discoveries about near-Earth asteroids as well as potentially an
interstellar probe.  In this respect, the recommendation from
the \textit{Origins, Worlds, and Life} Decadal Survey report for NSF
and NASA to conduct a study on ground-based planetary radar needs, and
the establishment of an NSF-NASA Deep Space Radar Assessment working
group, may represent an opportunity for such a facility to be constructed.

In the remainder of this section, we summarize briefly the world's
current planetary radar infrastructure, then summarize some of the
potential technologies that could be applied to produce a new radar
capability.

The world's planetary radar infrastructure is based on astronomical
and deep space telecommunications infrastructure.
The Goldstone Solar System Radar (GSSR), installed on the 70~m Deep
Space Station-14 (DSS-14) antenna at the DSN's Goldstone site, is part
of NASA's DSN.  The \hbox{GSSR} has a transmitter
power of~\hbox{450~kW} and is equipped with a sensitive receiver.
The Robert C.~Byrd {Green Bank Telescope (GBT)} is a sensitive
receiving antenna, though it does not have a transmitter
currently; there has been one demonstration of a (low-power)
transmitter on the \hbox{GBT} \citep{wha+22}.
The Southern Hemisphere Asteroid Radar Project (SHARP) consists of
the \emph{bistatic} combination of the 70~m Deep Space Station-43
(DSS-43) antenna at the DSN's Canberra Complex equipped with an 80~kW
transmitter and the Australia Telescope Compact Array (ATCA).  Not
shown is the 34~m DSS-13 antenna at the DSN's Goldstone Complex
equipped with an 80~kW transmitter, which can be used in a bistatic
manner with the \hbox{GBT}.  For the Southern Hemisphere, in some
cases, a 34~m DSN antenna at the Canberra Complex has been used in
place of DSS-43, and early tests used the Parkes Radio Telescope
rather than the \hbox{ATCA}; however, Parkes does not normally have a
receiver mounted that can receive at the frequencies at which DSS-43
(or other DSN Canberra Complex antennas) can
transmit.  \cite{2016AJ....152...99N} provide further technical
details and comparisons between the various facilities.
There have been recent demonstrations of radar observations of 
near-Earth asteroids by the bistatic combination of the 70~m DSS-63 at
the DSN's Madrid Complex serving as the transmitter and various
European radio astronomical antennas serving as the receivers.
Finally, while no longer operational, the 305~m-diameter Arecibo
Observatory was equipped with a powerful transmitter and sensitive
receiver.

The need for high sensitivity antennas and high transmitter powers is
driven by the radar equation
\begin{equation}
\mathrm{S}/\mathrm{N} \propto \frac{G_{\mathrm{RX}}P_{\mathrm{TX}}G_{\mathrm{TX}}}{R^4},
\label{eqn:radar}
\end{equation}
where S/N is the signal-to-noise ratio; $G_{\mathrm{RX}}$ and
$G_{\mathrm{TX}}$ are the gains of the receiving and transmitting
antennas, respectively; $P_{\mathrm{TX}}$ is the power of the
transmitter, and~$R$ is the range (distance) to the target.  For an
antenna of diameter~$D$, the gain is $G \propto D^2$.

As summarized in equation~(\ref{eqn:radar}), there are two key aspects
to a planetary radar system---transmitter power~$P_{\mathrm{TX}}$ and antenna gain~$G$.  A new radar facility could take advantage of developing technologies and potentially improve both capability and reliability in both of these areas.
\begin{description}
\item[Transmitter Power]%
  Fundamental to the GSSR are
  \emph{klystrons}\footnote{
  The Arecibo Observatory also had a klystron-based transmitter.}
  ---high-power, vacuum tube-based microwave
  amplifiers.  Klystrons operate by producing an electron beam that is
  then modulated to produce a radio frequency signal, often with a
  desired waveform.  While klystrons are a standard component of
  radars, some medical devices, and particle accelerators, the
  klystrons used by planetary radars are distinguished by their power
  levels, which can be orders of magnitude higher than all other
  applications.  As a consequence, they often have to be operated near
  the edge of instability, which can lead to relatively short
  lifetimes (of order a year).

  A potential technology development activity would be the development
  of even higher power but more reliable microwave amplifiers.  The
  GSSR obtains its full power by the phased combinations of the
  outputs of two klystrons.  One likely approach for higher power,
  more reliable microwave amplifiers would be more modular systems,
  consisting of many, lower power amplifiers the outputs of which are
  combined coherently.  Lower power amplifiers would be presumably
  more reliable, and a modular approach would offer the possibility of
  a graceful degradation.  A challenge with such an approach would be
  to maintain a high efficiency of coherent combination of the
  signals, as reflected power could damage the individual components.
  This challenge is particularly acute to obtain graceful degradation,
  as the loss of an individual amplifier would require a rapid
  response so that additional individual amplifiers are not damaged.

\item[Antenna Gain and Antenna Arrays]%
  Current single dish antennas are at such a scale that larger antennas
  are not a viable approach to obtain significant improvements in
  antenna gain.  For instance, no fully-steerable single dish radio
  antenna significantly larger than the GBT has since been constructed
  since the completion of the GBT in the 1990s.  The continued
  operation of these large single-dish antennas is therefore critical
  to a viable planetary radar capability.

  One potential approach for obtaining larger receiving
  gains~$G_{\mathrm{RX}}$ would be bistatic operations with future
  radio astronomical arrays.  For instance, in the southern
  hemisphere, the intermediate frequency component of
  the \textbf{Square Kilometre Array Phase~1 (SKA1-Mid)} is planned to
  be capable of operating at the transmitter frequencies of the DSN's
  Canberra Complex \citep{jl15}.  However, SKA1-Mid is planned to be
  sited in South Africa, while the DSN's Canberra Complex is in
  eastern Australia, limiting the amount of common sky visibility.  In
  the northern hemisphere, the \textbf{next-generation Very Large
  Array (ngVLA)} would offer both frequency coverage that overlaps
  with the GSSR transmitters and considerable mutual visibility with
  both \citep{bbmnl18}.

  Finally, while interferometric (``receive-only'') arrays are
  standard for radio astronomical telescopes and phased arrays are
  standard for terrestrial radar applications, phased interferometric
  arrays are not yet used
  for the transmit portion of planetary radar.  There have been
  limited experiments within the DSN of phasing transmit antennas to
  obtain higher effective transmit gains~$G_{\mathrm{TX}}$ (the DSN
  ``Uplink Array'').  These experiments have demonstrated the expected
  improvements in $G_{\mathrm{TX}}$ for coherent addition of antennas,
  but they have not been required to obtain the range and range-rate
  (Doppler) precision required for planetary radar.  Additional work
  would be necessary to obtain the required precisions, and additional
  antennas would be necessary to obtain comparable or larger transmit
  powers.  (For example, $5 \times 34$~m antennas each equipped with
  an 80~kW transmitter $=$ 70~m antenna equipped with a 500~kW
  transmitter.)

\end{description}

\chapterimage{CoverImages/pdf/Bibliography}

\phantomsection
\addcontentsline{toc}{chapter}{\textcolor{MidnightBlue}{Bibliography}}
\bibliographystyle{aasjournal}
\bibliography{bibliography}

\clearpage

\begingroup
\thispagestyle{empty}
\begin{tikzpicture}[remember picture,overlay]
  \node[inner sep=0pt] (background) at (current page.center) {%
    \includegraphics[width=\paperwidth]{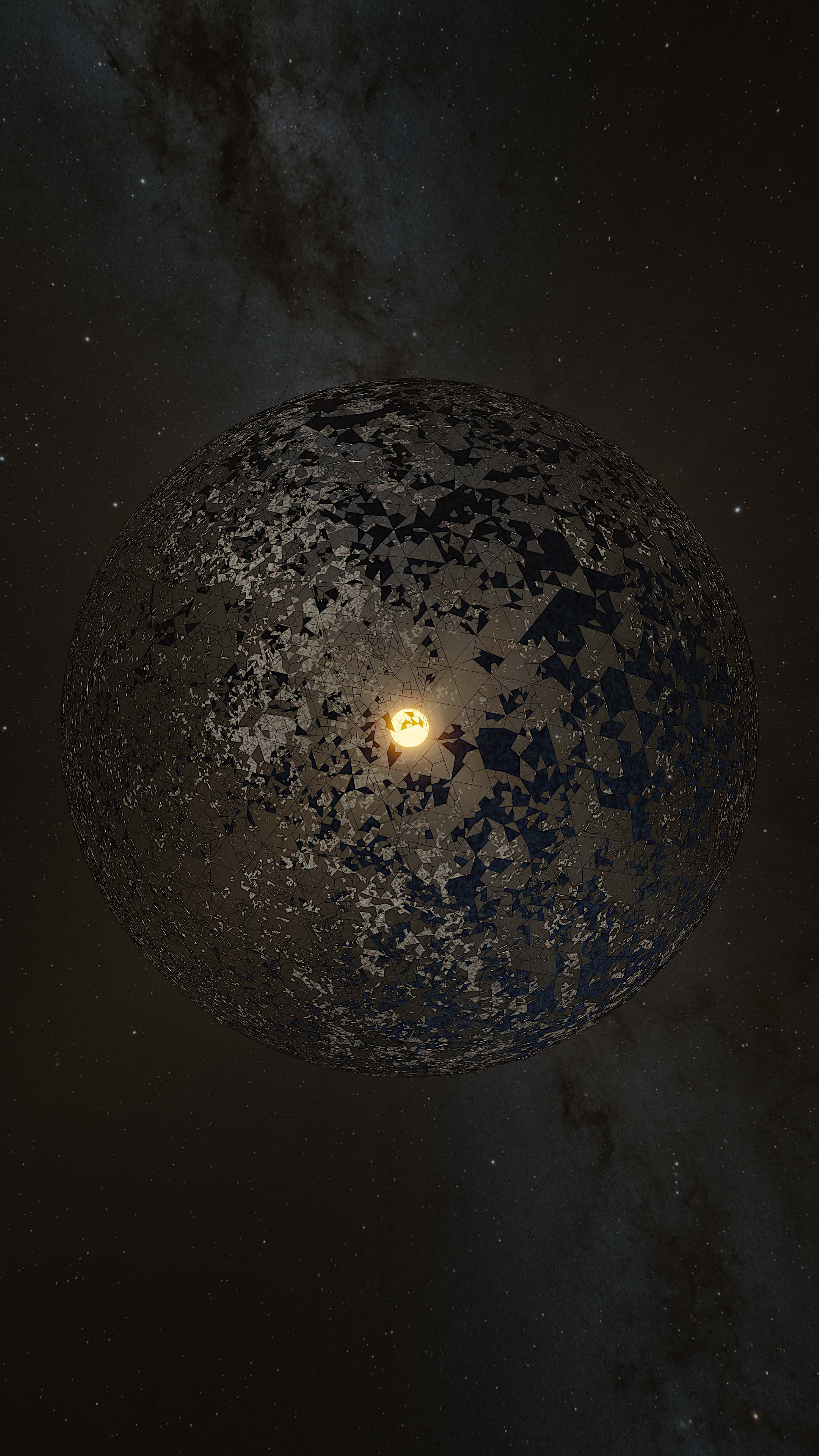}};
\end{tikzpicture}
\vfill
\endgroup

\end{document}